\documentclass{lmcs} 
\pdfoutput=1

\usepackage[utf8]{inputenc}
\usepackage{lastpage}
\lmcsdoi{17}{4}{13}
\lmcsheading{}{\pageref{LastPage}}{}{}%
{Dec.~30,~2020}{Nov.~25,~2021}{}


\usepackage{tikz}
\usetikzlibrary{arrows,shapes.geometric,shapes.misc,positioning,automata}
\usetikzlibrary{trees,decorations.pathmorphing,calc}
\usetikzlibrary{intersections}

\theoremstyle{plain} 

\usepackage{float}
\newfloat{snippet}{t}{lop}
\floatname{snippet}{Listing}

\usepackage{xifthen}
\usepackage{mathtools}
\usepackage{graphicx}
\usepackage{todonotes}
\usepackage{listings}
\usepackage{mathpartir}
\usepackage{booktabs}
\usepackage{placeins}
\usepackage{nicefrac}
\usepackage{hyperref}
\usepackage{cleveref}
\crefname{snippet}{listing}{Listing}

\definecolor{Green}{rgb}{0.0, 0.0, 0.8}

\newcommand{\metacomment}[1]{{#1}}

\sloppypar

\theoremstyle{definition}
\newtheorem{example}[thm]{Example}

\begin{document}

\title[Time Fluid Coordination]{Time-Fluid Field-Based Coordination through Programmable Distributed Schedulers}
\titlecomment{{\lsuper*}Extended version of paper ``Time-Fluid Field Based Coordination'',
appeared in ``Coordination Models and Languages'', 2020.
}

\author[D.~Pianini]{Danilo Pianini}[a]	
\address{Alma Mater Studiorum---Università di Bologna, Cesena (FC), Italy}	
\email{danilo.pianini@unibo.it}  
\email{roby.casadei@unibo.it}  
\email{mirko.viroli@unibo.it}  

\author[R.~Casadei]{Roberto Casadei}[a]	

\author[M.~Viroli]{Mirko Viroli}[a]	

\author[S.~Mariani]{Stefano Mariani}[b]	
\address{Università degli Studi di Modena e Reggio Emilia, Reggio Emilia (RE), Italy}	
\email{stefano.mariani@unimore.it}  

\email{franco.zambonelli@unimore.it}  
\author[F.~Zambonelli]{Franco Zambonelli}[b]	


\newcommand{\field}{\mu}


\newcommand{\builtinop}[3]{\llparenthesis #1 \rrparenthesis_{#2}^{#3}}
\newcommand{\filter}{F}
\newcommand{\BNFcce}{{\bf ::=}}
\newcommand{\BNFmid}{\;\bigr\rvert\;}

\newcommand{\PROGRAM}{\mathtt{P}}
\newcommand{\FUNCTION}{\mathtt{F}}
\newcommand{\e}{\mathtt{e}}
\newcommand{\fname}{\mathtt{d}}
\newcommand{\xname}{\mathtt{x}}
\newcommand{\yname}{\mathtt{y}}
\newcommand{\zname}{\mathtt{z}}
\newcommand{\bname}{\mathtt{b}}
\newcommand{\pe}{\mathtt{p}} 
\newcommand{\we}{\mathtt{w}} 

\newcommand{\name}{\tau}

\newcommand{\asuper}{\mathtt{s}}
\newcommand{\avalue}{\mathtt{v}}
\newcommand{\bvalue}{\mathtt{b}}
\newcommand{\obvalue}{\option{\bvalue}}
\newcommand{\lvalue}{\ell}
\newcommand{\olvalue}{\option{\lvalue}}
\newcommand{\nvalue}{\mathtt{n}}
\newcommand{\truevalue}{\mathtt{True}}
\newcommand{\falsevalue}{\mathtt{False}}
\newcommand{\zerovalue}{0}
\newcommand{\tvalue}{\mathtt{t}}
\newcommand{\hvalue}{\mathtt{h}}
\newcommand{\fvalue}{\phi}
\newcommand{\nolabel}{\_}
\newcommand{\fdom}[1]{\textit{dom}(#1)}
\newcommand{\oexec}{\epsilon}

\newcommand{\brvalue}{\textit{b}}
\newcommand{\obrvalue}{\option{\brvalue}}
\newcommand{\lrvalue}{\textit{l}}
\newcommand{\olrvalue}{\option{\lrvalue}}

\newcommand{\tvalueExt}[1]{(\lvalue_1,\ldots,\lvalue_{#1})}
\newcommand{\fvalueExt}[2]{\{#1\mapsto #2\}}
\newcommand{\fvalueOne}[2]{\{#1\mapsto #2\}}
\newcommand{\fvalueWrap}[1]{\{#1\}}
\newcommand{\fvaluewrapped}[2]{#1\mapsto #2}
\newcommand{\fvalues}[2]{#1\mapsto #2}
\newcommand{\memslot}[2]{#1:=#2}
\newcommand{\treeslot}[2]{#1\mapsto #2}
\newcommand{\emptyL}{\bullet}

\newcommand{\lvalueSet}{\mathcal{L}}
\newcommand{\nvalueSet}{\mathcal{N}}
\newcommand{\tvalueSet}{\mathcal{T}}
\newcommand{\hvalueSet}{\mathcal{H}}
\newcommand{\fvalueSet}{\Phi}
\newcommand{\defK}{\mathtt{def}}
\newcommand{\isK}{\mathtt{is}}
\newcommand{\nbrK}{\mathtt{nbr}}
\newcommand{\repK}{\mathtt{rep}}
\newcommand{\RepK}{\mathtt{REP}}
\newcommand{\ifK}{\mathtt{if}}
\newcommand{\elseK}{\mathtt{else}}

\newcommand{\defKK}[4]{\defK \; #1 \;#2 (#3) \;\isK\; #4}

\newcommand{\tname}{\mathtt{t}}

\newcommand{\ltrue}{\textit{t}}
\newcommand{\lfalse}{\textit{f}}


\newcommand{\wildcard}{{\cdots}} 
\newcommand{\option}[1]{\mathring{#1}} 
\newcommand{\evaluatedPlace}{{\place_\mathit{top}}} 
\newcommand{\re}{\textit{a}} 
\newcommand{\rv}{\mathtt{v}} 
\newcommand{\lre}{\textit{e}} 
\newcommand{\olre}{\option{\lre}} 
\newcommand{\orv}{\option{\rv}}  
\newcommand{\ob}{\textit{b}}  
\newcommand{\Env}{\Gamma}
\newcommand{\emptyEnv}{\bullet}
\newcommand{\Trees}{\Theta}
\newcommand{\emptyTrees}{\bullet}
\newcommand{\labelled}[2]{#1\!\!\cdot\!\!{#2}}
\newcommand{\labelledsmall}[2]{#1\cdot{#2}}
\newcommand{\labelledC}[2]{#1^{#2}}
\newcommand{\labelledCsmall}[2]{#1^{#2}}

\newcommand{\he}{h} 
\newcommand{\ohe}{\option{\he}} 
\newcommand{\ue}{\textit{u}} 
\newcommand{\uhe}{\option{\ue}} 

\newcommand{\unfoldK}{\mathtt{unfold}}

\newcommand{\newopsem}[5]{#1;#2;#3\vdash #4\rightarrow #5}
\newcommand{\opsem}[4]{#1;#2\vdash #3\rightarrow #4}
\newcommand{\opsemmany}[4]{#1;#2\vdash #3\rightarrow^* #4}
\newcommand{\opsemNF}[3]{#1;#2\vdash #3\not\rightarrow}

\newcommand{\self}{\mathtt{self}}


\newcommand{\actx}{\mathbb{A}}
\newcommand{\matches}{::}


\newcommand{\ctxapp}[2]{#1[#2]}
\newcommand{\ctxappfull}[3]{\ctxapp{#1}{#2}\langle #3\rangle}
\newcommand{\ctx}{\mathbb{C}}
\newcommand{\ctxr}{\mathbb{R}}
\newcommand{\ctxf}{\mathbb{N}}
\newcommand{\ctxc}{\mathbb{C}}
\newcommand{\ctxrt}{\mathbb{RT}}
\newcommand{\ctxt}{\mathbb{T}}
\newcommand{\ctxre}{\mathbb{RE}}
\newcommand{\ctxe}{\mathbb{E}}
\newcommand{\hole}{[]}
\newcommand{\place}{\langle\rangle}
\newcommand{\placefilled}[1]{\langle #1\rangle}
\newcommand{\inversectx}[2]{\spi{#1}(#2)}
\newcommand{\spi}[1]{\pi_{#1}}
\newcommand{\erase}[1]{|#1|}

\newcommand{\transition}[3]{
  \begin{array}{l@{\;}c}
    \stackrel{~}{{\tiny \textrm{[#1]}}} & #2 \\ \hline
    \multicolumn{2}{c}{#3}
  \end{array}
}
\newcommand{\transitiontwoprec}[4]{
  \begin{array}{l@{\qquad}c}
    & #2\\
    {\tiny \textrm{[#1]}} & #3 \\ \hline
    \multicolumn{2}{c}{#4}
  \end{array}
}
\newcommand{\nulltransition}[2]{
  \transition{#1}{}{#2}
}

\newcommand{\smallerskiptransition}{\\[-4pt]}
\newcommand{\smallskiptransition}{\\[0pt]}
\newcommand{\skipreduction}{\\}
\newcommand{\skiptransition}{\\[10pt]}
\newcommand{\bigskiptransition}{\\[15pt]}

\newcommand{\sta}{\textit{s}}
\newcommand{\stat}[4]{\langle#1,#2,#3,#4\rangle}
\newcommand{\topStat}[3]{\langle#1,#2,#3\rangle}
\newcommand{\tstat}[5]{#1\!::\!\langle #2,#3,#4,#5\rangle}
\newcommand{\sys}{\textit{N}}
\newcommand{\opar}{\;||\;}
\newcommand{\opard}{\oplus}
\newcommand{\lab}{\lambda}
\newcommand{\labtau}[1]{#1:\tau}
\newcommand{\labstart}[2]{#1\uparrow #2}
\newcommand{\labstop}[3]{#1\downarrow(#3) #2}
\newcommand{\upd}[2]{#1[#2]}
\newcommand{\replace}[2]{#2\rhd #1}
\newcommand{\proj}[2]{{#1}|_{#2}}
\newcommand{\topo}{\varSigma}
\newcommand{\envmap}[2]{#1\mapsto #2}

\newcommand{\fromMsg}[2]{\mathit{from}(#1,#2)}
\newcommand{\toMsg}[2]{\mathit{to}(#1,#2)}

\newcommand{\initNAME}{\textit{init}}
\newcommand{\init}[1]{\initNAME(#1)}
\newcommand{\ruleNameSize}[1]{{\scriptsize #1}}
\newcommand{\topT}[0]{\mathtt{INF}}
\newcommand{\senv}[1]{\mathtt{\##1}}
\newcommand{\sfuno}{f}
\newcommand{\deftt}[3]{\mathtt{def #1(#2) is #3}}
\newcommand{\onlydeftt}[2]{\mathtt{def #1(#2) is}}
\newcommand{\onlybody}[1]{\mathtt{#1}}
\newcommand{\spreadtt}[2]{\mathtt{{\char '173} #1 : #2 {\char '175}}}
\newcommand{\startt}{\mathtt{@}}

\newcommand{\Topo}{\tau}
\newcommand{\Sens}{\Sigma}
\newcommand{\Envi}{\textit{Env}}
\newcommand{\EnviS}[2]{#1:#2}
\newcommand{\Cfg}{N}
\newcommand{\Field}{\Psi}
\newcommand{\SystS}[2]{\langle #1;#2\rangle}
\newcommand{\devset}{I}
\newcommand{\devuniv}{\mathbb{N}}
\newcommand{\nopsem}[4]{#1\vdash #2\xrightarrow{#3} #4}
\newcommand{\nettran}[3]{#1\xrightarrow{#2} #3}
\newcommand{\mapupdate}[2]{#1[#2]}
\newcommand{\netArrIdStar}{\stackrel{\overline{\deviceId}}{\netArrStar}}
\newcommand{\localenv}{\gamma}
\newcommand{\globalenv}{\Gamma}
\newcommand{\envS}[2]{#1;#2}
\newcommand{\lopsem}[3]{#1\,\vdash #2\Downarrow #3}
\newcommand{\netArrIdFairStarN}[1]{\netArrIdStar_{#1}}
\newcommand{\initialV}[1]{#1_{\initial}}
\newcommand{\wfn}[1]{\textit{WFN}(#1)}
\newcommand{\wfe}[1]{\textit{WFE}(#1)}
\newcommand{\denottypemap}{\rightharpoonup}
\newcommand{\disjcup}{\biguplus}
\newcommand{\powerset}{\mathcal{P}}
\newcommand{\shift}[1]{\textbf{shift}(#1)}
\newcommand{\builtindenot}[2]{\mathcal{#1}\llbracket #2 \rrbracket}
\newcommand{\predevices}[1]{{#1}^{{}^-}\!\!\!}
\newcommand{\nbrdevice}[2]{{#1}^{#2}}
\newcommand{\repdevice}[1]{{#1}^-}
\newcommand{\decay}[0]{t_{\mathtt{d}}}
\newcommand{\FieldS}[0]{\mathcal{F}}
\newcommand{\PathS}[0]{\mathbf{P}}
\newcommand{\pathS}[0]{P}
\newcommand{\VarS}[0]{X}
\newcommand{\EventS}[0]{\mathbf{E}}
\newcommand{\eventS}[0]{E}
\newcommand{\eventId}[0]{\epsilon}
\newcommand{\timeS}[0]{t}
\newcommand{\posS}[0]{p}
\newcommand{\event}[3]{\langle #1,#2,#3\rangle}
\newcommand{\devF}[1]{\deviceId_{#1}}
\newcommand{\timeF}[1]{\timeS_{#1}}
\newcommand{\posF}[1]{\posS_{#1}}
\newcommand{\domS}[0]{D}
\newcommand{\DomS}[0]{\mathcal{D}}
\newcommand{\DomDevF}[2]{#1(#2)}
\newcommand{\DomTimeF}[2]{#1(#2)}
\newcommand{\DomMTimeF}[2]{#1^{-}(#2)}
\newcommand{\DomDomF}[2]{#1(#2)}
\newcommand{\DomDevTimeF}[3]{#1(#2,#3)}
\newcommand{\DomDevMTimeF}[3]{#1^{-}(#2,#3)}

\newcommand{\feS}[0]{\Phi}
\newcommand{\setVS}[0]{\textbf{V}}
\newcommand{\setTS}[0]{\textbf{T}}
\newcommand{\setCS}[0]{\textbf{C}}
\newcommand{\setFS}[0]{\mathbf{F}}

\newcommand\pto{\mathrel{\ooalign{\hfil$\mapstochar$\hfil\cr$\to$\cr}}}

\newcommand{\neighbour}[2]{\textit{neigh}(#1,#2)}

\newcommand{\denot}[1]{\mathcal{E}\llbracket{#1}\rrbracket}
\newcommand{\denotapp}[2]{\denot{#1}_{#2}}
\newcommand{\denotappsub}[3]{\denot{#1}_{#2}^{#3}}
\newcommand{\denotfun}[3]{\mathcal{L}\llbracket{#1}\rrbracket_{#2}^{#3}}
\newcommand{\denottype}[1]{\mathcal{T}\llbracket{#1}\rrbracket}
\newcommand{\denotval}[1]{\mathcal{V}\llbracket {#1} \rrbracket}
\newcommand{\denotexp}[3]{\mathcal{E}\llbracket {#1} \rrbracket_{#2}^{#3}}
\newcommand{\denotf}[2]{\lambda #1.#2}

\newcommand{\dvalue}[0]{\Phi}

\newcommand{\myeval}[4]{\epsilon(#1,#2,#3,#4)}

\newcommand{\fiecomp}[2]{#1[#2]}


\newcommand{\somevalue}{\mathtt{w}}
\newcommand{\anyvalue}{\mathtt{v}}
\newcommand{\anyvaluealt}{\mathtt{u}}
\newcommand{\anyvaluebis}{\mathtt{w}}
\newcommand{\anyvalueInNC}[2]{\anyvalue_{#1 (\textrm{in }#2)}}
\newcommand{\nullvalue}{\mathtt{null}}
\newcommand{\propervalue}{\mathtt{w}}
\newcommand{\numvalue}{\mathtt{n}}
\newcommand{\stringvalue}{\mathtt{s}}
\newcommand{\boolvalue}{\mathtt{b}}
\newcommand{\groundvalue}{\mathtt{g}}
\newcommand{\pairvalue}[2]{\langle#1,#2\rangle}

\newcommand{\snsname}{\mathtt{Sns}}

\newcommand{\pname}{\mathtt{p}}
\newcommand{\main}{\mathtt{main}}

\newcommand{\spreadK}{\mathtt{spread}}
\newcommand{\spreadTwo}[2]{\spreadK(#1:#2(*))}
\newcommand{\starK}{\mathtt{@}}
\newcommand{\progK}[2]{#1(\starK,#2)}
\newcommand{\spreadThree}[3]{\{#1:\progK{#2}{#3}\}}

\newcommand{\saaK}{\mathtt{grd}}
\newcommand{\saaTwo}[2]{\saaK(#1,#2)}
\newcommand{\saaThree}[3]{\saaK(#1,#2,#3)}

\newcommand{\lengthOf}[1]{\textit{length}(#1)}
\newcommand{\sizeOf}[1]{\textit{size}(#1)}
\newcommand{\signature}{\textit{signature}}
\newcommand{\signatureOf}[1]{\signature(#1)}
\newcommand{\bsWFTE}[4]{\textit{WFTE}(#1,#2,#3,#4)}

\newcommand{\nulltype}{\mathtt{any}}
\newcommand{\numtype}{\mathtt{number}}
\newcommand{\booltype}{\mathtt{boolean}}
\newcommand{\stringtype}{\mathtt{string}}
\newcommand{\pairtype}[2]{(#1\star#2)}

\newcommand{\lsempar}{[\![}
\newcommand{\rsempar}{]\!]}
\newcommand{\semOf}[1]{\lsempar{#1}\rsempar}
\newcommand{\lowerBound}{\textstyle{\bigwedge}}
\newcommand{\lowerBoundWith}[1]{\lowerBound_{#1}}

\newcommand{\suitableProse}{locally noetherian}
\newcommand{\suitableTypeText}{lnoe}
\newcommand{\suitableType}{\suitableTypeText}
\newcommand{\suitableTypeOf}[1]{\suitableType(#1)}
\newcommand{\suitableOpText}{suitable}
\newcommand{\SuitableOpText}{stabilising}
\newcommand{\suitableOp}{\SuitableOpText}
\newcommand{\suitableOpOf}[1]{\suitableOp(#1)}

\newcommand{\wfLt}{<}
\newcommand{\wfLtOf}[1]{\wfLt_{#1}}
\newcommand{\wfLe}{\le}
\newcommand{\wfLeOf}[1]{\wfLe_{#1}}

\newcommand{\wfGt}{>}
\newcommand{\wfGtOf}[1]{\wfGt_{#1}}
\newcommand{\wfGe}{\ge}
\newcommand{\wfGeOf}[1]{\wfGe_{#1}}

\newcommand{\neginf}{\mathtt{bottom}}
\newcommand{\neginfOf}[1]{\neginf(#1)}
\newcommand{\posinf}{\mathtt{top}}
\newcommand{\posinfOf}[1]{\posinf(#1)}
\newcommand{\initial}{\mathtt{default}}
\newcommand{\initialOf}[1]{\initial(#1)}
\newcommand{\topvalue}{\top}
\newcommand{\topvalueOf}[1]{\topvalue_{#1}}
\newcommand{\TypeDouble}{\mathtt{Double}}
\newcommand{\topvalueOfDouble}[1]{\topvalue_{#1}}

\newcommand{\minimize}{\textit{min}}
\newcommand{\minimizeOf}[1]{\minimize_{#1}}
\newcommand{\dom}{\textit{dom}}
\newcommand{\domOf}[1]{\dom(#1)}
\newcommand{\noetherian}{\textit{noetherian}}
\newcommand{\noetherianOf}[1]{\noetherianOf(#1)}
\newcommand{\noetherianOrd}{<}
\newcommand{\noetherianOrdOf}[1]{\noetherianOrd_{#1}}
\newcommand{\progressive}{\textit{Progressive}}
\newcommand{\progressiveOf}[1]{\progressive(#1)}
\newcommand{\sense}[2]{#1(#2)}

\newcommand{\netframe}{\Envi}
\newcommand{\mknetframe}[3]{(#1,#2,#3)}
\newcommand{\netframeName}{\textit{environment}}
\newcommand{\netframeOf}[1]{\netframeName(#1)}
\newcommand{\stablencOf}[1]{R_{#1}}

\newcommand{\deviceIdSet}{\textbf{D}}
\newcommand{\nodeSubSet}{\textbf{C}}
\newcommand{\stableNodeSet}{\textbf{S}}
\newcommand{\network}{\Cfg}
\newcommand{\netRestrict}[2]{#1|_{#2}}
\newcommand{\initialNetwork}{\mathcal{I}}
\newcommand{\stableNetwork}{R}
\newcommand{\mkgraph}[2]{(#1,#2)}
\newcommand{\mknetwork}[4]{(#1,#2,#4,#3)}
\newcommand{\nodeSet}{\textbf{N}}
\newcommand{\edgeSet}{\textbf{E}}
\newcommand{\stateMap}{\Delta}
\newcommand{\statePair}[2]{(#1,#2)}
\newcommand{\stateMapPair}[2]{(#1,#2)}
\newcommand{\snsFunMap}{\Sigma}
\newcommand{\topSnsFun}{\tau}
\newcommand{\InitialSnsFun}[1]{\textit{Initial}(#1)}
\newcommand{\snsFun}{\sigma}
\newcommand{\snsFunFor}[1]{\snsFun_{#1}}
\newcommand{\snsFunOf}[1]{\snsFunMap(#1)}
\newcommand{\snsFunMapPair}[2]{(#1,#2)}
\newcommand{\snsFunPair}[2]{(#1,#2)}
\newcommand{\mkedge}[2]{(#1,#2)}
\newcommand{\neigh}{\textit{neighbours}}
\newcommand{\neighOfIn}[2]{\neigh(#1,#2)}
\newcommand{\neighTE}{\textit{nvte}}
\newcommand{\neighTEOfIn}[3]{\neighTE(#1,#2,#3)}
\newcommand{\Disconnected}{\textit{Disconnected}}
\newcommand{\DisconnectedOfIn}[2]{\Disconnected(#1,#2)}
\newcommand{\codom}{\textit{codom}}
\newcommand{\codomOf}[1]{\codom(#1)}
\newcommand{\frontier}{\textit{frontier}}
\newcommand{\frontierOfIn}[2]{\frontier_{#1}(#2)}
\newcommand{\netopsemarrow}{\longrightarrow}
\newcommand{\netopsemarrowWith}[1]{\longrightarrow_{#1}}
\newcommand{\netopsemarrowStar}{\netopsemarrow^{\star}}
\newcommand{\netopsemarrowPlus}{\netopsemarrow^{+}}
\newcommand{\netopsem}[2]{#1\netopsemarrow#2}
\newcommand{\netopsemStar}[2]{#1\netopsemarrowStar#2}
\newcommand{\netopsemPlus}[2]{#1\netopsemarrowPlus#2}
\newcommand{\netopsemWith}[3]{#2\netopsemarrowWith{#1}#3}
\newcommand{\netopsemWithStar}[3]{#2\netopsemarrowWith{#1}#3}
\newcommand{\netopsemWithPlus}[3]{#2\netopsemarrowWith{#1}#3}
\newcommand{\netopsemRule}[3]{[\text{#1}]\inferrule{#2}{#3}}
\newcommand{\nullnetopsemRule}[2]{\nullsurfaceTyping{#1}{#2}}
\newcommand{\initialvtree}{\theta_{\main}}
\newcommand{\genvtree}{\ast}
\newcommand{\vtree}{\theta}
\newcommand{\vtreealt}{\eta}
\newcommand{\vtreeInNC}[2]{\theta_{#1( \textrm{in }#2)}}
\newcommand{\initialTraceFor}[1]{\evaluationTrace_{#1}}
\newcommand{\traceFor}[4]{\evaluationTrace_{(#1,#2,#3,#4)}}
\newcommand{\donotcare}{\_}
\newcommand{\ruleName}{\textbf{r}}
\newcommand{\ruleNameSet}{\textbf{R}}
\newcommand{\senstate}{\sigma}
\newcommand{\genmap}[2]{#1\rhd#2}
\newcommand{\tr}[1]{\langle#1\rangle}

\newcommand{\act}{\textit{act}}
\newcommand{\envact}{\textit{env}}
\newcommand{\netArr}{\rightarrow}
\newcommand{\netRed}[2]{#1\netArr#2}
\newcommand{\netArrAct}{\stackrel{\act}{\netArr}}
\newcommand{\netRedAct}[2]{#1\netArrAct#2}
\newcommand{\netArrWithAct}[1]{\stackrel{#1}{\netArr}}
\newcommand{\trueUpd}{\textsf{true}}
\newcommand{\falseUpd}{\textsf{false}}
\newcommand{\isUpd}{\textsf{b}}

\newcommand{\netArrStar}{\Longrightarrow}
\newcommand{\netArrActStar}{\stackrel{\overline{\act}}{\netArrStar}}
\newcommand{\netArrWithActStar}[1]{\stackrel{#1}{\netArrStar}}

\newcommand{\netRedStar}[2]{#1\netArrStar#2}
\newcommand{\netRedWithActStar}[3]{#2\netArrWithActStar{#1}#3}

\newcommand{\pres}{\not\epsilon}
\newcommand{\netArrPres}{\netArr_{\pres}}
\newcommand{\netArrPresStar}{\netArrStar}
\newcommand{\netArrActPres}{\netArrAct}
\newcommand{\netArrActPresStar}{\netArrActStar}
\newcommand{\netArrWithActPresStar}[1]{\stackrel{#1}{\netArrStar}}

\newcommand{\firing}{\textit{firing}}
\newcommand{\fair}{\textit{fair}}
\newcommand{\Fair}{\textit{Fair}}

\newcommand{\fairsym}{\star}
\newcommand{\netArrFairStar}{\netArrStar_{\fairsym}}
\newcommand{\netArrActFairStar}{\netArrActStar_{\fairsym}}
\newcommand{\netArrFairStarN}[1]{\netArrStar_{#1}}
\newcommand{\netArrActFairStarN}[1]{\netArrActStar_{#1}}
\newcommand{\netRedFairStar}[2]{#1\netArrFairStar#2}
\newcommand{\netRedActFairStar}[2]{#1\netArrActFairStar#2}
\newcommand{\netRedActFairStarN}[3]{#2\netArrActFairStarN{#1}#3}

\newcommand{\actDevIn}{\textsf{di}}
\newcommand{\actDevInOf}[1]{\envact}
\newcommand{\netArrDevInOf}[1]{\netArrWithAct{\actDevInOf{#1}}}
\newcommand{\netRedDevInOf}[3]{#2\netArrWithAct{\actDevInOf{#1}}#3}

\newcommand{\actDevOut}{\textsf{do}}
\newcommand{\actDevOutOf}[1]{\envact}
\newcommand{\netArrDevOutOf}[1]{\netArrWithAct{\actDevOutOf{#1}}}
\newcommand{\netRedDevOutOf}[3]{#2\netArrWithAct{\actDevOutOf{#1}}#3}

\newcommand{\actEdgeIn}{\textsf{ei}}
\newcommand{\actEdgeInOf}[1]{\envact}
\newcommand{\netArrEdgeInOf}[1]{\netArrWithAct{\actEdgeInOf{#1}}}
\newcommand{\netRedEdgeInOf}[3]{#2\netArrWithAct{\actEdgeInOf{#1}}#3}

\newcommand{\actEdgeOut}{\textsf{eo}}
\newcommand{\actEdgeOutOf}[1]{\envact}
\newcommand{\netArrEdgeOutOf}[1]{\netArrWithAct{\actEdgeOutOf{#1}}}
\newcommand{\netRedEdgeOutOf}[3]{#2\netArrWithAct{\actEdgeOutOf{#1}}#3}

\newcommand{\actSnsUpd}{\textsf{su}}
\newcommand{\actSnsUpdOf}[1]{\envact}
\newcommand{\netArrSnsUpdOf}[1]{\netArrWithAct{\actSnsUpdOf{#1}}}
\newcommand{\netRedSnsUpdOf}[3]{#2\netArrWithAct{\actSnsUpdOf{#1}}#3}

\newcommand{\actDevUpd}{\textsf{df}}
\newcommand{\actDevUpdOf}[1]{#1}
\newcommand{\netArrDevUpdOf}[1]{\netArrWithAct{\actDevUpdOf{#1}}}
\newcommand{\netRedDevUpdOf}[3]{#2\netArrWithAct{\actDevUpdOf{#1}}#3}

\newcommand{\resPair}[2]{#1,\,#2}
\newcommand{\emain}{\e_{\main}}
\newcommand{\opFun}{\epsilon}
\newcommand{\opFunOf}[1]{\semOf{#1}}
\newcommand{\opApply}[2]{#1(#2)}
\newcommand{\neighSet}{\Theta}
\newcommand{\labe}{\mathtt{l}}
\newcommand{\mklabe}[2]{#1[#2]}
\newcommand{\bsopsem}[5]{#1;#2;#3\vdash #4\Downarrow #5}
\newcommand{\deviceId}{\delta}
\newcommand{\deviceIdalt}{\kappa}
\newcommand{\evaluationDerivation}{\mathcal{D}}
\newcommand{\evaluationDerivationFor}[4]{\evaluationDerivation_{(#1,#2,#3,#4)}}
\newcommand{\abstractJud}[3]{\treeslot{#1}{#2\downarrow#3}}
\newcommand{\premise}{\mathbf{\pi}}
\newcommand{\premiseNum}[1]{\premise_{#1}}
\newcommand{\premiseNumOf}[2]{\premiseNum{#1}(#2)}
\newcommand{\vroot}{\mathbf{\rho}}
\newcommand{\vrootOf}[1]{\vroot(#1)}
\newcommand{\substitution}[2]{#1:=#2}
\newcommand{\applySubstitution}[2]{#1[#2]}
\newcommand{\trace}{\textit{trace}}
\newcommand{\traceOf}[1]{\trace(#1)}
\newcommand{\PropOne}{P1}
\newcommand{\PropTwo}{P2}

\newcommand{\nameInitRte}{\textit{init}}
\newcommand{\initRte}[1]{\nameInitRte(#1)}

\newcommand{\nameRC}{\textit{RC}}
\newcommand{\RC}[2]{\nameRC(#1,#2)}

\newcommand{\ttypeExt}[1]{(\ltype_1,\ldots,\ltype_{#1})}
\newcommand{\ftypeExt}[2]{\{#1\mapsto #2\}}

\newcommand{\device}{\mathtt{a}}

\newcommand{\type}{\textit{T}}
\newcommand{\ltype}{\textit{L}}
\newcommand{\ftype}{\textit{F}}
\newcommand{\rtype}{\textit{R}}
\newcommand{\stype}{\textit{S}}
\newcommand{\builtintype}{\textit{B}}

\newcommand{\btype}{\mathtt{bool}}
\newcommand{\ntype}{\mathtt{num}}
\newcommand{\ttype}{\mathtt{tuple}}

\newcommand{\OLDsurfaceTyping}[3]{
  \begin{array}{c}
    #2 \\ \hline #3
  \end{array} & \textrm{\ruleNameSize{[#1]}} \\[10pt]
}
\newcommand{\OLDnullsurfaceTyping}[2]{
  \begin{array}{c}
    #2
  \end{array} & \textrm{\ruleNameSize{[#1]}} \\[10pt]
}

\newcommand{\surfaceTyping}[3]{
  \begin{array}{l@{\;}c}
    \multicolumn{2}{c}{#2} \\ \hline \\[-9pt]
    \stackrel{{\tiny \textrm{[#1]}}}{~} & #3 
  \end{array}
}
\newcommand{\nullsurfaceTyping}[2]{
  \surfaceTyping{#1}{}{#2}
}

\newcommand{\surExpTypJud}[3]{#1\vdash #2 : #3}
\newcommand{\surFunTypJud}[2]{#2 \; \OK}
\newcommand{\surProTypJud}[2]{#2 \; \OK}

\newcommand{\runExpTypJud}[5]{#1;\,#2;#3 \vdash_{\textbf{r}} #4 : #5}
\newcommand{\treeTypJud}[5]{#1;\,#2;#3 \vdash_{\textbf{tree}} #4 : #5}

\newcommand{\subst}[2]{[#1 := #2]}

\newcommand{\treeOkNAME}{\textbf{wf}}
\newcommand{\treeOk}[1]{\treeOkNAME(#1)}

\newcommand{\treeAuxOkNAME}{\textbf{tree}_{\textbf{aux}}}
\newcommand{\treeAuxOk}[1]{\treeAuxOkNAME(#1)}

\newcommand{\exprOkNAME}{\textbf{expr}}
\newcommand{\exprOk}[1]{\exprOkNAME(#1)}

\newcommand{\exprAuxOkNAME}{\textbf{expr}_{\textbf{aux}}}
\newcommand{\exprAuxOk}[1]{\exprAuxOkNAME(#1)}

\newcommand{\treeErasedNAME}{\textbf{erased}}
\newcommand{\treeErased}[1]{\treeErasedNAME(#1)}

\newcommand{\True}{\textrm{True}}
\newcommand{\False}{\textrm{False}}
\newcommand{\MyIf}{\textrm{if}}
\newcommand{\Otherwise}{\textrm{otherwise}}
\newcommand{\AndMeta}{\textrm{and}}
\newcommand{\OrMeta}{\textrm{or}}
\newcommand{\Implies}{\textrm{implies}}
\newcommand{\Forall}{\textrm{for all}}

\newcommand{\OK}{\textsc{OK}}
\newcommand{\SurTypEnv}{\mathcal{A}}
\newcommand{\emptySurTypEnv}{\bullet}
\newcommand{\TopSurTypEnv}{\SurTypEnv_{\textit{top}}}
\newcommand{\typeofNAME}{\OStypEnv}
\newcommand{\typeof}[1]{\typeofNAME(#1)}
\newcommand{\typeofAt}[3]{\typeofNAME_{#1,#2}(#3)}

\newcommand{\domofNAME}{\mathbf{dom}}
\newcommand{\domof}[1]{\domofNAME(#1)}
\newcommand{\ranof}[1]{\textbf{ran}(#1)}

\newcommand{\lengthNAME}{\sharp}
\newcommand{\lengthof}[1]{\lengthNAME(#1)}


\newcommand{\prg}{\pi}
\newcommand{\prgB}[2]{#1[#2]}
\newcommand{\msg}{\mu}
\newcommand{\msgB}[2]{#1[#2]}
\newcommand{\sname}{\texttt{n}}
\newcommand{\strigger}{\texttt{trigger}}
\newcommand{\LField}{\Lambda}
\newcommand{\sopsem}[5]{#1;#2;#3\vdash #4\Downarrow^s #5}
\newcommand{\pmain}{{\pi_{\main}}}
\newcommand{\enabled}[1]{\texttt{sch}(#1)}
\newcommand{\wf}[2]{F_{#1}(#2)}

\begin{abstract}
  \noindent 
Emerging application scenarios, such as 
cyber-physical systems (CPSs),
the Internet of Things (IoT),
and edge computing,
call for coordination approaches addressing openness, self-adaptation, heterogeneity, and deployment agnosticism.
Field-based coordination is one such approach, promoting the idea of programming system coordination declaratively from a global perspective, in terms of functional manipulation and evolution in ``space and time'' of distributed data structures called \emph{fields}.
More specifically regarding time,
in field-based coordination
it is assumed that local activities in each device
are regulated by a fair and unsynchronised fixed clock working at the platform level.
In this work, we challenge this assumption, and propose an alternative approach where scheduling is programmed in a natural way (along with usual field-based coordination) in terms of \emph{causality fields}, each enacting a programmable distributed notion of a computational ``cause'' (why and when a field computation has to be locally computed) and how it should change across time and space.
Starting from low-level platform triggers, such causality fields can be organised into multiple layers, up to defining high-level, collectively-computed time abstractions, to be used at the application level.
This reinterpretation of the traditional view of time in terms of articulated causality relations allows us to express what we call ``time-fluid'' coordination, where scheduling can be finely tuned so as to select the triggers to react to, generally allowing to  adaptively balance performance (system reactivity) and cost (resource usage) of computations.
We formalise the proposed scheduling framework for field-based coordination in the context of the field calculus, discuss
an implementation in the aggregate computing framework, and finally evaluate the approach via simulation on several case studies.
\end{abstract}

\maketitle

\section{Introduction}\label{sec:intro}

Emerging application scenarios, such as
cyber-physical systems (CPSs),
the Internet of Things (IoT),
and edge computing,
call for software design approaches addressing openness, self-adaptation, heterogeneity, and deployment agnosticism \cite{Ass2017}.
To effectively address this issue, researchers strive to define increasingly higher-level concepts, reducing the ``abstraction gap'' with the problems at hand, e.g., by designing new languages and paradigms.
In the context of coordination models and languages, field-based coordination is one such approach \cite{mamei2006fields,TOCL2019,Coordination2018,BealIEEEComputer2015,DBLP:journals/corr/Lluch-LafuenteL16,ViroliCoordination2012}.
In spite of its many variants and implementations, field-based coordination roots in the idea of programming system coordination declaratively and from a global perspective, in terms of distributed data structures called (computational) \emph{fields}, which span the entire deployment in space (each device is \emph{situated} and holds a value) and time (each device continuously compute, updating its value). 

Regarding time, which is the focus of this paper, field-based coordination typically abstracts from it in two ways:
\emph{(i)} when a specific notion of local time is needed, this is accessed through a sensor as for any other environmental variable; and
\emph{(ii)} a specification (or aggregate program) is actually interpreted as a small computation chunk to be carried on in \emph{computation rounds}.
In each round a device:
\emph{(i)} sleeps for some time;
\emph{(ii)} gathers information about the state of the computation in the previous round, messages received by neighbours while sleeping, and contextual information (i.e.\ sensor readings); and
\emph{(iii)} uses such information to evaluate the coordination specification, storing state information in memory, producing an output value, and sending relevant information to neighbours.
So far, field-based coordination approaches considered computation rounds as being regulated by local clocks, typically asynchronous with respect to clocks in other devices, and independent from the actual outcomes of computations: altogether, they may be seen as resulting from a fixed fair distributed scheduler working at the platform level.
This assumption holds for many other distributed approaches to coordination (e.g.\ channel-based \cite{arbab_2004}, tuple-based \cite{Ciatto2019, tucson}, attribute-based \cite{abc}), but it has a number of consequences and limitations, both philosophical and pragmatic. 

From a philosophical point of view, it follows a pre-relativity view of time that meets general human perception, i.e., where time is absolute and independent of the actual dynamics of events.
This hardly fits with more modern views connecting time with a deeper concept of \emph{causality}~\cite{lobo2008nature}, as being only meaningful relative to the existence of events as in \emph{relational} interpretations of space-time~\cite{relational-quantum}, or even being a mere derived concept introduced by our cognition~\cite{timeless-quantum}---as in Loop Quantum Gravity~\cite{quantum-loop}.
From a practical point of view, consequences are mixed.
The key practical advantage for field-based coordination is simplicity.
First, the designer can abstract from time, leaving the scheduling issue to the underlying platform.
Second, the platform itself can simply impose local schedulers statically,
using fixed frequencies that mostly depend on the device computational power or energetic requirements.
Third, the execution in proactive rounds allows a device to discard messages received few rounds before the current one, thus considering non-proactive senders to have abandoned the neighbourhood,
and simply modelling the state of communication by maintaining the most recent message received from each neighbour.

However, there is a price to pay for such simplicity.
The first is that ``stability'' of the computation, namely, situations in which the field will not change after execution of a round, is ignored.
As a consequence, ``unnecessary'' computations may be performed, consuming resources (including, e.g., energy and bandwidth capacity), and thus reducing the \emph{efficiency} of the system.
Symmetrically, there is a potential \emph{responsiveness} issue: some computations may require to be executed more quickly under some circumstances.
For instance, consider a crowd monitoring and steering system for urban mass events as the one exemplified in~\cite{FGCS2017}:
in case the measured density of people gets dangerous,
a more frequent evaluation of the field delivering steering advices is likely to provide
more precise and timely recommendations. 
Similar considerations apply for example to the area of landslide monitoring \cite{Rosi11},
where long intervals of immobility are interspersed by sudden slope movements:
sensors sampling rate can and should be low most of the time,
but it needs to get promptly increased when slope changes.
This generally suggests a key unexpressed potential for field-based computation: the general ability to adaptively balance performance (system reactivity) and cost (usage of computational resources) of computations, and most importantly, to do so based on a high-level, domain-dependent concept of ``cause'' for computation.
For instance, the crowd and landslide monitoring systems should ideally slow down
(possibly, halt entirely)
the evaluation in case of sparse crowd density or absence of surface movements, respectively; however, they should start being more and more responsive with growing crowd densities or in case of landslide activation.

So, we observe that the scheduling of distributed rounds of computation should not be fixed, but rather \emph{(i)} dynamically depend on the outcome of the computation itself, \emph{(ii)} be based on high-level notions of ``cause'', which can result from space-time properties of computation, and \emph{(iii)} be programmable as a part of the coordination specification---and hence, possibly, using similar constructs.
This general idea can be captured in field-based coordination by modelling time (i.e., the clock that regulates executions of computations) as a \emph{causality field}, namely as a field programmable along with (and hence intertwined with) the usual coordination specification, dictating at each point in space-time where a computation round occurs, i.e., whether a given field computation has to be scheduled for execution or not. To allow for higher expressiveness, then, causality fields can be stacked in multiple layers, to model increasingly higher-level notions of time.
Programming causality along with coordination leads us to a notion of \emph{time-fluid coordination}, where it is possible to flexibly control the balance between performance and cost of system execution.

Accordingly, in this work we present a framework to support causality-driven field-based coordination in the context of the field calculus~\cite{TOCL2019}, a core calculus capturing the essential elements of so-called \emph{aggregate computing} \cite{BealIEEEComputer2015,JLAMP2019}. 
Technically, this is achieved by the idea of organising field calculus programs into a tree structure, where leafs intercept platform triggers and hence enact low-level causality fields; then, these are used by parent programs to enact higher-level causality fields until scheduling the top-level program, which is considered the application-level one.
At each level hence, a field calculus program is scheduled based on the outcome of its execution (self-regulation), on platform triggers (low-level, platform events), and on the execution outcome of leaf programs (called its schedulers, or also, guard policies).
%

More specifically, the contribution of this work (extending upon \cite{published}) can be summarised as follows:
\begin{itemize}
    \item first, we propose a model that enriches field-based coordination approaches with
the possibility to explicitly program the scheduling of coordination actions,
by enabling a \textit{functional description of causality}, relying on a tree-based composition to express high-level notions of time;
    \item second, we formally define the operational semantics of such a model in the context of the field calculus, as an extension and variation of its network semantics in \cite{TOCL2019};
    \item third, we evaluate the proposed model in the context of aggregate programming, showcasing its expressiveness and practicality in achieving improved efficiency and, in some cases, performance.
\end{itemize}


The remainder of this work is structured as follows.
\Cref{sec:back} frames the work with respect to the existing literature on the topic.
\Cref{sec:mechanism} informally describes the proposed model, discussing its goals, motivations, and potential implications.
\Cref{sec:opsem} formalises the proposed time-fluid model.
\Cref{sec:acimpl} presents a prototype implementation in the framework of aggregate computing, along with practical programming examples.
\Cref{sec:eval} evaluates the model and the prototype via simulation of relevant case studies.
Finally, \Cref{sec:conclusion} discusses future directions and concludes the work.

\section{Background and Related Work}\label{sec:back}

Time and synchronisation have always been key issues in the area of distributed and pervasive computing systems. In fact, the absence of a globally shared physical clock among nodes makes it impossible to rely on absolute notions of time, and thus makes it hard to evaluate distributed properties. 

To observe and evaluate the distributed state of a computation, logical clocks can be adopted \cite{lamport1978} to realise a sort of causally-driven notion of time: the ``passing time'' of a distributed computation (that is, the ticks of logical clocks) directly expresses causal relations between distributed events.
As a consequence, any observation of a distributed computation that respects such causal relations, independently of the relative speeds of processes, is a consistent one \cite{babaoglu1993}.
Our proposal somewhat goes in that direction, by trying to make coordination activities abstract from the passing of physical time and rather be guided by some domain-specific notion of time.

Many algorithms exist for computing in a distributed way aggregation functions over data
(e.g., local variables or locally sensed values)
held by a set of distributed processes \cite{Jel05}.
However, when such data dynamically vary with time,
the aggregation functions must be periodically recomputed,
which introduces the problem of identifying the appropriate frequency at which to recompute them.
Most solutions typically adopt a predefined frequency \cite{Jel05,Bicocchi12},
but this strategy does not take into account that such frequency should be properly tuned to match the dynamics of the data being aggregated.
Also, synchronising processes to let them organise aggregation in successive rounds
is far from easy in asynchronous systems lacking a shared physical clock \cite{helary2000,fischer1983}.
Our proposal starts from similar problems,
but is specifically conceived for distributed field-based aggregation,
and  accounts for the strict relations between the spatial and temporal dimensions that exist in situated computations.

Specific timing problems also arise in the area of wireless sensor networks. There, acquiring a globally shared notion of time (as accurate as possible) is of fundamental importance \cite{sundararaman2005clock} to capture accurate snapshots of the distributed phenomena under observation.
However, global synchronisation also serves energy saving purposes.
In fact, when not monitoring or not communicating, the nodes of the network should go to sleep to avoid energy waste, but this implies that to exchange monitoring information with each other they must periodically wake-up in a synchronised way.
In most of existing proposals, though, this is done in awakening and communicating rounds of fixed duration, which makes it impossible to adapt to the actual dynamics of the phenomena under observation. 
Several proposals exist for adaptive synchronisation in wireless sensor networks \cite{ageev2008towards,kho2009,ho2018}, dynamically changing the sampling frequency (and hence frequency of communication rounds) so as to adapt to the dynamics of the observed phenomena.
For instance, in the case of crowd monitoring systems,
it is likely that people (e.g., during an event) stay nearly immobile for most of the time,
then suddenly start moving (e.g., at the end of the event).
Similarly, in the area of landslide monitoring, the situation of a slope is stable for most of the time, with periodic occurrences of (sometimes very fast) slope movements.
In these cases, waking up the nodes of the network periodically would not make any sense and would waste a lot of energy.
Nodes should rather sleep most of the time, and wake up only upon detectable slope movements. 

Such adaptive sampling approaches challenge the underlying notion of time, but they tend to focus on the temporal dimension only, by adapting to the dynamics of a phenomena as locally perceived by the nodes.
Instead, there is need of going further, by making it possible to adapt in time and space as well: not only how fast a phenomenon changes in time, but how fast it propagates and induces causal effects in space.
For instance, in the case of landslide monitoring or crowd monitoring, this amounts to adapting to the dynamics of local perceived movements to the overall propagation speed of such movements across the monitored area.

Besides sensor networks,
the issue of adaptive sampling has recently landed in the broader area of IoT systems and applications \cite{trihinas2018},
again with the primary goal of optimising energy consumption of devices while not losing relevant phenomena under observation.
However,
in these contexts,
such optimisations typically take place in a centralised (cloud) \cite{traub2017} or semi-decentralised (fog) way \cite{lee2018},
which again disregards spatial issues and the strict space-time relations of phenomena.

Since coordination models and languages typical address a crosscutting concern of distributed systems, they are historically concerned with the notion of time in a variety of ways.
For instance, time is addressed in space-based coordination since Javaspaces \cite{javaspaces-book}, and corresponding foundational calculi for time-based Linda \cite{LindaJavaSpaces,DBLP:journals/entcs/LindenJ07}: the general idea is to equip tuples and query operations with timeouts, which can be interpreted either in terms of global or local clocks.
The problem of abstracting the notion of time became crucial when coordination models started addressing self-adaptive systems, and hence openness and reactivity.
In \cite{DBLP:conf/sac/MenezesW06, Mariani2012235}, it is suggested that a tuple may eventually fade, with a rate that depends on a usefulness concept measuring how many new operations are related to such tuple.
In the biochemical tuple-space model \cite{VC-COORD2009}, tuples have a time-dynamic ``concentration'' driven by stochastic coordination rules embedded in the data-space.

Field-based coordination emerged as a coordination paradigm for self-adaptive systems focusing more on ``space'' rather than ``time'', in works such as TOTA \cite{MameiZ09}, field calculus \cite{TOCL2019,Coordination2018}, and fixpoint-based computational fields \cite{DBLP:journals/corr/Lluch-LafuenteL16}.
However, the need for dealing with time is a deep consequence of dealing with space, since propagation in space necessarily implies ``evolution'' along time (as travelling in space requires time).
These approaches tend to abstract from the scheduling dynamics of local field evolution, in various ways.
In TOTA, the update model for distributed ``fields of tuples'' is an asynchronous event-based one: anytime a change in network connectivity is detected by a node, the TOTA middleware provides for triggering an update of the distributed field structures so as to immediately reflect the new situation.
In the field calculus and aggregate computing \cite{BealIEEEComputer2015}, as already mentioned, an external, proactive clock is typically used.
In \cite{DBLP:journals/corr/Lluch-LafuenteL16} this issue is mostly neglected since the focus is on the ``eventual behaviour'', namely the stabilised configuration of a field, as in \cite{TOMACS2018}.
For all these models, scheduling of updates is always transparent to the application/programming level, so the application designer cannot intervene on the relationship between the passing of time and coordination actions, hence cannot possibly optimise communications, energy expenses, and reactivity depending on the dynamics of the application at hand.

\newcommand{\policyname}[0]{G}
\newcommand{\sensorsname}[0]{S}
\newcommand{\messagename}[0]{M}
\newcommand{\eventname}[0]{T}
\newcommand{\programname}[0]{P}
\newcommand{\program}[1][]{$\mathbf{P}_{#1}$}
\newcommand{\expression}[1][]{$\mathbf{E}_{#1}$}
\newcommand{\messages}[0]{$\mathcal{\messagename}$}
\newcommand{\sensors}[0]{$\mathcal{\sensorsname}$}
\newcommand{\events}[0]{$\mathcal{\eventname}$}
\newcommand{\policy}[1][]{$\mathbf{\policyname}_{#1}$}

\tikzset{-,
  ev/.style={circle,draw, inner sep=1.8pt, minimum size=2pt, outer sep=0.5mm},
  point/.style={circle,fill=black, inner sep=1.8pt, minimum size=2pt, outer sep=0.5mm},
  pointp/.style={rectangle,fill=black, inner sep=2pt, minimum size=2pt, outer sep=0.5mm},
  event/.style={circle,fill=red, inner sep=3pt, minimum size=2pt, outer sep=0.5mm},
  past/.style={event, fill=ddarkgreen},
  present/.style={event,fill=gray},
  future/.style={event,fill=blue},
  locality/.style={circle,fill=white, inner sep=2pt, minimum size=2pt, outer sep=0.5mm},
  locality1/.style={locality,fill=orange!50},
  locality2/.style={locality,fill=gray!50},
  locality3/.style={locality,fill=blue!40},
  locality4/.style={locality,fill=black!30!green},
  locality5/.style={locality,fill=purple!50},
  d/.style={circle,fill=gray, inner sep=2pt, minimum size=2pt},
  da/.style={circle,draw,label={[left]$1$}}, 
  db/.style={circle,draw,label={[left]$2$}}, 
  dc/.style={circle,draw,label={[left]$3$}}, 
  lbl/.style={font=\footnotesize},
  state/.style={circle,draw}, 
  statein/.style={circle,draw,line width={2pt}},
  stateout/.style={circle,draw,line width={1pt},fill=black!16},
  leadstov/.style={draw,decorate,decoration={snake, amplitude=1.95mm, post=lineto, post length=2mm, segment length=1cm, pre length=0.4cm},->,>=stealth'}, 
  leadsto/.style={draw,decorate,decoration={snake, amplitude=0.35mm, post=lineto, post length=2mm, segment length=1.3mm},->,>=stealth'}, 
  leadstopast/.style={leadsto,draw=ddarkgreen},
  leadstopresent/.style={leadsto,draw=black},
  leadstofuture/.style={leadsto,draw=blue},
  frastagliato/.style={draw,decorate,decoration={snake, amplitude=0.45mm, post=lineto, post length=0mm, segment length=2.3mm},->,>=stealth'},
  appgraph/.style={rectangle,draw,minimum size=1.5cm},
  appgraphc/.style={ellipse,draw,minimum size=1.5cm},
  program/.style={rectangle,draw,radius=0.1cm},
  scheduler/.style={rounded rectangle,draw,radius=0.1cm,font=\small},
  evpast/.style={green},
  evpresent/.style={red},
  evfuture/.style={blue},
  evconc/.style={gray},
}

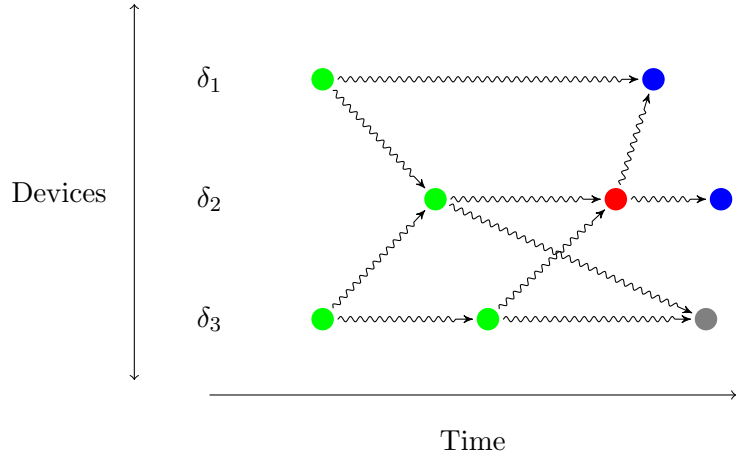
\begin{figure}
\begin{tikzpicture}
\node[] (d1l) [] {$\delta_1$};
\node[] (d2l) [below=of d1l] {$\delta_2$};
\node[] (d3l) [below=of d2l] {$\delta_3$};

\node[event,evpast] (d1e1) [right=of d1l] {};
\node[event,evfuture] (d1e2) [right=4cm of d1e1] {};
\node[event,evpast] (d2e1) [right=2.5cm of d2l] {};
\node[event,evpresent] (d2e2) [right=2cm of d2e1] {};
\node[event,evfuture] (d2e3) [right=1cm of d2e2] {};
\node[event,evpast] (d3e1) [right=of d3l] {};
\node[event,evpast] (d3e2) [right=1.8cm of d3e1] {};
\node[event,evconc] (d3e3) [right=2.5cm of d3e2] {};

\draw[leadsto] (d1e1) -- (d1e2);
\draw[leadsto] (d2e1) -- (d2e2);
\draw[leadsto] (d3e1) -- (d3e2);
\draw[leadsto] (d3e2) -- (d3e3);
\draw[leadsto] (d1e1) -- (d2e1);
\draw[leadsto] (d3e1) -- (d2e1);
\draw[leadsto] (d2e1) -- (d3e3);
\draw[leadsto] (d2e2) -- (d1e2);
\draw[leadsto] (d2e2) -- (d2e3);
\draw[leadsto] (d3e2) -- (d2e2);

\draw[<->] (-1,-4) -- node[xshift=-1cm]{Devices} (-1,1);
\draw[->](0,-4.2) -- node[yshift=-0.6cm]{Time} (7,-4.2);
\end{tikzpicture}
\caption{Example of an event structure. Each node is an event (i.e., a computation round happening at a given device), and the curly arrows denote the causal relationship between events resulting from communication, which defines a partial order on the events. 
Given a reference event (red),
 through the $\leadsto$ relationship
 it is possible to define its ``causal past'' event cone (green),
 ``causal future'' event cone (blue),
 and concurrent events (gray).
Conventionally, we assume that the events horizontally aligned to the device labels $\delta_i$ are rounds performed by those devices.
Along the arrows, information can be transferred, modelling state persistence in a given device or communication between different devices:
in this example, $\delta_1$ and $\delta_3$ both communcate bidirectionally with $\delta_2$.
Moreover, since devices might be situated in space, and the event structure captures a distributed, platform-level notion of time, the events can denote space-time locations and, together with the corresponding computed values, a space-time computational field (modelling, e.g., a temperature field, a warning field, a field of suggestions for crowd dispersal, etc.).}
\label{fig:event-struct}
\end{figure}

\section{Time-fluid field-based coordination}\label{sec:mechanism}

In this section, we introduce a model for time-fluid field-based coordination.
The core idea of our proposed approach is to leverage field-based coordination itself
for maintaining a \emph{causality field} that drives the dynamics of computations of the application-level fields.
Our discussion is in principle applicable to any field-based coordination framework,
however, for the sake of clarity, we here focus on the general framework originated from the field calculus \cite{TOCL2019}.

Execution of programs in the field calculus are assumed to result in so-called \emph{computational fields}, which are mappings from \emph{events} to computational values, where an event is seen as a space-time point (as in physics), namely, as the execution of a computation round at a given device. 
Communication can happen between events, which are hence called \emph{neighbour events}; so, the execution of a system of computing and interacting devices
 can be modelled as a directed acyclic graph of events,
 also known as an \emph{event structure}~\cite{mattern1988virtualtime}---see \Cref{fig:event-struct} for a graphical example.
Locally, an event is typically scheduled by the middleware platform sustaining field computations, and is hence considered as a low-level (asynchronous, distributed) ``clock'', on top of which higher-level, fluid notions of time can be defined.
In the case of a landslide monitoring distributed application, for instance, the events could be due to sensors' readings measuring whether the terrain is moving, how fast, etc., and the corresponding computational field could be the probability of a landslide being triggered in the next $T$ minutes.

\subsection{A time-fluid model}

Considering a field calculus program \program{}, each of its rounds can be thought of as consuming:
\textit{i)} a set of valid messages received from neighbours, $M\in\mathcal{\messagename}$;
and \textit{ii)} some contextual information $S\in\mathcal{\sensorsname}$, usually obtained via so-called sensors.
In the case of landslide monitoring, $S$ could include measurements about terrain movements, and $M$ could include messages about how neighbouring sensors measured such movements in their areas.
The platform or middleware in charge of executing field calculus programs has to decide whether at a given event it should  launch the next evaluation round of \program{}, also providing valid values for \messages{} and \sensors{}.
Note that in general the platform could execute many programs concurrently.

In order to support causality-driven coordination,
we first require the platform to associate to each event a \emph{(local event) trigger},
representing the cause for that event to happen, typically involving some kind of change at the application level (in \messages{}) or at the physical level (in \sensors{}).
Typical examples of triggers include ``a new message has arrived'', ``a given sensor provides a new value'' (as in landslide monitoring), or ``1 second has passed''.
We here denote by $\mathcal{\eventname}$ the set of all possible local event triggers the platform can manage.

As a second step, we introduce a specific type of field calculus program \policyname, called a \emph{guard policy}, which will be used as a scheduler for a ``parent'' field computation: as such it will be expressed in the same language of \program{}, as detailed in next sections.
More specifically, whenever evaluated across space and time, a guard policy can be locally viewed as a function $f_G$ of the kind 
$f_G\colon{}(\mathcal{\sensorsname}, \mathcal{\messagename})\to\mathcal{V}$.
Namely, a policy has the same input as any field computation, and can produce any output (values in $\mathcal{V}$).

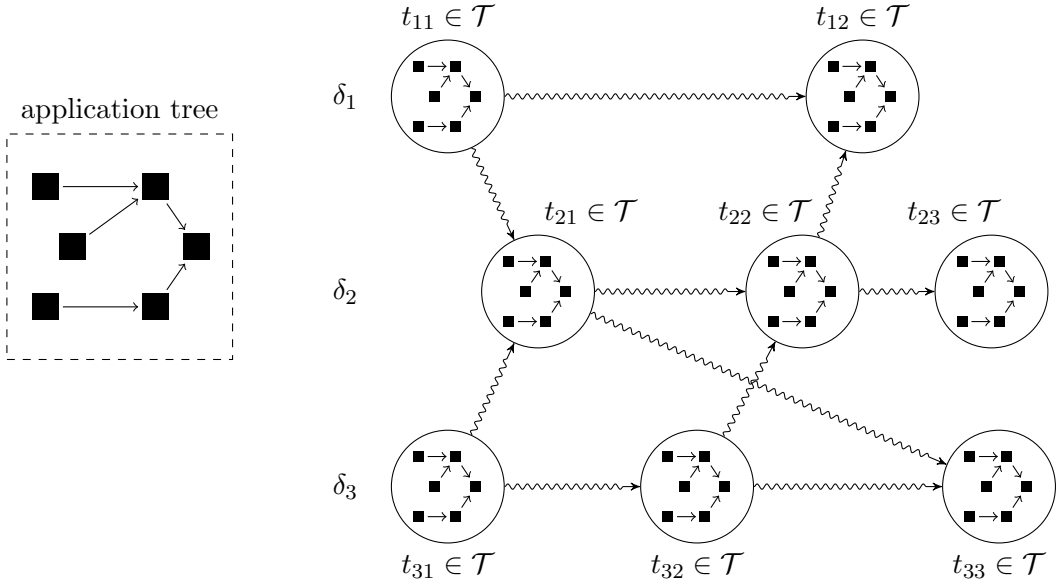
\begin{figure}
\begin{tikzpicture}[node distance=2cm]
\newcommand{\drawgraphin}[1]{\drawappgraphin{#1}{0.25cm}{0.4cm}{0.25}{pointp}}
\newcommand{\drawappgraphin}[5]{
\node[#5] (#1p1) [right=#4 of #1.west,yshift=#3] {};
\node[#5] (#1p2) [right=#2 of #1p1] {};
\node[#5] (#1p3) [right=-0.1*#2 of #1p1,yshift=-#3] {};
\node[#5] (#1p4) [right=1.2*#2 of #1p3] {};
\node[#5] (#1p5) [right=#4 of #1.west,yshift=-#3] {};
\node[#5] (#1p6) [right=#2 of #1p5] {};
\draw[->] (#1p1) -- (#1p2);
\draw[->] (#1p3) -- (#1p2);
\draw[->] (#1p5) -- (#1p6);
\draw[->] (#1p6) -- (#1p4);
\draw[->] (#1p2) -- (#1p4);
}

\node[rectangle,draw,dashed,black,minimum width=3cm,minimum height=3cm] (legend) at (-3cm,-2cm) [label={above:{application tree}}] {};
\drawappgraphin{legend}{1cm}{0.8cm}{0.3cm}{pointp,inner sep=5pt}

\node[] (d1l) [] {$\delta_1$};
\node[] (d2l) [below=of d1l] {$\delta_2$};
\node[] (d3l) [below=of d2l] {$\delta_3$};

\node[appgraphc] (d1e1) [right=0.3cm of d1l,label={above:{$t_{11}\in\mathcal{T}$}}] {};
\node[appgraphc] (d1e2) [right=4cm of d1e1,label={above:{$t_{12}\in\mathcal{T}$}}] {};
\node[appgraphc] (d2e1) [right=1.5cm of d2l,label={[xshift=0.7cm]above:{$t_{21}\in\mathcal{T}$}}] {};
\node[appgraphc] (d2e2) [right=2cm of d2e1,label={[xshift=-0.5cm]above:{$t_{22}\in\mathcal{T}$}}] {};
\node[appgraphc] (d2e3) [right=1cm of d2e2,label={[xshift=-0.5cm]above:{$t_{23}\in\mathcal{T}$}}] {};
\node[appgraphc] (d3e1) [right=0.3cm of d3l,label={below:{$t_{31}\in\mathcal{T}$}}] {};
\node[appgraphc] (d3e2) [right=1.8cm of d3e1,label={below:{$t_{32}\in\mathcal{T}$}}] {};
\node[appgraphc] (d3e3) [right=2.5cm of d3e2,label={below:{$t_{33}\in\mathcal{T}$}}] {};

\drawgraphin{d1e1}
\drawgraphin{d1e2}
\drawgraphin{d2e1}
\drawgraphin{d2e2}
\drawgraphin{d2e3}
\drawgraphin{d3e1}
\drawgraphin{d3e2}
\drawgraphin{d3e3}

\draw[leadsto] (d1e1) -- (d1e2);
\draw[leadsto] (d2e1) -- (d2e2);
\draw[leadsto] (d3e1) -- (d3e2);
\draw[leadsto] (d3e2) -- (d3e3);
\draw[leadsto] (d1e1) -- (d2e1);
\draw[leadsto] (d3e1) -- (d2e1);
\draw[leadsto] (d2e1) -- (d3e3);
\draw[leadsto] (d2e2) -- (d1e2);
\draw[leadsto] (d2e2) -- (d2e3);
\draw[leadsto] (d3e2) -- (d2e2);
\end{tikzpicture}
\caption{This is an example of an event structure
(structurally equivalent to that of \Cref{fig:event-struct}) where each event (circle) is an evaluation of a tree of field calculus programs (the tree of black squares shown in the dashed box on the left). The labels on the nodes indicate the platform triggers causing the events to happen.
The application tree has a root program (the right-most node)
 and schedulers as children and within its sub-trees; the nodes with no incoming arrows are scheduled merely on the basis of platform triggers.}
\label{fig:event-struct-fluid}
\end{figure}

Differently from the work in \cite{published}, where a field computation could be assigned a single guard policy, here we extend this approach by allowing many guard policies to be associated to an actual field computation, and since guard policies are particular kinds of field computations as well, we also allow guard policies to be stacked at multiple layers, forming a \emph{tree structure}.
So, essentially, platform triggers schedule low-level guard policies, the leafs of the tree, which are field computations enacting low-level (causality-based) ``clocks''.
Such clocks can be combined together and feed higher-level guard policies so as to enact higher-level ``clocks'', recursively.
Ultimately, such clocks will then schedule evaluation of the top-level field computation, the root of the tree, which is the application-level one.
In our landslide monitoring example, platform triggers may be the events related to the aforementioned sensors readings, low-level guard policies may use such readings to decide whether to compute the probability of an imminent landslide or not (which is a field computation), and higher levels scheduling policies could use such probability to compute a landslide frontier and visualise it on a map.
Additionally, the result of evaluation of a guard policy at a given time can also be used as a feedback to schedule its next evaluation: this is used all the times the dynamics of evolution of a certain field computation should be used to self-regulate its actual timing.
Consider \Cref{fig:event-struct-fluid} 
for a graphical example.

The mechanisms by which the result of a guard policy affects scheduling are captured by the following function, defined at the platform level, called the \emph{causality function}:
\[f_{C}\colon{}(\mathcal{T},\mathcal{V},\mathcal{V}_1,\ldots,\mathcal{V}_n)\mapsto \{0,1\}\]
In a given round, and for a given program \programname~(hence also for guard policies), it takes the trigger of the current round in the first argument, the result of evaluating \programname~at the previous round (feedback) in second argument, and then the results of evaluating all its $n$ lower-level schedulers, and returns a Boolean stating whether this program is to be scheduled for evaluation at the current round.
Hence, function $f_{C}$ maps any event to Booleans, hence denoting a Boolean field acting as a \emph{causality field} for \programname, namely, enacting a programmable, distributed and self-regulated clock achieved by the collective behaviour of all the involved devices.

In the proposed framework, a field of low-level triggers is processed by a hierarchical set of schedulers, ultimately defining a top-level causality-field that schedules the application-level computation.
This mechanism thus overall introduces a structured guard mediating between the evolution of 
\emph{platform triggers} and the actual execution of application rounds,
allowing for fine control over the actual temporal dynamics,
as exemplified in \Cref{sec:examples}.

Crucially,
the ability to sense context
(namely, the contents of \sensors{})
and to express event triggers
(namely, the possible contents of \events{})
has a large impact on the expressivity of the proposed model.
In general, it is reasonable to assume that a platform or middleware hosting a field computation can generate events due to triggers that include changes to any value of \sensors{} (this allows the computation to be reactive to changes in the device perception,
or, symmetrically speaking, makes such changes the \emph{cause} of the computation) and that triggers flipping their value from \texttt{false} to \texttt{true} can model timers, making the classic time-driven approach a special case of the proposed framework (see \Cref{ssec:timer-based-sched} for a concrete example).

\subsection{Consequences}

The above informal introduction to our proposed model allows us to emphasise in more detail some of the implications on expressiveness of field-based coordination.

\subsubsection{Programming \emph{the} space-time and propagating causality}

As soon as we let an application affect its own
execution policy,
we are effectively programming \emph{the} time
(instead of \emph{in} time, as is typically done in field-based coordination):
evaluating a field computation at different frequencies actually amounts at modulating the perception of time from the application standpoint.
For instance, sensors' values may be sampled more often or more sparsely, affecting the perception that the application has of its operating environment along the time scale.
In turn, as stemming from the distributed nature of the communicating system at hand,
such an adaptation along time would immediately cause adaptation across space too,
by affecting the communication rate of devices,
hence the rate at which events and information spread across the network.
It is worth emphasising that this a consequence of embracing a notion of time founded on causality.
In fact,
as we are aware of computational models adaptive to the time fabric,
as mentioned in \Cref{sec:back},
we are not aware of any model allowing \emph{programming} the perception of time at the application level.


\subsubsection{Adapting to causality}

Being able to affect the space-time fabric as described above necessarily requires the capability of being \emph{aware} of the space-time fabric in the first place. When the notion of space-time is crafted upon the notion of causality between events, such a form of awareness translates to awareness of the \emph{dynamics of causal relations} among events.
Under this perspective, the application is no longer adapting to the passage of time and the extent of space, but to the temporal and spatial distribution of causal relations among events.
In other words, the application is able to ``chase'' events not only as they travel across time and space, but also as their ``traveling speed'' changes.
For instance, whenever in a given region of space some event happens more frequently,
devices operating in the same area may compute more frequently as well,
increasing the rate of communications among devices in that region,
thus leading to an overall better recognition of the quickening dynamics of the phenomenon under observation.
%

\subsubsection{Controlling situatedness}

The ability to control both the above mentioned capabilities at the application level
enables fine control over the \emph{degree of situatedness} exhibited by the overall system,
along two dimensions:
the ability to decide the granularity at which event triggers should be perceived;
and the ability to decide how to adapt to changes in events dynamics.
In modern distributed and pervasive systems
the ability to quickly react to changes in environment dynamics are of paramount importance \cite{PSC2013}.
For instance, in the mentioned case of landslide monitoring,
as anomalies in measurement increase in frequency, intensity, and geographical coverage,
the monitoring application should match the pace of the accelerating dynamics.

\subsubsection{Co-causal field computation}
\label{s:cocausality}
On the practical side,
associating field computations to programmable scheduling policies
brings both advantages and risks
(as most extensions to expressiveness do).
One important gain in \emph{expressiveness} is the ability to let field computations affect the scheduling policy
of other field computations,
as in the example of crowd steering or landslide monitoring:
the denser some regions get,
the faster will the steering field be computed;
the more intense vibrations of the ground get, the more frequently monitoring is performed.
On the other hand
(provided that field computations can perform \emph{actuation}
which in turn results into changes in sensor values)
this opens the door to \emph{circular dependencies} among field computations and scheduling policies.
Although the scheduling graph is always acyclic, in fact, inter-round communication could be carried out through sensors and actuators.
This situation may lead to undesirable global behaviours, such as deadlocks or livelocks;
however, such circular dependencies must be explictly programmed: they appear by design, not by chance.
Hence, the programmer should design carefully the scheduling policies of the system at hand not to incur in these problematic situations---a care that should be taken when programming in any expressive language.
In this regard, ``fragments'' of composable Protelis programs which are ``safe'' to run can be detected,
as something similar has already been done with field calculus~\cite{TOMACS2018}.
Finally, for any case in which circular dependencies are actually desired by the designer,
simulations can be used for assessing (i.e., with a reasonable confidence)
whether the system does not incur in unpredicted behaviours.

%

%

\subsubsection{Pure reactivity and its limitations}
Technically,
replacing a scheduler guided by a fixed clock with one triggering computations as consequence of events,
turns the system from time-driven to \emph{event-driven}.
In principle, this makes the system \emph{purely reactive}:
the system is idle unless some event trigger happens.
Depending on the application at hand, this may be a blessing or a curse:
since pro-activity is lost, the system is chained to the dynamics of event triggers,
and cannot act on its own will.
Of course,
it is easy to overcome such a limitation:
assuming that a clock is available in the pool of event triggers
makes pro-activity a particular case of reactivity,
where the tick of the clock dictates the granularity.
Furthermore,
since guard policies allow the specification of retroactive feedback on their scheduling, 
the designer can always design a ``fall-back'' plan relying on expiration of a timer:
for instance, it is possible (and reasonable) to express a policy such as
``trigger as soon as event $\epsilon$ happens, or timer $\tau$ expires, whichever comes first''.
%
%
\begin{figure}[!t]{
 \framebox[1\textwidth]{
 $\begin{array}{l}
 \textbf{a) Field calculus syntactic elements:}\\
\begin{array}{lr@{\hspace{1.7cm}}|@{\hspace{1.7cm}}lr}
\anyvalue \qquad\qquad & {\footnotesize \mbox{field calculus value}}  & 
\e \qquad\qquad &   {\footnotesize \mbox{field calculus expression}} \\
\vtree &   {\footnotesize \mbox{field calculus value-tree}} & 
\deviceId &   {\footnotesize \mbox{device unique identifier}} \\
\sname &  {\footnotesize \mbox{sensor name}} \\
\end{array}\\[5pt]
\hline\\[-8pt]
\textbf{b) Field calculus abstracted semantics:}\\
\begin{array}{lcl@{\hspace{5.5cm}}r}
\bsopsem{\deviceId}{\envmap{\overline\delta}{\overline\vtree}}{\envmap{\overline\sname}{\overline \anyvalue}}{\emain}{\vtree}
& & & {\footnotesize \mbox{round semantics judgment}} 
\end{array}\\[5pt]
\hline\\[-8pt]
\textbf{c) Local and scheduling configurations:}\\
\begin{array}{lcl@{\hspace{7.5cm}}r}
\prg &  \BNFcce & \prgB{\e}{\overline\prg}& {\footnotesize \mbox{scheduling program}} \\
\msg &  \BNFcce & \msgB{\vtree}{\overline\msg}& {\footnotesize \mbox{exported message}} \\
\LField & \BNFcce & \envmap{\overline\deviceId}{\overline\msg} & {\footnotesize \mbox{local status field}} \\
\senstate & \BNFcce & \envmap{\overline\sname}{\overline \anyvalue}  &   {\footnotesize \mbox{local sensor state}} \\
\end{array}\\[5pt]
\hline\\[-8pt]
%
\textbf{d) Scheduling semantics:} \hfill
  \boxed{\sopsem{\deviceId}{\LField}{\senstate}{\prg}{\msg}}
  \\[0.2cm]
\vspace{0.5cm}
\begin{array}{c}
\surfaceTyping{ROUND}{
	\{\forall i,\; \sopsem{\deviceId}{\envmap{\overline\deviceId}{\overline\msg^i}}{\senstate}{\prg_i}{\msg_i},\; \msg_i = \msgB{\vtree^o_i}{\overline{\msg}^0} \}\quad \left\{\begin{array}{l}
                             \vtree = \vtree_j \qquad\textrm{if~~}\neg\enabled{\vtree_j,\overline\vtree^o,\senstate(\strigger)}\\
                             \bsopsem{\deviceId_j}{\envmap{\overline\delta}{\overline\vtree}}{\senstate}{\e}{\vtree} \hfill\textrm{otherwise}\\
                           \end{array}\right.
	\!\!\!\!
 }{
\sopsem{\deviceId_j}{\envmap{\overline\deviceId}{\msgB{\overline\vtree}{\overline\msg^1,\ldots,\overline\msg^n}}}{\senstate}{\prgB{\e}{\overline\prg}}{\msgB{\vtree}{\overline\msg}} 
}\\[10pt]
\end{array}\\
\hline\\[-8pt]
\textbf{e) System configurations:}\\[5pt]
\begin{array}{lcl@{\hspace{7.5cm}}r}
\devset & \BNFcce & \overline\deviceId & {\footnotesize \mbox{neighbourhood}} \\
\Topo & \BNFcce &  \envmap{\overline\deviceId}{\overline\devset}    &   {\footnotesize \mbox{topology field}} \\
\Sens & \BNFcce &  \envmap{\overline\deviceId}{\overline\senstate}    &   {\footnotesize \mbox{sensor field}} \\
\Field & \BNFcce &  \envmap{\overline\deviceId}{\overline\LField}    &   {\footnotesize \mbox{status field}} \\
\Envi & \BNFcce &  \EnviS{\Topo}{\Sens}    &   {\footnotesize \mbox{environment}} \\
\Cfg & \BNFcce &  \SystS{\Envi}{\Field}    &   {\footnotesize \mbox{network configuration}} \\
%
\end{array}\\[5pt]
\hline\\[-8pt]
\textbf{f) Environment well-formedness:}\\[5pt]
\begin{array}{l}
\wfe{\EnviS{\Topo}{\Sens}} \textrm{~~holds iff~~} \domof{\Topo}=\domof{\Sens} \textrm{~~and~~} \{\delta\}\subseteq \Topo(\deviceId) \subseteq \domof{\Sens}, \forall \deviceId \in \domof{\Sens} \\

\end{array}\\[5pt]
\hline\\[-10pt]
\textbf{g) Network semantics:} \hfill
  \boxed{\nettran{\Cfg}{\act}{\Cfg}}
  \\[10pt]
\begin{array}{c}
\surfaceTyping{N-FIR}{
                  \Envi=\EnviS{\Topo}{\Sens}\qquad{\sopsem{\deviceId}{\Field(\deviceId)}{\Sens(\deviceId),\envmap{\strigger}{\anyvalue_t}}{\pmain}{\msg}}
                 \qquad
                 \Field_1=\envmap{\Topo(\deviceId)}{\{\envmap{\deviceId}{\msg}\}}}
                 {\nettran{\SystS{\Envi}{\Field}}{\deviceId:\anyvalue_t}{\SystS{\Envi}{\mapupdate{\Field}{\Field_1}}}
                 }
\skiptransition\\[0.05cm]
\surfaceTyping{N-ENV}
                 {\qquad \wfn{\Envi'}\qquad \Envi'=\EnviS{\envmap{\overline\deviceId}{\overline\devset}}{\envmap{\overline\deviceId}{\overline\senstate}} \qquad
                  \Field_0=\envmap{\overline\deviceId}{\envmap{\overline\devset}{\msg_\perp^\pmain}}
                 }
                 {\nettran{\SystS{\Envi}{\Field}}{\envact}{\SystS{\Envi'}{\mapupdate{\Field_0}{\Field}}}
                 }
\skiptransition\\[0.05cm]
\end{array}\\
\end{array}$}
} 
 \caption{Operational semantics for time-fluid coordination in the field calculus.} \label{fig:networkSemantics}
\end{figure}

\clearpage 
\section{Formal Semantics in Field Calculus}\label{sec:opsem}
We now formalise our proposal of time-fluid field-based coordination.
It is based on the field calculus \cite{TOCL2019}, the prominent formal framework to capture the essential aspects of computational fields.
This calculus is at the core of implementations such as Protelis \cite{PianiniSAC2015,BealIEEEComputer2015}, which will in turn be used in the next section to explain the impact on aggregate programming.

\Cref{fig:networkSemantics} provides the whole formalisation, as a variation and extension of the network semantics for the field calculus presented in \cite{TOCL2019}.
Starting from a black-box description of the field calculus syntax and semantics, and ending up with the operational semantics of network evolution, each part is described in the following.

The formalisation to come adopts standard conventions used in other calculi for programming languages (like Featherweight Java and its descendants \cite{FJ}):
we use the overbar notation to represent sequences, such that if $\e$ is a meta-variable over expressions, then a sequence of $n$ expressions is denoted $\overline\e$ or equivalently $\e_1,\ldots,\e_n$,
and a list of zero elements is sometimes denoted by symbol $\emptyL$.
The notation is then abused using the overbar notation over 2 (or more) symbols in a term to mean a sequence of terms constructed out of those 2 (or more) sequences, and this is typically used to model functions as sequence of mappings between pairs of terms.
As an example, a field will be written as 
$\envmap{\overline\deviceId}{\overline\anyvalue}$ to mean $\envmap{\deviceId_1}{\anyvalue_1},\ldots,\envmap{\deviceId_n}{\anyvalue_n}$.
This models a function $\phi$ that associates a device identified by $\deviceId_j$ to value $\anyvalue_j$ (for any $j$), hence we shall also use the natural function application notation $\phi(\deviceId_j)$ to mean $\anyvalue_j$.
%
We then use the update operator $\mapupdate{\phi}{\phi'}$ for functions, to mean the function obtained by updating any mapping already present in $\phi$ with a possible new value as indicated in $\phi'$.
For instance, $\mapupdate{(\envmap{\deviceId_1}{\anyvalue_1},\envmap{\deviceId_2}{\anyvalue_2})}{\envmap{\deviceId_2}{\anyvalue'_2},\envmap{\deviceId_3}{\anyvalue_3}}$ gives $\envmap{\deviceId_1}{\anyvalue_1},\envmap{\deviceId_2}{\anyvalue'_2}$.

\subsection{Field calculus syntactic elements and abstracted semantics}\label{sec-fc-semantics}

The key idea of the field calculus is to represent the behaviour of a distributed program in term of a functional expression, whose denotational semantics is defined in terms of a resulting \emph{computational field}, namely, a map from space-time events (specific moments of time in which a specific device performs a computation) to computational values (the result of such computation).
More specifically,
such expressions
(ranged over by $\e$)
comprise usual mechanisms of function definition and call,
use of built-in operators
(for working with arithmetic, logic, and useful data structures),
access to local sensors
(whose name is ranged over by $\sname$, each providing values ranged over by $\anyvalue$),
as well as field operators to deal with evolution in time (\texttt{rep}),
interaction with neighbours (\texttt{nbr}),
and space-time branching (\texttt{if}).
The actual syntax of field calculus is of no interest here, for it is orthogonal to the management of schedulers we ought to formalise; though, the examples and case studies presented later in this paper will make use of its incarnation in the Protelis language, conveniently described in the next section.

Expressions are to be understood as being repetitively and asynchronously evaluated in each device of the network, that is, in computation rounds. 
A network is considered equipped with a dynamically evolving and reflexive \emph{neighbourhood} relation, supported at the platform level: a device can send messages only to its current neighbours.
When a round occurs at a device with id $\deviceId$, a context is available that includes messages received from its neighbours (only the last message from each neighbour is considered) and locally sensed information: the expression is evaluated against such a context (namely, is affected by it) and an output message is produced, which is assumed to be broadcasted to all the neighbours in turn.
Crucially, such output messages -- called \emph{value-trees}, or \emph{vtrees} in short, ranged over by $\vtree$ -- are structured as a tree of values, essentially reflecting the dynamic unfolding of expression evaluation.
For instance, the vtree of expression $\e_1 + \e_2$ is a tree whose root is tagged with the resulting value $\anyvalue$ of expression evaluation, and whose two sub-trees are the vtrees of $\e_1$ and $\e_2$, recursively---and similarly for any other built-in operator or construct of the language.
A device gathers the vtrees of neighbours shipped with messages (there including the vtree of its own latest local round) into a so-called \emph{value-tree environment} (a mapping $\envmap{\overline\deviceId}{\overline\vtree}$ from neighbours to vtrees).
Since a value-tree environment is available during expression evaluation, 
 it is then possible to deal with fields expressing evolution over time and communication with neighbours.
Additionally, to support the branching mechanism of field calculus \cite{TOCL2019},
it is possible to distinguish,
for each sub-expression,
which neighbours evaluated it or not due to branching,
correctly restricting observation of values in neighbours (as of construct \texttt{nbr})\footnote{This latter notion, called \emph{alignment}, does not play a crucial role in this paper, but is key to enable safe composition of expressions in field calculus:
 the idea is that different branches of computation are unrelated and, given a certain point in the computation path,
 only the devices that reached that point during evaluation
 are actually allowed to interact (these devices are said to be \emph{aligned} in that vtree position).
}.
The result of an expression evaluation (the root of the vtree) at each device at each moment of time is what defines the computational field resulting from ``distributed execution'' of that expression.

These mechanisms are precisely captured by the field calculus round semantics, defined by a judgment of the form $\bsopsem{\deviceId}{\envmap{\overline\deviceId}{\overline\vtree}}{\envmap{\overline\sname}{\overline \anyvalue}}{\emain}{\vtree}$ to be read as ``expression $\emain$ evaluates to value-tree $\vtree$ on device $\deviceId$ with respect to the value-tree environment
$\envmap{\overline\deviceId}{\overline\vtree}$ and sensor state $\envmap{\overline\sname}{\overline\anyvalue}$''.
This judgment is defined by syntax-directed rules of a big-step operational semantics---the reader can refer to \cite{TOCL2019} for details.

\newcommand{\tempsns}{\ensuremath{\mathtt{temp}}}
\metacomment{
\begin{example}[Minimum temperature in neighbourhood]
\label{example:min-temp}
Consider a simple program expression, $\pmain = \mathtt{fold}(\infty,\mathtt{min},\mathtt{nbr}(\tempsns))$, where each device in the network
 computes the minimum temperature value
 sensed in its neighbourhood,
 assuming every device has a built-in sensor $\tempsns$ available
 for local temperature sensing.
It works by accumulating the neighbours' temperature values
 with built-in function $\mathtt{fold}$
 applied 
 to the minimum function $\mathtt{min}$
 and starting value $\infty$;
 neighbour values are collected through field calculus construct $\mathtt{nbr}(e)$, where $e$ is the expression whose value is to be shared.
As per the field calculus big-step operational semantics,
 program $\pmain$ evaluates to a vtree as follows.
\begin{minipage}{\textwidth}
\centering
\newcommand{\op}[1]{\ensuremath{#1}}
\begin{tikzpicture}[-,node distance=0.5cm and 1.5cm, node/.style={rectangle,draw}, node2/.style={draw}]
\node[] (d1l) [] {$\delta_2$};
\node[] (d2l) [below=of d1l] {$\delta_1$};
\node[] (d3l) [below=of d2l] {$\delta_3$};

\node[event,evpast] (d1e1) [right=of d1l,label=above:{$\eventId_{2,1}$}] {};
\node[event,evpast] (d2e0) [right=0.5cm of d2l,label=above:{$\eventId_{1,1}$}] {};
\node[event] (d2e1) [right=1.5cm of d2e0,label=above:{$\eventId_{1,2}$}] {};
\node[event,evpast] (d3e1) [right=of d3l,label=below:{$\eventId_{3,1}$}] {};

\draw[leadsto] (d1e1) -- (d2e1);
\draw[leadsto] (d2e0) -- (d2e1);
\draw[leadsto] (d3e1) -- (d2e1);

  \node[] (a7) [left=5cm of d1l] {\texttt{fold}};
  \node[node] (a8) [below left=0.3cm of a7] {$\infty$};
  \node[node] (a9) [below=0.3cm of a7] {$\min$};
  \node[node] (a10) [below right=0.3cm of a7] {$nbr$};
  \node[node] (a11) [below=0.3cm of a10] {$\tempsns$};

  \path 
    (a7) edge [] node [] {} (a8)
    (a7) edge [] node [] {} (a9)
    (a7) edge [] node [] {} (a10)
    (a10) edge[] (a11)
    ;
\end{tikzpicture}
\end{minipage}
After a device evaluates $\pmain$,
 it shares the resulting vtree
 with neighbours for supporting coordination
 and contributing to the ``aggregate'' system behaviour. 
Now, suppose we have three devices $\deviceId_1, \deviceId_2, \deviceId_3$
 and to evaluate the program
 in event $\eventId_{1,2}$ (the second event of $\deviceId_1$),
 which can use information from
 $\eventId_{1,1}$ (the previous event of the same device)
 as well as
 $\eventId_{2,1}$ and $\eventId_{3,1}$ (the first event of devices $\deviceId_2$ and $\deviceId_3$)---see the example event structure above.
Suppose such input information consists of the mapping $\overline{\deviceId} \mapsto \overline{\vtree}$ from neighbour devices to vtrees where 
$\deviceId_1 \mapsto \vtree_{1,1} = 14\tr{\infty,\texttt{min},(\deviceId_1 \mapsto 14)\tr{14\tr{\tempsns}}}$ (the output of the previous computation), 
$\deviceId_2 \mapsto \vtree_{2,1} = 15\tr{\infty,\texttt{min},(\deviceId_2 \mapsto 15)\tr{15\tr{\tempsns}}}$,
$\deviceId_3 \mapsto \vtree_{3,1} = 12\tr{\infty,\texttt{min},(\deviceId_3 \mapsto 12)\tr{12\tr{\tempsns}}}$, 
 and that the $\tempsns$ sensor yields $13$ in $\eventId_{1,2}$.
Then, we have the big-step
``$\bsopsem{\deviceId_1}{\{\deviceId_1 \mapsto \vtree_{1,1},\deviceId_2 \mapsto \vtree_{2,1},\deviceId_3 \mapsto \vtree_{3,1}\}}{\{\tempsns \mapsto 13\}}{\emain}{
12\tr{\infty,\texttt{min},(\deviceId_1 \mapsto 13, \deviceId_2 \mapsto 15, \deviceId_3 \mapsto 12)\tr{13\tr{\tempsns}}}
}$'', where the root value $12$ is the output of the computation.
\end{example}
}

\subsection{Local scheduling configuration and semantics}\label{sec-calculus-device-semantics}

We now describe the formalisation of scheduling as proposed in this paper, extending on top of the field calculus standard syntax and semantics.
A global field-calculus scheduled program $\prg$ has the form $\prgB{\e}{\overline\prg}$: it consists of a standard expression $\e$ scheduled by zero, one, or many scheduling programs $\overline\prg$.
Therefore, a program is essentially a tree of expressions where the root expression is used to provide the result of field computation, child nodes are expressions used as schedulers for parents (possibly organised in many layers), and finally leafs have the form $\prgB{\e}{\emptyL}$---they are expressions directly scheduled by the platform as usual in field calculus.
Given that a scheduled program is a tree of expressions, and evaluation of each expression is essentially isolated, devices will now exchange not just a vtree, but rather an exported message $\msg$ with a corresponding structure $\msgB{\vtree}{\overline\msg}$, that is, a tree of elements $\vtree$, each obtained by node-wise application of the field calculus operational semantics recalled in \Cref{sec-fc-semantics} to the tree.
We are hence ready to define the operational semantics of scheduling, which is given in terms of a judgment ``$\sopsem{\deviceId}{\LField}{\senstate}{\prg}{\msg}$'' to be read as ``scheduling program $\prg$ evaluates to the exported message $\msg$ on device $\deviceId$ with respect to the local status field $\LField$ and sensor state $\senstate$'', where a local status field $\LField$ is a mapping $\envmap{\overline\deviceId}{\overline\msg}$ from neighbours to messages, and sensor state $\senstate$ is a mapping $\envmap{\overline\sname}{\overline\anyvalue}$ as usual.

Informally, the idea is that, at any round, expressions will all be \emph{considered} for evaluation bottom-up in the tree.
Evaluation of an expression will actually take place (that is, its field computation is actually scheduled), depending on a \emph{scheduling context} composed by
(i) the results of evaluations of its low-level scheduler programs,
(ii) the result of evaluation of the same expression at its latest round, and
(iii) information about the trigger that caused the platform to start this round.
If scheduling is activated, the expression is evaluated as usual and the resulting vtree will be added to the export message; if it is not activated, the latest vtree is simply reused without re-evaluation.

Formalisation of these two cases is provided by rule [ROUND] in \Cref{fig:networkSemantics} (d).
It deals with a program $\prgB{\e}{\overline\prg}$ executed against a neighbourhood $\overline\deviceId$ where each device $\deviceId_k$ previously sent message $\msgB{\vtree_k}{\msg^1_k,\ldots,\msg^n_k}$---hence altogether the local status field is written $\envmap{\overline\deviceId}{\msgB{\overline\vtree}{\overline\msg^1,\ldots,\overline\msg^n}}$.
Then, it proceeds evaluating each scheduler $\prg_i\in\overline\prg$ recursively, each yielding its part of the exported message $\msg_i$ whose top-value tree is named $\vtree^o_i$: note that to do so we should enter the local field status accordingly, using $\envmap{\overline\deviceId}{\overline\msg^i}$ for any $i$.
Following, we check whether the main expression $\e$ of current program $\prgB{\e}{\overline\prg}$ can be evaluated:
this is achieved by predicate $\enabled$, whose evaluation on all events realise the causality field for top level expression $\e$ as discussed in previous section.
It takes the vtrees obtained from evaluation of schedulers, the vtree of current device at latest round (the current device is $\deviceId_j$ in list $\overline\deviceId$, hence its latest vtree is $\vtree_j$ as can be extracted from the local status field), and the value read from $\strigger$ sensor.
Note that we abstract from this predicate: it depends on the specific incarnation of the model, which should dictate how to extract and combine meaningful information for scheduling from the aforementioned scheduling context.

Rule [ROUND]  then differentiates based on whether the predicate prevents evaluation or not.
In the first case, simply the latest vtree $\vtree_j$ is used for the exported message, since no evaluation of $\e$ takes place; in the second case instead, expression $\e$ is evaluated using standard field calculus approach as described in previous subsection, which gives $\vtree$ as result, and which is then used to create the resulting export message $\msgB{\vtree}{\overline\msg}$.

\subsection{System configuration and Network semantics}\label{sec-calculus-network-semantics}

On top of the scheduling semantics, dictating which expressions of the entire scheduling program get actually evaluated at each round, and what is the shape of the exported message, it is possible to derive an operational semantics of the overall distributed system behaviour, directly following the approach presented in \cite{TOCL2019}---which we tailor to the end of formalising scheduling. 
As with standard field calculus, it is then assumed that a single aggregate program $\pmain$ exists.
As shown in \Cref{fig:networkSemantics}(e), we let $\devset$ range over set of devices (typically forming a neighbourhood), and introduce three kinds of fields used to define the snapshot of a system configuration at a given time: $\Topo$ is a field of neighbours (a map from devices to their neighbours), representing system topology; $\Sens$ is a field of sensor state (a map from devices to sensor states $\senstate$), representing information sensed from the environment; and $\Field$ is a global status field (a map from devices to local status field $\LField$), representing the pending messages available across the system, and hence, the actual status of field computations over at a given time.
We shall write $\msg_\perp^\pmain$ to denote a default (empty) message compliant with program $\pmain$: it will be automatically available in the global status field when a new device enters the network.
A network configuration $\Cfg$ is hence a pair of an environment $\Envi$ and a global status field $\Field$, where the environment is itself a pair of topology and sensor fields.

As described in \Cref{fig:networkSemantics}(f) an environment is considered well-formed if the sensor field has the same domain of the topology field, and a topology is reflexive and closed under such a domain.
The operational semantics, defined in \Cref{fig:networkSemantics}(g) is constructed so as to guarantee that well-formedness of the environment is preserved across transitions.
This is defined by two rules, one modelling scheduling ([N-FIR]) and one changes in the environment ([N-ENV]).

Rule [N-FIR] is labelled $\deviceId:\anyvalue_t$, modelling scheduling of a round at device $\deviceId$ with trigger $\anyvalue_t$.
It performs one round of evaluation of $\pmain$ using the local context of $\deviceId$ ($\Field(\deviceId);\Sens(\deviceId)$), and adding to the sensor state the mapping $\envmap{\strigger}{\anyvalue_t}$.
To model emission of the resulting export message $\msg$ to neighbours, we update the global status field with a new component $\Field_1$, which adds to each neighbour of $\deviceId$ the term $\envmap{\deviceId}{\msg}$, representing message $\msg$ received from $\deviceId$.

Rule [N-ENV] is labelled $\envact$, and models a generic change in the environment (topology or sensors), moving from $\Env$ to any $\Env'$ that is well-formed.
Let the new topology be the mapping $\envmap{\overline\deviceId}{\overline\devset}$, it first constructs a complete global status field $\Field_0$ with compliant empty messages $\msg_\perp^\pmain$ available everywhere needed, namely, each device has one such message per neighbour.
We then update $\Field_0$ with the previous global status field $\Field$.


\newcommand{\timesns}{\mathtt{time}}

\metacomment{
\begin{figure}[h!]
\newcommand{\fired}{red}
\begin{tikzpicture}
\newcommand{\drawgraphin}[4]{\drawappgraphin{#1}{0.25cm}{0.4cm}{0.25}{#2}{#3}{#4}{\tiny}}
\newcommand{\drawappgraphin}[8]{
\node[pointp,#5] (#1p1) [right=1.2*#4 of #1.west,yshift=0.4*#3,label={#8$e_1$}] {};
\node[pointp,#6] (#1p2) [right=1.2*#4 of #1.west,yshift=-0.4*#3,label=below:{#8$e_2$}] {};
\node[pointp,#7] (#1p4) [above right=-0.1cm and 1*#4 of #1p2,label=right:{#8$e_0$}] {};
\draw[->] (#1p1) -- (#1p4);
\draw[->] (#1p2) -- (#1p4);
}

\node[rectangle,draw,dashed,black,minimum width=3cm,minimum height=3cm] (legend) at (-3cm,-2cm) [label={above:{application tree}}] {};
\drawappgraphin{legend}{1cm}{0.8cm}{0.6cm}{inner sep=5pt}{inner sep=5pt}{inner sep=5pt}{};

\node[pointp,inner sep=5pt] (nschedprog) [below=0.5cm of legend.south west,label=right:{Not scheduled}] {};
\node[pointp,inner sep=5pt,\fired] (schedprog) [below=0.5cm of nschedprog,,label=right:{Scheduled}] {};

\node[] (d1l) [] {$\delta_1$};
\node[] (d2l) [below=1.5cm of d1l] {$\delta_2$};
\node[] (d3l) [below=1.5cm of d2l] {$\delta_3$};

\node[appgraphc] (d1e1) [right=0.3cm of d1l,label={above:{$\anyvalue_{t1},12,57$}}] {};
\node[appgraphc] (d1e2) [right=4cm of d1e1,label={above:{$\anyvalue_{t2},13,58$}}] {};
\node[appgraphc] (d2e1) [right=1.5cm of d2l,label={[xshift=0.3cm]above:{$\anyvalue_{t2},15,33$}}] {};
\node[appgraphc] (d2e2) [right=2cm of d2e1,label={[xshift=-0.5cm]above:{$\anyvalue_{t2},15,34$}}] {};
\node[appgraphc] (d2e3) [right=1cm of d2e2,label={above:{$\anyvalue_{t1},15,35$}}] {};
\node[appgraphc] (d3e1) [right=0.3cm of d3l,label={below:{$\anyvalue_{t1},14,60$}}] {};
\node[appgraphc] (d3e2) [right=1.8cm of d3e1,label={below:{$\anyvalue_{t3},15,61$}}] {};
\node[appgraphc] (d3e3) [right=2.5cm of d3e2,label={below:{$\anyvalue_{t2},15,62$}}] {};

\drawgraphin{d1e1}{\fired}{}{}
\drawgraphin{d1e2}{}{\fired}{\fired}
\drawgraphin{d2e1}{}{\fired}{\fired}
\drawgraphin{d2e2}{}{\fired}{}
\drawgraphin{d2e3}{\fired}{}{}
\drawgraphin{d3e1}{\fired}{}{\fired}
\drawgraphin{d3e2}{}{\fired}{}
\drawgraphin{d3e3}{}{\fired}{\fired}

\draw[leadsto] (d1e1) -- 
(d1e2);
\draw[leadsto] (d2e1) -- (d2e2);
\draw[leadsto] (d3e1) -- (d3e2);
\draw[leadsto] (d3e2) -- (d3e3);
\draw[leadsto] (d1e1) -- (d2e1);
\draw[leadsto] (d3e1) -- (d2e1);
\draw[leadsto] (d2e1) -- (d3e3);
\draw[leadsto] (d2e2) -- (d1e2);
\draw[leadsto] (d2e2) -- (d2e3);
\draw[leadsto] (d3e2) -- (d2e2);
\end{tikzpicture}
\caption{Application tree and event structure for \Cref{example:net-and-sched}. The labels of events denote sensor values for the three sensors $\strigger$, $\tempsns$, $\timesns$.}
\label{fig:sched-semantics-example}
\end{figure}
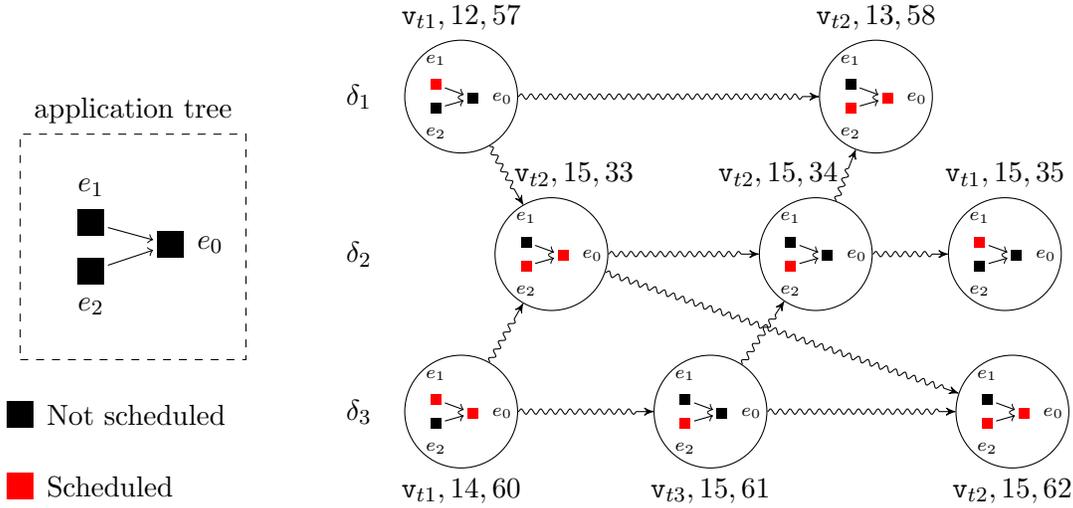
\begin{example}[Network semantics and scheduling]
\label{example:net-and-sched}
Consider the event structure in \Cref{fig:sched-semantics-example}.
This (and any other) event structure
 can be modelled as a small-step evolution 
 of a global state,
 also keeping track of the payload of messages and
 data available in any event (not shown graphically).
For instance, the topology $\Topo$ of a global time instant 
 corresponding to the first event of $\deviceId_2$ may be 
 given by $\{ \deviceId_1 \mapsto \{ \deviceId_1, \deviceId_2 \},
 \deviceId_2 \mapsto \{ \deviceId_1, \deviceId_2, \deviceId_3 \}, \deviceId_3 \mapsto \{ \deviceId_2, \deviceId_3 \} \}$.
Rule [N-ENV] can be used to evolve topology, sensor state, and status field to represent the state of a system 
 and a device's local context
 corresponding to a particular event.

Each event represents an application of rule [N-FIR],
 which can fire 
 if the sensor $\strigger$ local to a device provides a value.
Suppose that, beside $\strigger$, the devices 
 have a temperature sensor $\tempsns$ and a timestamp sensor $\timesns$.
The sensor field $\Sens$ may initially be as follows:
$\{ \deviceId_1 \mapsto \{ \strigger \mapsto \anyvalue_{t1}, \tempsns \mapsto 12, \timesns \mapsto 57 \},
\deviceId_2 \mapsto \{ \tempsns \mapsto 15, \timesns \mapsto 33 \},
\deviceId_3 \mapsto \{ \strigger \mapsto \anyvalue_{t1}, \tempsns \mapsto 14, \timesns \mapsto 60 \} \}$,
showing that $\deviceId_1$ and $\deviceId_3$ can fire (but not $\deviceId_2$ yet).

Then, consider a scheduled program $\prg_{main}=\prgB{\e_0}{\e_1,\e_2}$.
The platform supports two local triggers, i.e., sensor $\strigger$ returns either $\anyvalue_{t1}$ or $\anyvalue_{t2}$.
\begin{itemize}
\item $\e_1$ is fired by a ``local timestamp update'' trigger $\anyvalue_{t1}$ and computes a vtree $\vtree_1$ whose root is $\truevalue$ if $\timesns\mod~60$ equals $0$ or $\falsevalue$ otherwise (hence yielding a field which is $\truevalue$ once a minute);
\item $\e_2$ is fired by either a ``message received'' trigger $\anyvalue_{t2}$ or a ``tempature has changed'' trigger $\anyvalue_{t3}$, and computes a vtree $\vtree_2$ whose root is 
$\truevalue$ if  the minimum temperature value in the neighbourhood (cf. \Cref{example:min-temp}) has changed from the last round;
\item $\e_0$ may be a program that outputs a log message about the minimum temperature in the neighbourhood,
 to be scheduled if any of the outputs of $\e_1$ or $\e_2$ is $\truevalue$.
\end{itemize}
In particular, the scheduling predicate $\enabled{\vtree_j,\overline\vtree^o,\senstate(\strigger)}$ can be as follows:
$$
\begin{dcases}
\truevalue & \text{if}~\overline\vtree^o \neq \emptyset \land \text{any of}~ \overline\vtree^o ~\text{yields}~ \truevalue\\
\truevalue & \text{if}~\overline\vtree^o = \emptyset \land 
\senstate(\strigger) \in \{ \anyvalue_{t1}, \anyvalue_{t2}, \anyvalue_{t3} \} \\
\falsevalue & \text{otherwise}
\end{dcases}
$$
It prescribes the triggering of
 top-level schedulers 
 by the aforementioned low-level triggers, 
 and the triggering of the downstream program 
 based on the results of such top-level schedulers.
Then,
 for a certain event (rule [N-FIR]),
 the scheduling predicate is used
 to determine which programs of the program tree are to be evaluated (rule [ROUND]), leading e.g. to the executions shown in \Cref{fig:sched-semantics-example}.
 
\end{example}
}

\section{Time-fluid Aggregate Computing}
\label{sec:acimpl}

The proposed model, formalised in \Cref{sec:opsem}, has been implemented within the framework of aggregate computing~\cite{BealIEEEComputer2015,JLAMP2019}, which is based on the computational fields abstraction, as detailed in this section.
For the implementation,
we leveraged \emph{Alchemist}~\cite{PianiniJOS2013},
an extensible simulator with
 pre-existing support for
the Protelis programming language~\cite{PianiniSAC2015}
and the Scala-internal DSL ScaFi~\cite{casadei2020isola-fscafi}.
Specifically, 
 we developed an extension of the simulator 
 supporting the definition of trees of scheduling programs
 using the same aggregate programming language used for the actual software specification.
The framework has been open-sourced, released at a public repository\footnote{\url{https://github.com/DanySK/Experiment-2020-LMCS-TimeFluid}},
extensively evaluated in three paradigmatic cases, covered in \Cref{sec:eval}.

In this section,
we first briefly introduce the basics of programming in the Protelis language (\Cref{sec:protelisprimer}), for the sake of self-containedness.
Then, we describe how the formalised model 
 is implemented into an Alchemist-Protelis time-fluid incarnation (\Cref{sec:incarnation}).
Finally, we provide examples of time-fluid aggregate program specifications to showcase the expressive power of the proposed approach (\Cref{sec:examples}).

\subsection{A short Protelis primer}
\label{sec:protelisprimer}

\definecolor{assignment}{rgb}{0.5, 0.5, 0.0}
\lstdefinelanguage{Protelis}{ 
  emph={NEGATIVE_INFINITY, TIMER, SENSOR,MESSAGE_RECEIVED,MESSAGE_TIMEOUT,TICK,EVERY_SECOND}, emphstyle=\color{Green},
  basicstyle=\lst@ifdisplaystyle\footnotesize\fi\ttfamily,
  breaklines=true,                 
  commentstyle=\color{gray},    
  deletekeywords={},            
  escapeinside={\%*}{*)},          
  extendedchars=true,              
  frame=single,	                   
  keepspaces=true,                 
  language=Java,                 
  morekeywords=[1]{def ,env,let ,self,rep,module ,it,nbr},            
  keywordstyle=[1]\color{purple},       
  keywordstyle=[2]\color{purple},       
  numbers=left,                    
  numbersep=5pt,                   
  numberstyle=\tiny\color{black}, 
  rulecolor=\color{black},         
  showspaces=false,                
  showstringspaces=false,          
  showtabs=false,                  
  stepnumber=1,                    
  stringstyle=\color{blue},     
  tabsize=2,	                   
  alsoletter={:},
}
\lstset{language=Protelis}
\lstset{xleftmargin=13pt} 

Protelis is an incarnation of the field calculus, in terms of a purely functional,
higher-order,
interpreted,
and dynamically typed
aggregate programming language interoperable with Java.
This Protelis language primer is intended as a quick reference for understanding the subsequent examples---more details can be found in \cite{PianiniSAC2015}.
%
%

Programs are written in \emph{modules},
and are composed of any number of \emph{function definitions} and of an optional \emph{main script}, as exemplified in \Cref{pt:module}.
\begin{snippet}[h]
\begin{lstlisting}[language=Protelis]
// src/main/protelis/some/namespace.pt
// Module declaration
module some:namespace
// Imports
import java.util.HashSet;
// Function definitions
def f1(a,b,c) { /* ... */ }
def f2() = // ...
// (Optional) main script
f1(1,2,3) + f2()
\end{lstlisting}
\caption{The structure of a Protelis module.}
\label{pt:module}
\end{snippet}

Declaration \lstinline[language=Protelis]!module some:namespace! defines a new module whose fully-qualified name is
\lstinline[language=Protelis]!some:namespace!.
Modules' functions can be imported locally using the \lstinline[language=Protelis]!import! keyword followed by the
fully qualified module name.
The same keyword can be used to import Java members, with 
\lstinline[language=Protelis]!org.protelis.Builtins!,
\lstinline[language=Protelis]!java.lang.Math!, and
\lstinline[language=Protelis]!java.lang.Double!
being imported implicitly by default.
Similarly to other dynamic languages such as Ruby and Python,
in Protelis
top level code outside any function is considered to be the main script.
%
%
Definition \lstinline[language=Protelis]!def f(a, b) { code }! introduces a new function named \lstinline[language=Protelis]!f!
with two arguments \lstinline[language=Protelis]!a! and \lstinline[language=Protelis]!b!.
Upon invocation, the function body \lstinline[language=Protelis]!code! -- consisting of a series of expressions -- is executed, and the function returns the value of the last evaluated expression.
In case the function has a single expression, a shorter, Scala-/Kotlin-style syntax is allowed:
\lstinline[language=Protelis]!def f(a, b) = expression!.
\emph{Anonymous functions} are written with a syntax reminiscent of Kotlin and Groovy:
\lstinline[language=Protelis]!{ a, b, -> code }! evaluates to an anonymous function
with two parameters and \lstinline[language=Protelis]!code! as body.
%
%
Protelis also shares with Kotlin the \emph{trailing lambda convention}:
if the last parameter of a function call is an anonymous function,
then it can be placed outside the parentheses.
If the anonymous function is the only argument to that call, the parentheses can be omitted entirely;
the calls depicted in \Cref{pt:trailing-lambda} are in fact equivalent.
\begin{snippet}[h]
\begin{lstlisting}[language=Protelis]
[1, 2].map({ a -> a + 1 }) // returns [2, 3]
[1, 2].map() { a -> a + 1 } // returns [2, 3]
[1, 2].map { a -> a + 1 } // returns [2, 3]
\end{lstlisting}
\caption{Trailing lambda convention in Protelis.}
\label{pt:trailing-lambda}
\end{snippet}

The \lstinline[language=Protelis]!let v = expression! statement adds a variable named \lstinline[language=Protelis]!v! to the local name space, associating its value to the value of the \lstinline[language=Protelis]!expression! evaluation.
Square brackets delimit tuple literals: \lstinline[language=Protelis]![]! evaluates to an empty tuple,
\lstinline[language=Protelis]![1, 2, "foo"]! to a tuple of three elements with two numbers and a string.
Methods can be invoked with the same syntax of Java:
\lstinline[language=Protelis]!obj.method(a, b)!
tries to invoke method \lstinline[language=Protelis]!member! on the result of evaluation of expression
\lstinline[language=Protelis]!obj!, passing the results of the evaluation of expressions \lstinline[language=Protelis]!a!
and \lstinline[language=Protelis]!b! as arguments.
Special keywords \lstinline[language=Protelis]!self! and \lstinline[language=Protelis]!env!
allow access to contextual information:
\lstinline[language=Protelis]!self! exposes sensors via direct method call
(typically leveraged for system access),
while \lstinline[language=Protelis]!env! allows dynamic access to sensors by name
(hence supporting more dynamic contexts).
An example of their use is presented in \Cref{pt:self-env}.
\begin{snippet}[h]
\begin{lstlisting}[language=Protelis]
import java.lang.System.out
let currentTime = self.getCurrentTime()
let temperature = env.get("temperature")
out.println("Temperature " + temperature + " at time " + currentTime)
\end{lstlisting}
\caption{Example interaction with the runtime: the Java standard output is imported, the local timer is accessed, a sensor is read, and finally both these values are printed.}
\label{pt:self-env}
\end{snippet}

Then, Protelis supports the field calculus constructs as well,
whose semantics is recalled in \Cref{sec-fc-semantics}.
The \lstinline[language=Protelis]!rep (v <- initial) { code }! expression enables stateful computation
by associating \lstinline[language=Protelis]!v!
with either the previous result of the \lstinline[language=Protelis]!rep! evaluation,
or with the value of the \lstinline[language=Protelis]!initial! expression.
The \lstinline[language=Protelis]!code! block is then evaluated,
and its result is returned (and used as value for \lstinline[language=Protelis]!v! in the subsequent round).
For instance, in \Cref{pt:round-count} a field of local round counts is maintained.

\begin{snippet}[h]
\begin{lstlisting}[language=Protelis]
rep(k <- 0) { k + 1 } // Initially 0, increment and store in each round
\end{lstlisting}
\caption{A counter of locally executed rounds.}
\label{pt:round-count}
\end{snippet}

The \lstinline[language=Protelis]!if(condition) {then} else {otherwise}! expression
requires \lstinline[language=Protelis]!condition! to evaluate to a Boolean value;
if such value is \lstinline[language=Protelis]!true!, the \lstinline[language=Protelis]!then! block is evaluated and
the value of its last expression returned,
while if the value of \lstinline[language=Protelis]!condition! is \lstinline[language=Protelis]!false!
the \lstinline[language=Protelis]!otherwise! code block gets executed,
and the value of its last expression returned.
Notably, \lstinline[language=Protelis]!rep! expressions that find themselves in a non-evaluated branch
lose their previously computed state,
hence restarting the state computation from the initial value.
This behaviour is peculiar of the field calculus semantics,
where the branching construct is lifted to a distributed operator with the meaning of domain segmentation~\cite{TOCL2019}.
For instance, the expression in \Cref{pt:pendulum} builds an ``integer pendulum'' bouncing indefinitely from \texttt{1} to \texttt{-1} and vice versa.
The logic is the following:
\begin{itemize}[leftmargin=48pt]
 \item[\textit{round 0:}] \lstinline[language=Protelis]!old! is initially \lstinline[language=Protelis]!0!,
so the \lstinline[language=Protelis]!else! branch of the \lstinline[language=Protelis]!if! is selected.
Evaluation of the inner \lstinline[language=Protelis]!rep! block sets \lstinline[language=Protelis]!count! to \lstinline[language=Protelis]!0!,
then evaluates \lstinline[language=Protelis]!count + 1! returning \lstinline[language=Protelis]!1!,
which is also the result of the evaluation of the \lstinline[language=Protelis]!if! expression and thus of the outer \lstinline[language=Protelis]!rep!'s body
and of the overall program. \emph{Result:} \texttt{1}.
 \item[\textit{round 1:}] \lstinline[language=Protelis]!old!'s value is \lstinline[language=Protelis]!1! from the previous iteration:
the ``then'' branch of the \lstinline[language=Protelis]!if! expression is selected,
and the \lstinline[language=Protelis]!rep!'s body has no previous value, so \lstinline[language=Protelis]!count! is set to 0.
Evaluation of the \lstinline[language=Protelis]!rep!'s body yields \lstinline[language=Protelis]!-1!,
which is again also the result of the evaluation of the \lstinline[language=Protelis]!if! expression and thus of the body of the outer \lstinline[language=Protelis]!rep!
and of the complete program. \emph{Result:} \texttt{-1}.
 \item[\textit{round 2:}] \lstinline[language=Protelis]!old! is \lstinline[language=Protelis]!-1! from the previous round,
the \lstinline[language=Protelis]!else! branch is selected as in \emph{round 0},
but it did not compute in the previous round, so the value of \lstinline[language=Protelis]!count! is re-initialized to \lstinline[language=Protelis]!0!,
causing the same behaviour of the first iteration, with the program returning \lstinline[language=Protelis]!1!. \emph{Result:} \texttt{1}.
\end{itemize}
\begin{snippet}[h]
\begin{lstlisting}[language=Protelis]
rep(old <- 0) { // begin from zero
  if (old > 0) {
    rep (count <- 0) { count - 1 }
  } else {
    rep (count <- 0) { count + 1 }
  }
}
\end{lstlisting}
\caption{A ``pendulum'' swinging at each round between \texttt{1} and \texttt{-1}.}
\label{pt:pendulum}
\end{snippet}
The other fundamental field calculus construct is \lstinline|nbr|, which captures communication with neighbours
 in both directions at once.
Given an expression \lstinline|nbr(e)| 
 the device evaluates expression \lstinline|e|,
  producing a value that will be shared with its neighbours,
 and the whole construct evaluates to a map from neighbours to corresponding evaluations of \lstinline|e| there.
Such maps (also called neighbouring fields) are then typically collapsed into a single value via
reduction and folding operations
functions such as \lstinline|foldMin|, \lstinline|foldMax|, and the like.
A simple but fundamental example involving \lstinline|nbr|
 is the implementation of a \emph{gradient}~\cite{audrito2017ult} depicted in \Cref{pt:distanceTo},
 i.e., a self-healing field of minimum distances from any device to source devices---which will also be exercised in the following sections.
It works as follows: source devices
(i.e., those for which field \lstinline|source| is \lstinline|true|)
return \lstinline|0|
(a source is at distance zero from itself),
 whereas non-source devices
 return the minimum value 
 of the gradient value \lstinline|d| shared by neighbours
 augmented by the corresponding distance.
This algorithm adapts to changes in the source set
 and in the connectivity structure as devices move or enter/quit the system,
 eventually stabilising to the correct values after some transient.

\begin{snippet}[h]
\begin{lstlisting}
def distanceTo(source, metric) {
  rep(d <- POSITIVE_INFINITY) {
    let shortestPathViaNeighbours = foldMin( // must be always executed
        POSITIVE_INFINITY, // base value
        nbr(d) + metric()
    )
    if (source) { 0 } else { shortestPathViaNeighbours }
  }
}
\end{lstlisting}
\caption{
A simple self-healing gradient in Protelis. Note that \lstinline|nbr(d) + metric()| must be computed outside of the \lstinline|if| branches:
information is sent and received when \lstinline|nbr| gets executed,
so even sources must run that instruction.
}
\label{pt:distanceTo}
\end{snippet}

\subsection{A Protelis Incarnation for Time-Fluid Scheduling}
\label{sec:incarnation}
In this section,
 we discuss how the model formalised in \Cref{sec:opsem}
 can be instantiated,
 and cover our implementation in Alchemist-Protelis.
Indeed, the operational semantics in \Cref{fig:networkSemantics}
 abstracts over certain aspects,
 which must be filled in by a proper implementation.
Specifically,
 an implementation must decide upon the following elements:
\begin{enumerate}
\item what are the triggers (i.e., the values $\anyvalue_t$ assumed by sensor $\strigger$) that cause a local round to be executed (i.e.,  an activation of rule [N-FIR]);
\item how the scheduling tree is specified, i.e., how the application developer can define a whole program $\prg$ in terms of a main script $\e$ together with its dependencies to other scheduling programs $\overline{\prg}$;
\item how the predicate $\enabled$ is implemented and specified, to effectively control the logic of scheduling upon the structure defined as per the previous point.
\end{enumerate}

Alchemist~\cite{PianiniJOS2013} is an event-driven simulator 
 where simulations are usually defined in a declarative way
%
 defining a network of nodes
 (there implementing the \emph{neighbouring} relation---see \Cref{sec-fc-semantics}),
 and, for each node, the script(s) to be executed 
 when a certain platform event (trigger) occurs.
In the proposed model, the script is a tree of program nodes, where a \emph{program node} is a named Protelis module
 and an arrow $\prg\to\prg'$ between program nodes defines a dependency.
So, a program node maps to $\prg$, i.e., to a $\e[\overline{\prg}]$,
 where the corresponding Protelis module maps to the field calculus program expression $\e$, and
 the dependencies on other program nodes capture 
 the relationship to children $\overline{\prg}$.
Therefore, following the operational semantics,
 it is sufficient to specify the Protelis module
 at the root of the tree
 for considering, in a round, the scheduling of the entire tree.
This way, the Protelis modules with no dependencies
 will be considered for execution first,
 subsequently those whose children nodes (i.e., schedulers)
 have already been considered, and finally the root module.

Considering a module for execution means 
 checking predicate $\enabled{\vtree,\overline\vtree^o,\senstate(\strigger)}$ for it.
It is the responsibility of the platform
 to call such a predicate 
 with actual parameters.
This could actually be implemented in several ways: e.g.,
 as a single function for the whole program tree,
 or as a separate function per individual module.
In the following, we use the latter approach
 and propose an API where each Protelis module 
 can specify its scheduling predicate 
 using a special function \lstinline|scheduler|
 whose inputs are obtained by parameter injection. For instance, in \Cref{pt:predicate}
the module \lstinline|a:b|, when considered for execution,
 is effectively ran if its \lstinline|scheduler| function
 returns \lstinline|true|.
Annotation \lstinline|@Input| is used to bind sensor values, e.g., corresponding to the \lstinline|trigger|
 that activated the round,
 the previous value of module \lstinline|a:b| itself (or \lstinline|null| on the first execution),
 and the previous value of another module \lstinline|c:d| (or \lstinline|null| on the first execution), respectively.
Notice that the logic of \lstinline|scheduler| is purely local---i.e., no aggregate constructs are admitted there.
Annotation \lstinline|@Changed| binds the parameter to a Boolean 
 stating whether the sensor value has changed
 from the last evaluation of the \lstinline|scheduler| function;
 this is a shorthand to avoid the creation of change-tracking scheduler modules using \lstinline|rep| (cf. \Cref{ssec:timer-based-sched}).
Note that by statically analysing all the Protelis programs referenced from the main script,
it is possible to infer the program tree
from source code files
with no additional configuration.

\begin{snippet}[h]
\begin{lstlisting}[language=Protelis]
module a:b
def scheduler(
  @Input("trigger") t,
  @Input("a:b") old,
  @Input("c:d") someSchedulerOutput,
  @Changed("c:d") didChange
) = // local scheduling logic
              
// function declarations
// main expression
\end{lstlisting}
\caption{Per-module implementation of predicate $\enabled{\vtree,\overline\vtree^o,\senstate(\strigger)}$.}
\label{pt:predicate}
\end{snippet}

\subsection{Examples of Time-Fluid Aggregate Computing}
\label{sec:examples}

In this section,
 we provide examples adopting our scheduling approach
 for time-fluid aggregate computations.
%
%
In the following,
triggers provided by the platform 
(i.e., the values returned by sensor $\strigger$---see \Cref{sec-fc-semantics})
will be highlighted in {\color{blue}blue}.
%
%
As we will see,
 it is often the case that the \lstinline|scheduler| functions
 (proxies for the calculus parameter $\enabled$)
 are essentially n-ary logical disjunctions (ORs) of their inputs (when Booleans) or simple predicates on these:
 this is natural, since a computation generally needs to run
 as soon as \emph{at least one} of its potential causes actually happened.

\newcommand{\EventFun}{\ensuremath{\Downarrow}}

\subsubsection{Timer-based scheduling}\label{ssec:timer-based-sched} 
In our first example,
we show a policy recreating the classic, timer-based execution model,
thus demonstrating how this approach subsumes the original execution model of field computations~\cite{TOCL2019}.

Consider a chain of events (and the corresponding triggers) local to one device such as the following.
\begin{center}
\begin{tikzpicture}[node distance=1cm and 2cm]
\footnotesize
\node[] (e0) [] {$\ldots$};
\node[ev] (e1) [right=of e0,label={above:{\lstinline|SENSOR("position")|}}] {$e_1$};
\node[ev] (e2) [right=of e1,label={below:{\lstinline|TICK|}}] {$e_2$};
\node[ev] (e3) [right=of e2,label={above:{\lstinline|SENSOR("temperature")|}}] {$e_3$};
\node[ev] (e4) [right=of e3,label={below:{\lstinline|SENSOR("position")|}}] {$e_4$};
\node[] (e5) [right=of e4] {$\ldots$};

\draw[leadsto] (e0) -- (e1);
\draw[leadsto] (e1) -- (e2);
\draw[leadsto] (e2) -- (e3);
\draw[leadsto] (e3) -- (e4);
\draw[leadsto] (e4) -- (e5);
\end{tikzpicture}
\end{center}
%
%
Assuming the platform exposes a sensor for obtaining the current local timestamp in seconds, it is possible to define a timer-like scheduler through a program as in \Cref{pt:timer}.
\begin{snippet}[h]
\begin{lstlisting}[language=Protelis]
module timefluid:every_second

// Applies condition to current and to the last value that made condition
// true, returning the truth value and storing the new state if needed
def updated(init, current, condition) =
  rep(state <- [init, true]) {
    let old = state.first()
    if (condition(current, old)) { [current, true] } else { [old, false] }
  }.get(1)

updated(-1, self.getCurrentTime()) { now, last -> now-last>=1 }
\end{lstlisting}
\caption{Timer-like scheduler.}
\label{pt:timer}
\end{snippet}

\noindent There is no declared scheduler, so it runs in every event.
Its output is a Boolean field mapping a Boolean value to each event.
We use notation $\EventFun$ in the following picture 
 to indicate the input (portion) and output of an event corresponding to a certain program execution.
\begin{center}
\begin{tikzpicture}[node distance=1cm and 2cm]
\footnotesize
\node[] (e0) [] {$\ldots$};
\node[ev] (e1) [right=of e0, label={[align=center]above:{\lstinline|getCurrentTime()|$~\to~55$\\$\EventFun$}},
label={[align=center]below:{$\EventFun$\\\lstinline|true|}}] {$e_1$};
\node[ev] (e2) [right=of e1, label={[align=center,label distance=0.6cm]above:{\lstinline|getCurrentTime()|$~\to~56$\\$\EventFun$}},
label={[align=center]below:{$\EventFun$\\\lstinline|true|}}] {$e_2$};
\node[ev] (e3) [right=of e2,label={[align=center]above:{\lstinline|getCurrentTime()|$~\to~56$\\$\EventFun$}},
label={[align=center]below:{$\EventFun$\\\lstinline|false|}}] {$e_3$};
\node[ev] (e4) [right=of e3,label={[align=center,label distance=0.6cm]above:{\lstinline|getCurrentTime()|$~\to~60$\\$\EventFun$}},
label={[align=center]below:{$\EventFun$\\\lstinline|true|}}] {$e_4$};
\node[] (e5) [right=of e4] {$\ldots$};

\draw[leadsto] (e0) -- (e1);
\draw[leadsto] (e1) -- (e2);
\draw[leadsto] (e2) -- (e3);
\draw[leadsto] (e3) -- (e4);
\draw[leadsto] (e4) -- (e5);
\end{tikzpicture}
\end{center}
Notice that you could not actually schedule ``every second'' 
 if the underlying platform events 
 are not triggered with a second or sub-second frequency;
 moreover, here we only generally consider soft real-time tasks.
So, for instance, a downstream program that needs to run at most once per second will not be scheduled in round $e_3$.
Such a program can be defined with a scheduling predicate logic that merely reuses the output of module \lstinline|timefluid:every_second| as in \Cref{pt:reuse}.
\begin{snippet}[h]
\begin{lstlisting}[language=Protelis]
module timefluid:some_program

// Special function 'scheduler' maps input scheduling fields to Boolean
def scheduler(@Input("timefluid:every_second") new_t) = new_t

// main expression
\end{lstlisting}
\caption{A scheduling predicate logic that merely reuses the output of module \lstinline|timefluid:every_second|.}
\label{pt:reuse}
\end{snippet}
Function \lstinline|scheduler| is used by the platform to control the scheduling of \lstinline|some_program|:
it is invoked with the set of triggers of the current event,
and other input parameters corresponding to the upstream scheduling programs.

An alternative would be to define a base timer trigger \lstinline|EVERY_SECOND| at the platform level that is issued every second.
\begin{center}
\begin{tikzpicture}[node distance=1cm and 2cm]
\footnotesize
\node[ev] (e1) [label={above:{\lstinline|EVERY_SECOND|}}] {$e_1$};
\node[ev] (e2) [label={[align=center]below:{\lstinline|SENSOR("xxx")|}},right=of e1] {$e_2$};
\node[] (e3) [right=of e2] {$\ldots$};
\node[ev] (e4) [label={above:{\lstinline|EVERY_SECOND|}},right=of e3] {$e_4$};
\node[ev] (e5) [label={[align=center]below:{\lstinline|MESSAGE_TIMEOUT|}},right=of e4] {$e_5$};

\draw[leadsto] (e1) -- (e2);
\draw[leadsto] (e2) -- (e3);
\draw[leadsto] (e3) -- (e4);
\draw[leadsto] (e4) -- (e5);
\end{tikzpicture}
\end{center}
Once we have a user-defined program or platform-level timer, derived timers could be defined as in \Cref{pt:derived-timers}.
\begin{snippet}[h]
\begin{lstlisting}[language=Protelis]
module timefluid:every_minute

import platform.Triggers.EVERY_SECOND

// OPTION A) restricting the domain of a high-level scheduler
def scheduler(@Input("timefluid:every_second") s) = s
// OPTION B) restricting the domain of a low-level, platform scheduler
def scheduler(@Input("trigger") t) = t == EVERY_SECOND 

// returns true once every 60 timer ticks
rep(old <- -1) { old + 1 } % 60 == 0
\end{lstlisting}
\caption{Example timers derived from domain restriction.}
\label{pt:derived-timers}
\end{snippet}


\subsubsection{Reacting to changes in the local context}
On the opposite side of the spectrum of possible policies we have purely reactive execution:
the local field computation is performed only if
there is a change in the value of any available sensors (\lstinline[language=Protelis]!SENSOR(".*")!);
if a message with new information is received (\lstinline[language=Protelis]!MESSAGE_RECEIVED!);
or if a message is discarded from the neighbour knowledge base (\lstinline[language=Protelis]!MESSAGE_TIMEOUT!),
for instance because the sender of the original message is no longer available---as in \Cref{pt:reactive}.
%
%
\begin{snippet}[h]
\begin{lstlisting}[language=Protelis]
module timefluid:some-program

import platform.Triggers.*

def scheduler(@Input("trigger") t) =
  t == SENSOR(".*") || // Regular expression ".*" matches any sensor name
  t == MESSAGE_RECEIVED("timefluid:some-program") ||
  t == MESSAGE_TIMEOUT("timefluid:some-program")
  
// main expression
\end{lstlisting}
\caption{A purely reactive scheduling policy.}
\label{pt:reactive}
\end{snippet}

\noindent In general,
 reacting to sensors provides a way to wake a computation up,
 and reacting to messages enables such an activation to spread around the system
 as well as to keep the system computing until the output stabilises.

\subsubsection{Aggregate computations as schedulers: crowd density estimation driving alerting and crowd steering}
Now,
we articulate a case in which the result of an aggregate computation is the cause for another computation to get triggered.
Monitoring crowds in large events is a typical scenario for field-based coordination, as mentioned in \Cref{sec:intro}. There, two tasks are usually performed: monitoring people density across the event area, to detect presence and growth of crowds, and steering people away from such crowds to avoid further increase in density and potential fatal events.
%
Both tasks can be implemented in aggregate computing via appropriate fields: one for the estimated density, one for steering people.
For the sake of efficiency, we would like to update the crowd steering field only when there is a noticeable change in the perceived density of the surroundings.

To do so, in \Cref{pt:scr} we write a Protelis program
leveraging the SCR pattern~\cite{Coordination2019-sgcg}
to partition space in regions 300 meters-wide and compute the average crowd density within them.
\begin{snippet}[h]
\begin{lstlisting}[language=Protelis]
module timefluid:steering:density
import ...

def scheduler(@Input("trigger") t) =
  t == SENSOR("people_count") ||
  t == MESSAGE_RECEIVED("timefluid:steering:density") ||
  t == MESSAGE_TIMEOUT("timefluid:steering:density")
  
let distToLeader = distanceTo(S(300)) // network partitioning
// sum of all the perceived people
let count = summarize(distToLeader, env.get("people_count"), sum)
// computes an upper bound to the radius
let radius = summarize(distToLeader, distToLeader, max)
count / (2 * PI * radius)//approximates crowd density as (people count/area)
\end{lstlisting}
\caption{Local crowd density estimation through the SCR pattern.}
\label{pt:scr}
\end{snippet}

\noindent Functions \lstinline[language=Protelis]!S!
(network partitioning with the desired grain),
\lstinline[language=Protelis]!summarize!
(aggregation of data over a spanning tree and partition-wide broadcast of the result),
and \lstinline[language=Protelis]!distanceTo!
(computation of distance)
come from the Protelis-lang library shipped with Protelis~\cite{SASO2017-protelislang}.
%
%
Now that density computation is in place,
the platform reifies its final result as a local sensor,
which can in turn be used to drive another computation,
framed in \Cref{pt:dependent},
that determines when the steering computation should be scheduled.
%
%
\begin{snippet}[h]
\begin{lstlisting}[language=Protelis]
module timefluid:steering:steering-scheduler

def scheduler(@Changed("timefluid:steering:density") d) = d

changed(exponentialBackOff(env.get("timefluid:steering:density"),0.1)){
  cur, old -> abs(cur - old) > 0.5
}  
\end{lstlisting}
\caption{A computation depending on the current local crowd density estimate.}
\label{pt:dependent}
\end{snippet}
%
%
%

\noindent There, a low pass filter \texttt{exponentialBackOff} avoids to get the program running in case of spikes
(e.g. due to the density computation re-stabilisation).
Note that access to the density computation is realised by accessing a sensor
with the same name of the \lstinline[language=Protelis]!module! containing the density evaluation program,
thus reifying a causal chain between field computations.
Finally, the steering program in \Cref{pt:steering} would predicate on the corresponding scheduler;
these modules altogether define the following application tree.
\begin{snippet}[h]
\begin{lstlisting}[language=Protelis,escapechar=|]
module timefluid:steering:steering

def scheduler(@Input("timefluid:steering:steering-scheduler") sched) = sched

// Steering logic (see, e.g., |\cite{BealIEEEComputer2015}|)
\end{lstlisting}
\caption{The scheduling of the final crowd steering program.}
\label{pt:steering}
\end{snippet}

\noindent We represent Protelis modules as square nodes.
Sometimes, we may also explicitly show their corresponding \lstinline|scheduler| functions, as rounded rectangles left-adjacent to the program nodes.
We also denote potential triggers as labels.
Solid arrows denote potential causal dependencies between Protelis programs. Dashed arrows denote potential causal dependencies \emph{on triggers}.
Therefore, arrows capture the inputs to \lstinline|scheduler| functions.
Dotted lines, instead, are used to connect parts of the picture to informal comments.

\begin{center}
\begin{tikzpicture}[node distance=1cm and 1.5cm]
\node[program] (density) [] {\texttt{density}};
\node[scheduler] (densityS) [left=0.0cm of density] {$s_1$};
\node[] (sensor) [left=of density] {\texttt{people\_count}};
\node[program] (steeringsched) [right=of density] {\texttt{steering-scheduler}};
\node[scheduler] (steeringschedS) [left=0.0cm of steeringsched] {$s_2$};
\node[program] (steering) [right=of steeringsched] {\texttt{steering}};
\node[scheduler] (steeringS) [left=0.0cm of steering] {$s_3$};
\draw[->,dashed] (sensor) -- (densityS);
\draw[->] (density) -- coordinate[midway](m1) (steeringschedS);
\draw[->] (steeringsched) -- coordinate[midway](m2) (steeringS);

\node[] (lbl0) [below=of densityS,align=center] {schedules on\\people count change};
\node[] (lbl1) [below=of steeringschedS,align=center] {schedules on\\density change};
\node[] (lbl2) [below=of steeringS,align=center] {schedules when\\input is true};

\node[] (msg11) [above left=1.2cm and 0.2cm of densityS] {\texttt{MESSAGE\_RECEIVED}};
\node[] (msg12) [above right=0.5cm and -1cm of densityS] {\texttt{MESSAGE\_TIMEOUT}};


\draw[dotted] (lbl0) -- (densityS);
\draw[dotted] (lbl1) -- (steeringschedS);
\draw[dotted] (lbl2) -- (steeringS);

\draw[->,dashed] (msg11) -- (densityS);
\draw[->,dashed] (msg12) -- (densityS);
\end{tikzpicture}
\end{center}
%

%
%

\subsubsection{Self-scheduling: the case of a gradient program that self-determines whether to slow down or speed up}
As formalised in the previous section, other than due to triggers and lower-level schedulers, a program could self-schedule itself.
For instance, consider the program in \Cref{pt:self-scheduling} that computes a gradient (see \Cref{sec:protelisprimer})
 and then analyses the gradient output to determine
 whether it should slow down or not.
\begin{snippet}[h]
\begin{lstlisting}[language=Protelis]
module timefluid:gradient
import protelis:state:time
import ...

def scheduler(@Input("trigger") t,
              @Input("timefluid:every_minute") every_minute,              
              @Input("timefluid:gradient") g) =
  // run at least once every minute
  every_minute ||
  // run faster when self says so
  (t == "every_second" && g.get(1) == "fast")

let g = distanceTo(env.get("source"))
[g, if (isSignalStable(g, THRESHOLD)) { "slow" } else { "fast" }]
\end{lstlisting}
\caption{The scheduling of the final crowd steering program.}
\label{pt:self-scheduling}
\end{snippet}

If the gradient value \lstinline|g| does not change for some time \lstinline|THRESHOLD|
(tracked by the library function \lstinline|isSignalStable|),
then the computation can be slowed down:
this instruction is reified as a string (\lstinline|"slow"| in this case) that is returned as the second element of the 2-element-tuple output.
Then, the matcher function \lstinline|scheduler|
 leverages two platform-level timers, \lstinline|every_second| and \lstinline|every_minute|, to set the scheduling frequency according to the self-instruction (\lstinline|g.get(1)|).
\begin{center}
\begin{tikzpicture}[node distance=0.5cm and 1cm]
\node[program] (g) [] {\texttt{gradient}};
\node[] (sensor1) [align=center,above left=of g] {\texttt{every-second}};
\node[] (sensor2) [align=center,below left=of g] {\texttt{every-minute}};
\draw[->,dashed] (sensor1) -- (g.170);
\draw[->,dashed] (sensor2) -- (g.190);
\draw[->] (g.330) -- ++(0,-0.6) -- ++(-1.1,0) 
	node [midway,below] {$(g,s)$} -- (g.210);
\end{tikzpicture}
\end{center}
This pattern can promptly switch the computation frequency from \lstinline|"fast"| to \lstinline|"slow"|,
since it re-evaluates its state every second,
while switching from \lstinline|"slow"| to \lstinline|"fast"|
can be only done at the frequency set for the \lstinline|"slow"| progression
(in this case, every minute).

\section{Evaluation}\label{sec:eval}

In this section,
we leverage our implementation to get insights on how a time-fluid version of a field-based coordination system compares
to a classic time-driven version,
showing that the proposed approach can in many relevant cases achieve \emph{both} better performance
(lower error)
and lower global resources usage.
Then, we discuss how the time-fluid version of a field-based coordination system can be finely tuned to respond to the amount of changes in a system,
enabling the designer to focus on the desired performance,
and having the system autonomously slow down (or stop altogether) when the ``amount of change'' happening is sufficiently low.

\subsection{Experimental setup}

We consider three different setups, that we will shortly refer to with the \emph{gradient}, \emph{moving}, and \emph{channel} experiments.
In all the three scenarios,
we leverage a distance estimate between each node and a target (a \emph{gradient}~\cite{audrito2017ult}---see \Cref{sec:protelisprimer}).
In the former two cases, the gradient is also the object of our experiments,
while in the latter multiple distance estimations are exploited to build a broadcast~\cite{SASO2017-protelislang},
and a spatially-redundant, self-adaptive communication channel~\cite{TAAS2017} between two communicating endpoints.
Distance estimation is central in our evaluation
as it is one of the most common building blocks over which other,
more elaborate forms of coordination get built~\cite{naco2013,TOMACS2018,FGCS2020-scr}.
Computing distance from a source without a central coordinator in arbitrary networks is a representative application of aggregate computing,
for which several implementations exist~\cite{TOMACS2018,audrito2017ult}.
In this work,
since the goal is to explore the behaviour of the time-fluid program version
rather than the efficiency of the distance estimation algorithm itself,
we use a variant of the adaptive Bellman-Ford~\cite{bellmanford} algorithm
exploiting the recent \texttt{share} primitive~\cite{LMCS2020-share},
even though it is known not to be the most efficient
implementation for the task at hand~\cite{audrito2017ult}.

\noindent The baseline for assessing our proposal is a \emph{classic} implementation in the framework of aggregate computing:
time-driven, unsynchronised, and fair scheduling of rounds set at 1Hz.
We compare the classic approach with a time fluid version whose structure will be presented in the experiment-specific sections that follow.
The time-fluid version is purely \emph{reactive}, and its performance depends on two parameters:
\begin{itemize}
 \item[$\epsilon$] (\emph{tolerance}), measured in meters, is the distance required for a node to consider itself to have moved from its previous position;
 \item[$\lambda$] (\emph{mean arrival frequency}), measured in Hz, represents the reciprocal of the mean time required by a network message to be prepared, sent to a neighbour, and decoded by such neighbour.
\end{itemize}
Higher values of $\epsilon$ lead to less frequent evaluation by the time fluid version of the system,
at the price of a greater expected error.
Values of $\lambda$ mostly depend on the performance of devices composing the system and of the communication network.
We suppose devices to have similar performance,
and we model the network latencies to be Weibull-distributed
as suggested by the literature on traffic modelling~\cite{weibull}.
We set the Weibull distribution shape $\alpha=1$ and scale $\beta=\lambda^{-1}$;
this way, the expected value for samples of such distribution is $\mathbb{E}=\beta=\lambda^{-1}$.
In other words, with this network model,
we have a direct link between the value of $\lambda$ and the mean time
required to deliver a message, which will be $\lambda^{-1}$ seconds.
In the remainder of this work, we will refer to $\lambda^{-1}$ to indicate the network performance,
which can be interpreted as the mean time required for a message to be delivered and processed by the receiver.

Devices are deployed differently in each example,
and details of each deployment are presented in the subsection detailing each experiment.
In all experiments, though, mobile nodes share the same constant speed $\|\vec{v}\|$,
which is among the controlled variables.
\begin{table}[t]
    \begin{center}
        \begin{tabular}{| c | c | c |}
        \toprule
        Name & Description & Values\\
        \midrule
        $\lambda^{-1}$ & mean network latency & $\{0.01, 0.03, 0.1, 0.3, 1\}\, s$ \\
        \hline
        $\epsilon$ & distance tolerance & $\{0, 0.01, 0.03, 0.1, 0.3, 1, 3\}\, m$ \\
        \hline
        $\|\vec{v}\|$ & mobile nodes' speed & $\{0, 0.01, 0.03, 0.1, 0.3, 1, 3\}\, \nicefrac{m}{s}$ \\
        \hline
        algorithm & implementation type & classic, time-fluid \\
        \hline
        scenario & position and program of devices & gradient, moving, channel \\
        \bottomrule
        \end{tabular}
        \caption{Independent variables for the evaluation.}
        \label{t:vars}
    \end{center}
\end{table}
All controlled (independent) variables are summarised in \Cref{t:vars}.
For each combination in the Cartesian product of the variables' values,
we perform multiple simulation runs
(50 for the channel scenario, and 100 for the gradient and moving scenarios),
changing the simulation seed.

\begin{table}[t]
    \begin{center}
        \begin{tabular}{| c | c | c |}
        \toprule
        Name & Description & Notes \\
        \midrule
        $\delta$ & error of a single device w.r.t an oracle & \\
        \hline
        $\rho$ & rounds executed by a single device &  \\
        \hline
        $\mathbb{E}(\delta)$ & mean error w.r.t. an oracle & proxy for the global error \\
        \hline
        $\mathbb{E}(\rho)$ & mean round count & proxy for resource cost \\
        \hline
        $\sigma(\rho)$ & standard deviation of round count & proxy measure of asymmetry \\
        \hline
        $\int_0^t\mathbb{E}(\delta)\,dt\cdot\mathbb{E}(\rho)$ & cumulative error times round count & proxy for efficiency \\
        \bottomrule
        \end{tabular}
        \caption{Metrics for the evaluation.}
        \label{t:metrics}
    \end{center}
\end{table}

Our metrics are summarised in \Cref{t:metrics}.
We measure the error $\delta$ of any device with respect to an oracle.
In the case of the gradient,
the error is the distance between the correct value as provided by an oracle and the value as computed by the executing algorithm;
in the case of the channel,
$\delta=0$ if the algorithm being executed and the oracle agree on whether a device is part of the channel,
and $\delta=1$ otherwise.
We count, for each device, how many rounds have been executed ($\rho$).
For the classic version of the algorithm,
every round contributes to the count with a unitary increment of the value.
For the time-fluid versions,
a round increments $\rho$ only if the target algorithm is effectively scheduled for execution and performed,
the evaluation of upstream schedulers is not considered.
Starting from these two raw metrics, we derive the following ones:
\begin{itemize}
 \item $\mathbb{E}(\delta)$: mean error across all devices, as a proxy for the global network error;
 \item $\mathbb{E}(\rho)$: mean count of rounds across all devices, as a proxy for global resource usage (more rounds imply more network communications and power usage);
 \item $\sigma(\rho)$: standard deviation of rounds across all devices, as a proxy for asymmetry in execution among devices,
 with higher values indicating that some devices run more frequently than others---higher standard deviations also witness a more noticeable difference between the time-fluid and the classic version
 (whose $\sigma(\rho)$ is the one closer to zero because most devices run at a similar frequency);
 \item $\int_0^t\mathbb{E}(\delta)\,dt\cdot\mathbb{E}(\rho)$: we multiply the mean cumulative count of rounds (resource usage) by the cumulative mean error of the network, as a proxy for efficiency (how much the error is reduced with respect to the additional rounds executed).
\end{itemize}

The simulation has been implemented in Alchemist~\cite{PianiniJOS2013},
writing the aggregate programs in Protelis~\cite{PianiniSAC2015}.
Data has been processed with Xarray~\cite{xarray},
and charts have been produced via matplotlib~\cite{matplotlib}.
All the experiments and the production of the charts presented in this work has been automated,
open-sourced, documented, and made available for easy reproduction\footnote{\url{https://github.com/DanySK/Experiment-2020-LMCS-TimeFluid}}~\cite{experiments}.
Besides the source code,
we provide detailed instructions,
as well as all the means to execute the experiment in a containerised environment
(including a pre-configured Kubernetes pod, in case the reader has access to a cluster that can speed up the computation).

\begin{figure}
\begin{tikzpicture}
\node[program] (p1) [align=center] {Has position\\{}changed at least of $\epsilon$?};
\node[program] (p2) [below=of p1,align=center] {Has neighbourhood\\{}changed?};
\node[program] (p3) [below=of p2,align=center] {Has the gradient value\\{}of a neighbour\\{}changed at least of $\epsilon$?};

\node[] (poschange) [left=of p1] {\texttt{position-change}};
\node[] (nbr) [left=of p2] {\texttt{neighbours}};
\node[] (newmsg) [left=of p3] {\texttt{new-message("gradient")}};

\node[program] (g) [right=of p2] {\texttt{gradient}};

\draw[->,dashed] (poschange) -- (p1);
\draw[->,dashed] (nbr) -- (p2);
\draw[->,dashed] (newmsg) -- (p3);
\draw[->] (p1) -- (g);
\draw[->] (p2) -- (g);
\draw[->] (p3) -- (g);

\end{tikzpicture}
\caption{Application tree for the \emph{gradient} experiment.}
\label{fig:gradient-experiment}
\end{figure}
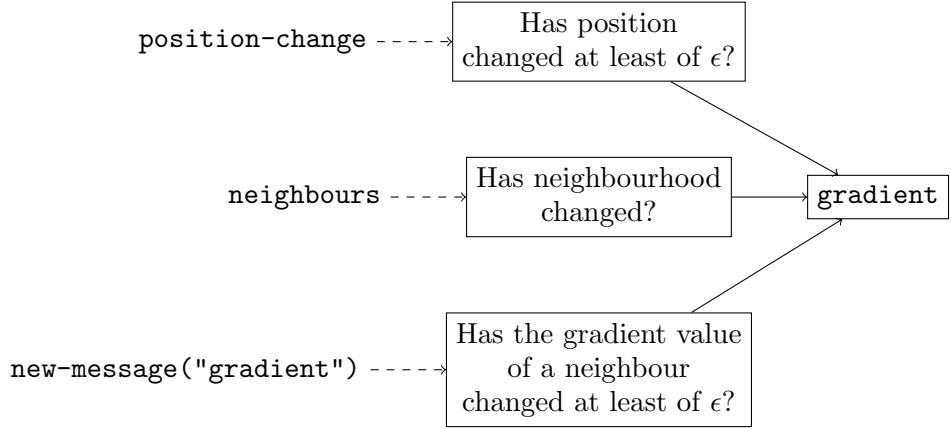

\subsubsection{Gradient experiment}

\begin{figure}[t]
 \begin{center}
  \includegraphics[width=.486\textwidth]{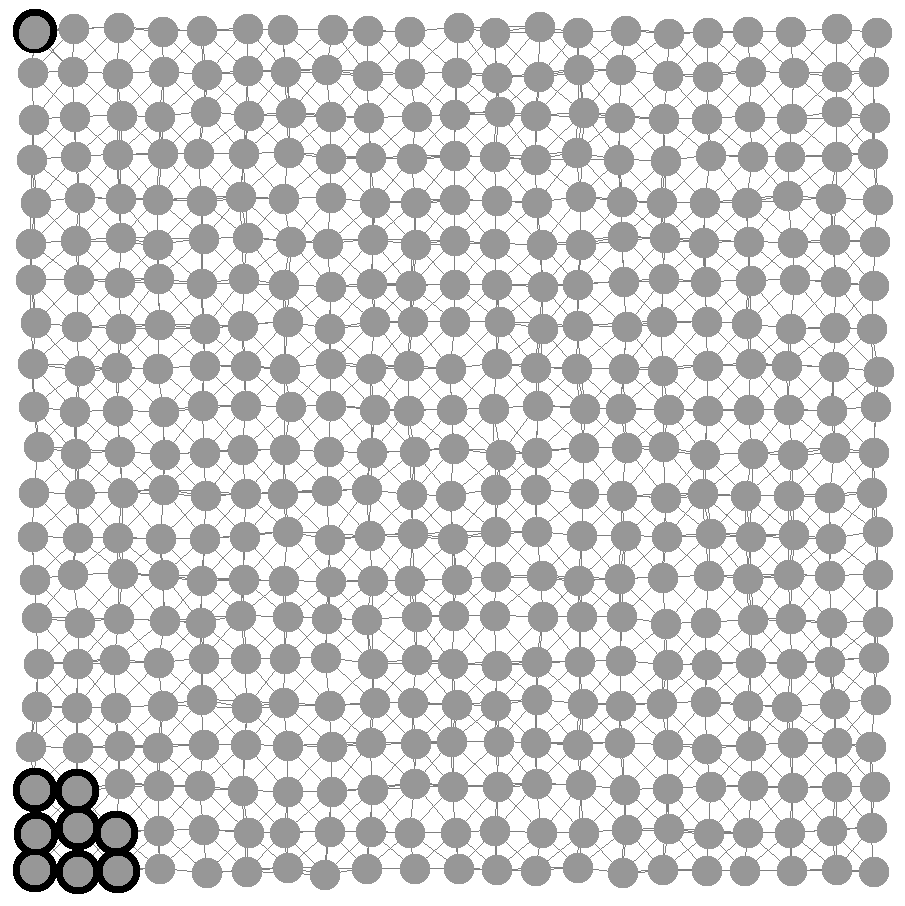}
  \includegraphics[width=.486\textwidth]{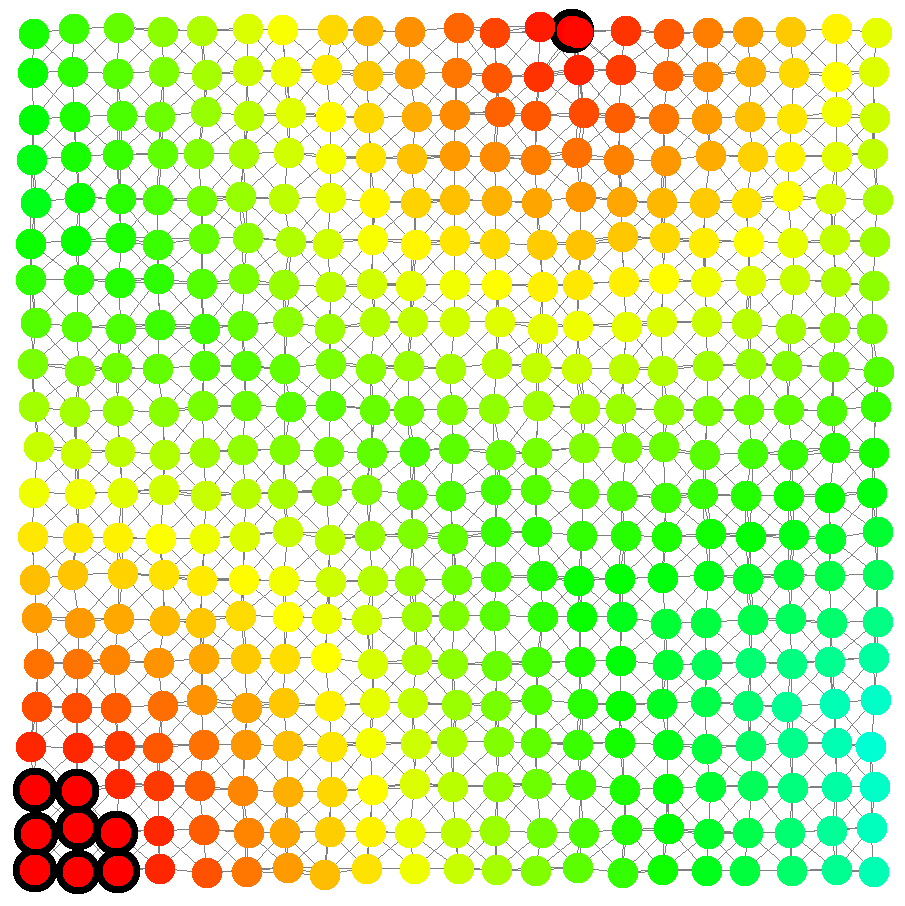}
 \end{center}
 \caption{
Snapshots of the simulation running in the gradient scenario.
Nodes are initially displaced in an irregular grid, with and additional device on the top left corner.
This device and those at the bottom left corner, depicted with a black border, are sources.
When running, the additional node on top oscillates between the left and right edges of the arena,
and the gradient must adjust accordingly.
We use colors to denote the gradient values:
 warmer colors denote lower gradient values (i.e., smaller distances, or greater proximity to a source),
 whereas cooler colors denote higher gradient values
 (i.e., bigger distances, or greater remoteness from sources).
}
 \label{snapshots:gradient}
\end{figure}

The goal of this setup is to measure how the time-fluid versions of the field-based coordination compare to a classic implementation
when a part of the network is stable,
and another is instead constantly changing due to the movement of a small subset of its components.
The time-fluid program is graphically shown in \Cref{fig:gradient-experiment}.

The simulation is executed in a 21x21 irregular grid of devices displaced in a square arena,
each located randomly within a disc centred on the corresponding position of a regular grid;
and a single mobile node positioned to the top left of the network,
free to move at velocity $\vec{v}$ whose speed $\|\vec{v}\|$ is constant,
and whose direction reverses when it reaches the leftmost and rightmost limits of the arena.
The mobile device and the leftmost devices at bottom are ``sources'',
and the goal for each device is to estimate the distance to the closest source.
Snapshots of the deployment and its ongoing simulation are depicted in \Cref{snapshots:gradient}.

\subsubsection{Moving experiment}
\label{experiment:moving}

\begin{figure}[t]
 \begin{center}
  \includegraphics[width=.486\textwidth]{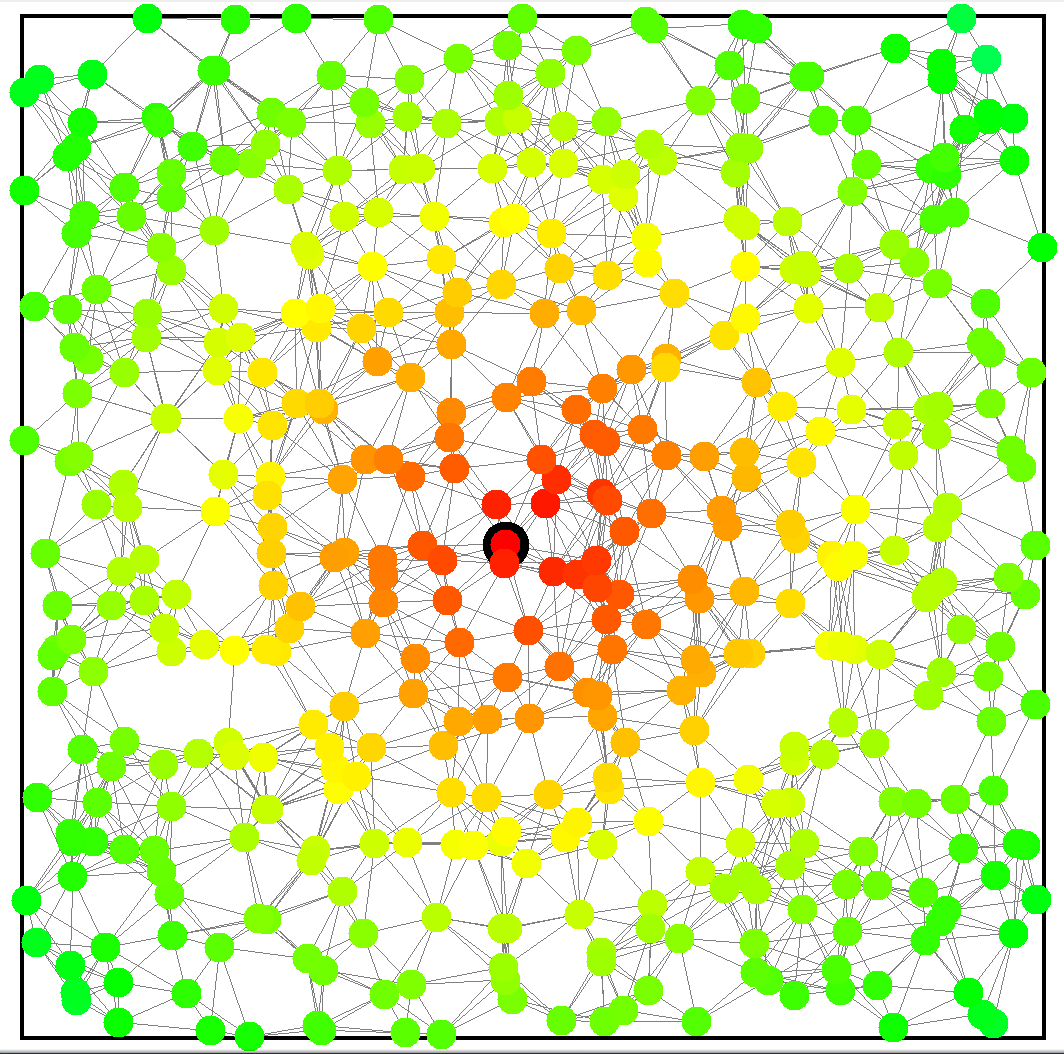}
  \includegraphics[width=.486\textwidth]{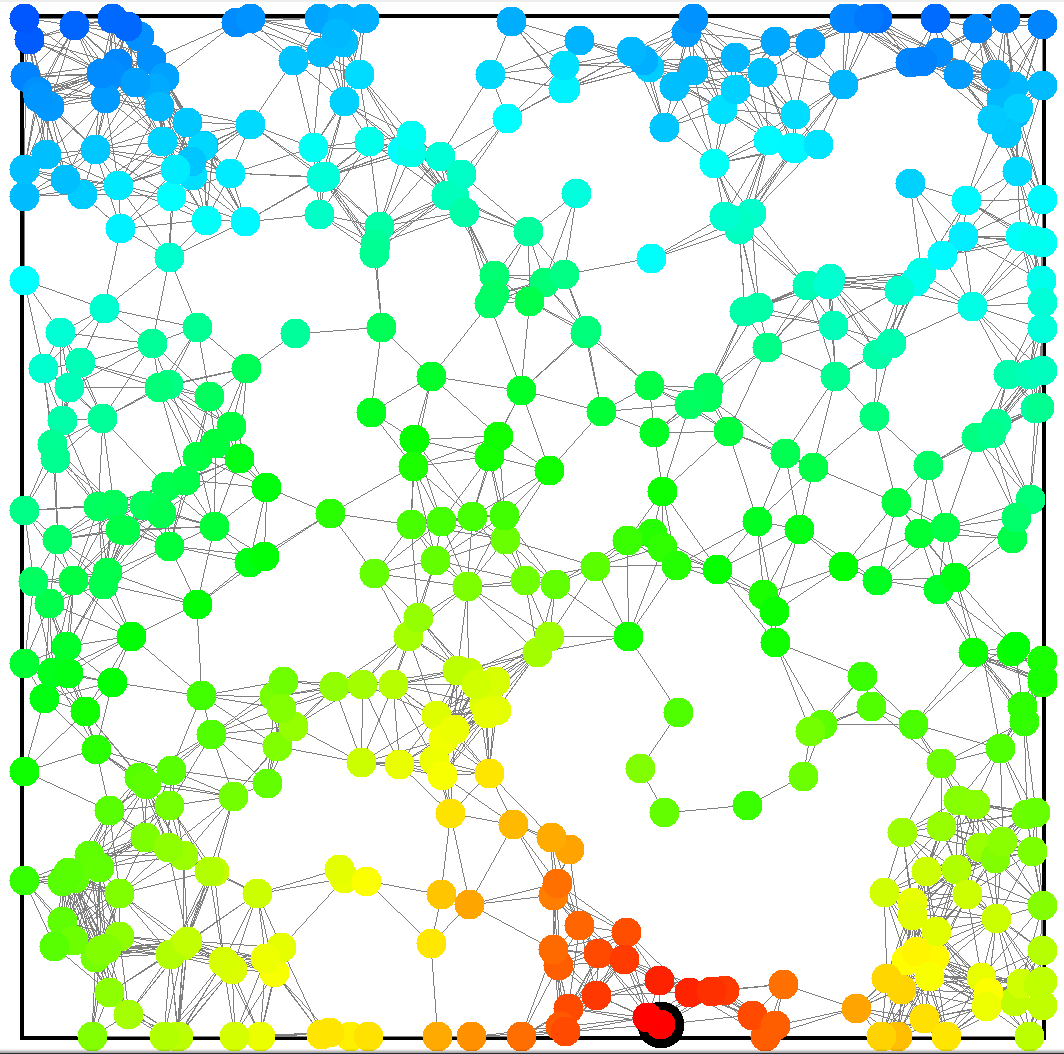}
 \end{center}
 \caption{
Snapshots of the simulation running in the moving scenario.
Nodes are randomly displaced in a square arena.
A device at the centre of the network is a source, depicted with a black border.
When running, all nodes move (including the source),
and the gradient must adjust accordingly.
}
 \label{snapshots:moving}
\end{figure}

This experiment's goal is to analyze how time-fluid and classic coordination compare when the system is entirely dynamic.
In particular,
we expect this case to show that the time-fluid version can autonomously tune the frequency at which devices compute
depending on what error is considered to be ``acceptable'' (in our case, the value of $\epsilon$).
Devices are initially displaced in a 21x21 irregular grid, enclosed in a square arena.
An additional device, marked as source, is located initially at the centre of the arena.
Devices are free to move within such arena with velocity $\vec{v}$;
speed $\|\vec{v}\|$ is constant within every experiment (see \Cref{t:metrics}),
while direction is determined with a uniform probability
and changed when nodes hit the boundary of the arena
or after they walked a length
determined by samples of a Pareto distribution with
scale $k=\nicefrac{1}{2}$ and
shape $\alpha=1$---thus creating a variant of Lévy Walks~\cite{levy-walks} for a physically constrained environment.
This kind of walks have selected as they reasonably approximate the walking behaviour of human beings~\cite{levywalk-pedestrians}.
Snapshots of the deployment and its ongoing simulation are depicted in \Cref{snapshots:moving}.

\begin{figure}
\begin{tikzpicture}[scale=0.8, transform shape]
\newcommand{\drawgrad}[1]{
\node[program] (#1p1) [below=0cm of #1,align=center] {Has position\\{}changed at least of $\epsilon$?};
\node[program] (#1p2) [below=of #1p1,align=center] {Has neighbourhood\\{}changed?};
\node[program] (#1p3) [below=of #1p2,align=center] {Has the gradient value\\{}of a neighbour\\{}changed at least of $\epsilon$?};

\node[] (#1poschange) [left=of #1p1,align=center] {Position\\{}changed?};
\node[] (#1nbr) [left=of #1p2,align=center] {\texttt{nbrs}};
\node[] (#1newmsg) [left=of #1p3,align=center] {New message\\{}
for \texttt{gradient}?};

\node[program] (#1g) [right=of #1p2] {\texttt{gradient}};

\draw[->,dashed] (#1poschange) -- (#1p1);
\draw[->,dashed] (#1nbr) -- (#1p2);
\draw[->,dashed] (#1newmsg) -- (#1p3);
\draw[->] (#1p1) -- (#1g);
\draw[->] (#1p2) -- (#1g);
\draw[->] (#1p3) -- (#1g);
}

\node[] (anchor1) [] {};
\drawgrad{anchor1};
\node[] (anchor2) [below=of anchor1p3] {};
\drawgrad{anchor2};

\node[program] (g1changed) [below=of anchor1g] {changed?};
\node[program] (g2changed) [above=of anchor2g] {changed?};

\node[program] (distbetween) [below right=of g1changed] {\texttt{distanceBetween}};
\node[program] (chan) [right=of distbetween] {\texttt{channel}};

\draw[->] (anchor1g) -- (g1changed);
\draw[->] (anchor2g) -- (g2changed);
\draw[->] (g1changed) -- (distbetween);
\draw[->] (g2changed) -- (distbetween);
\draw[->] (distbetween) -- (chan);
\draw[->] (g1changed) -- (chan);
\draw[->] (g2changed) -- (chan);
\end{tikzpicture}
\caption{Application tree for the \emph{channel} experiment.}
\label{fig:channel-apptree}
\end{figure}
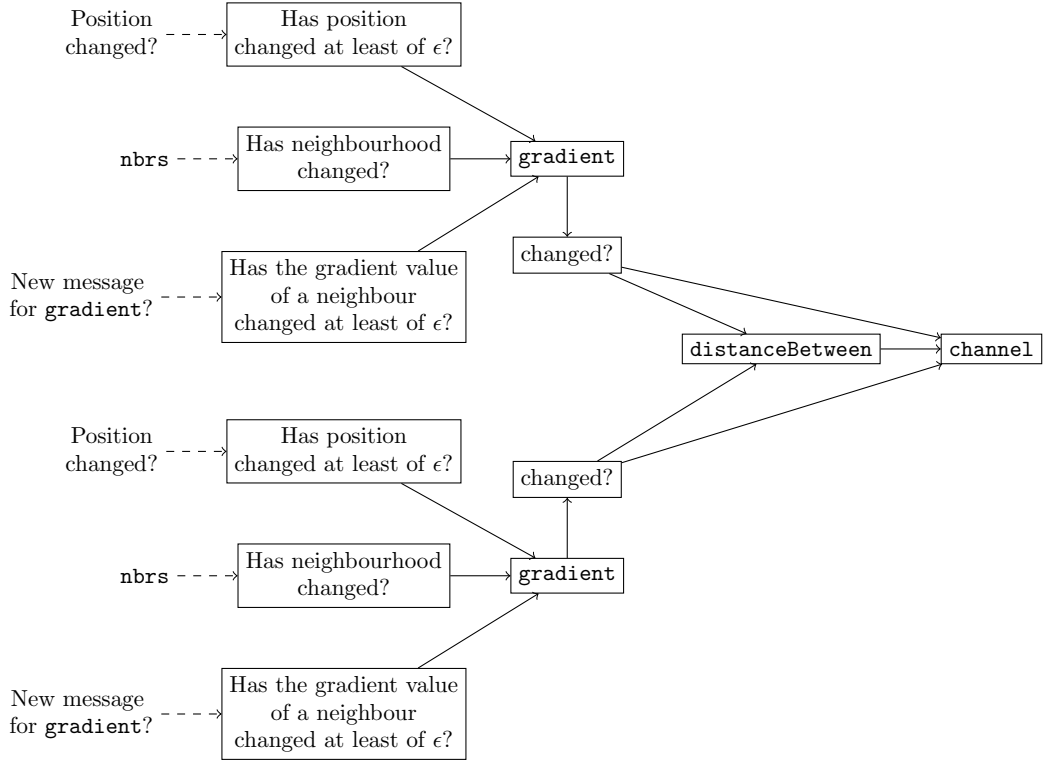

\subsubsection{Channel experiment}

\begin{figure}[t]
 \begin{center}
  \includegraphics[width=.94\textwidth]{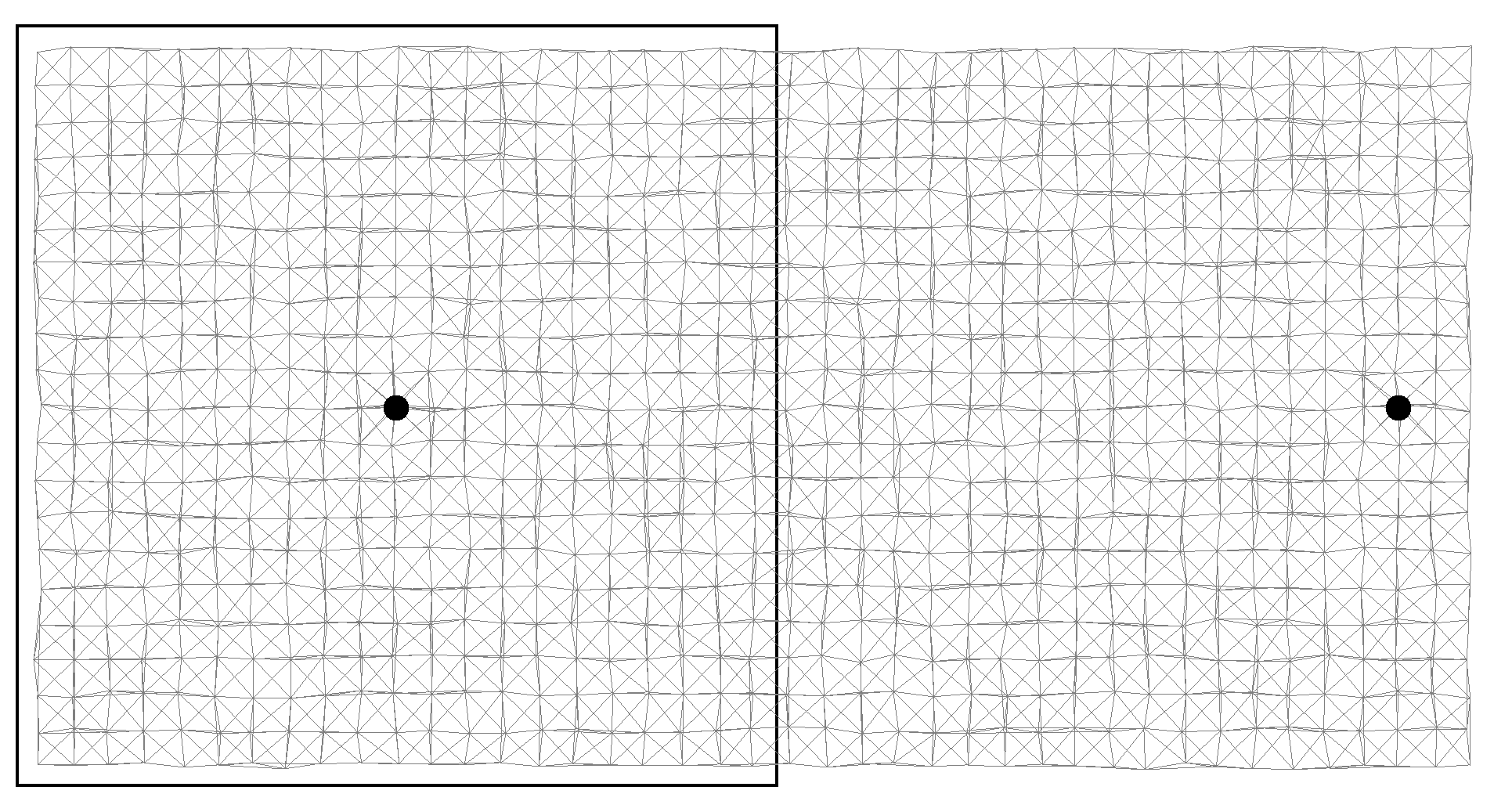}
  \includegraphics[width=.94\textwidth]{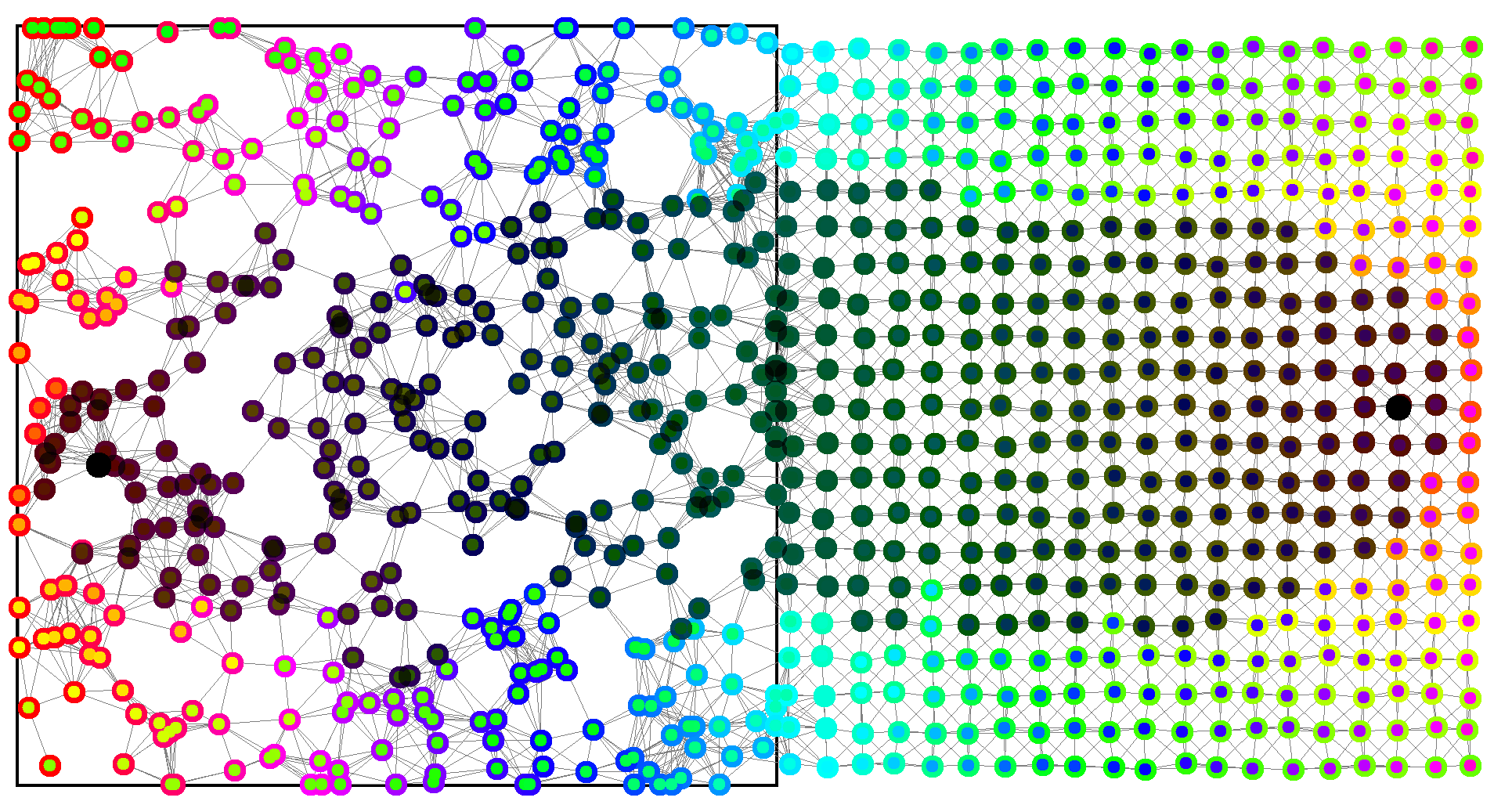}
 \end{center}
 \caption{
Snapshots of the simulation running in the channel scenario.
Nodes are randomly displaced in a rectangular irregular grid.
The left half of such grid is enclosed in a square arena, where devices are free to roam.
The two devices at the centre of each half of the network are the communication ends of the channel (marked in black).
When running, nodes within the square arena move (including the source),
and the system must keep the redundant channel up.
The dark (or shadowed) nodes in the bottom snapshot are the devices that make up (or ``belong to'') the channel, i.e., for which the \lstinline|channel| function returns \lstinline|true|.
}
 \label{snapshots:channel}
\end{figure}

This experiment is meant to investigate if and how the effects of time-fluid computations ``stack''
when a more elaborate setup is in place.
To this end, in this experiment we do not just run a gradient,
but we build a redundant communication channel between two devices.
\noindent{} This is done in a purely distributed setup by:
\begin{enumerate}
 \item propagating a gradient for each of the devices willing to establish the channel, thus computing for each device $i$ two distance values $d_i^{0}$ and $d_i^{1}$, respectively measuring distance from one communication end (labelled device $0$ and $1$);
 \item propagating along another gradient (i.e., broadcasting) the distance that one of the two communication ends perceives from the other, e.g., $d_0^1$ (or, equivalently in Euclidean folds, $d_1^0$); and
 \item considering as part of the channel all those nodes for which $d_i^{0} + d_i^{1} < d_0^1 + w$, with $w$ being the \emph{width} of the channel.
\end{enumerate}
In Protelis this can be coded as in \Cref{pt:channel}.

\begin{snippet}[h]
\begin{lstlisting}
def distanceBetween(a,b) = broadcast(a, distanceTo(b))
def channel(a,b,w) = distanceTo(a)+distanceTo(b) <= distanceBetween(a,b)+w
\end{lstlisting}
\caption{A simple channel pattern in Protelis.}
\label{pt:channel}
\end{snippet}

\noindent This algorithm lends itself to a decomposition in its building blocks
(two gradients and a gradient plus a broadcast),
and thus to a time-fluid modelling that is responsive to changes in the perceived distances, as shown in \Cref{fig:channel-apptree},
where the \texttt{distanceBetween} block is implemented with the homonym function of the Protelis-lang library~\cite{SASO2017-protelislang}.

In order to create a challenging environment,
we displace devices in a 42x21 irregular grid, and we enclose the left-hand half of them within a square arena,
letting communication cross such a boundary.
Devices outside the square arena remain still, while devices within the arena move with the same Lévy Walks described for the moving experiment in \Cref{experiment:moving}.
Snapshots of the deployment and its ongoing simulation are depicted in \Cref{snapshots:channel}.

\subsection{Classic and time-fluid coordination compared}

\begin{figure}[t]
 \begin{center}
  \includegraphics[width=.486\textwidth]{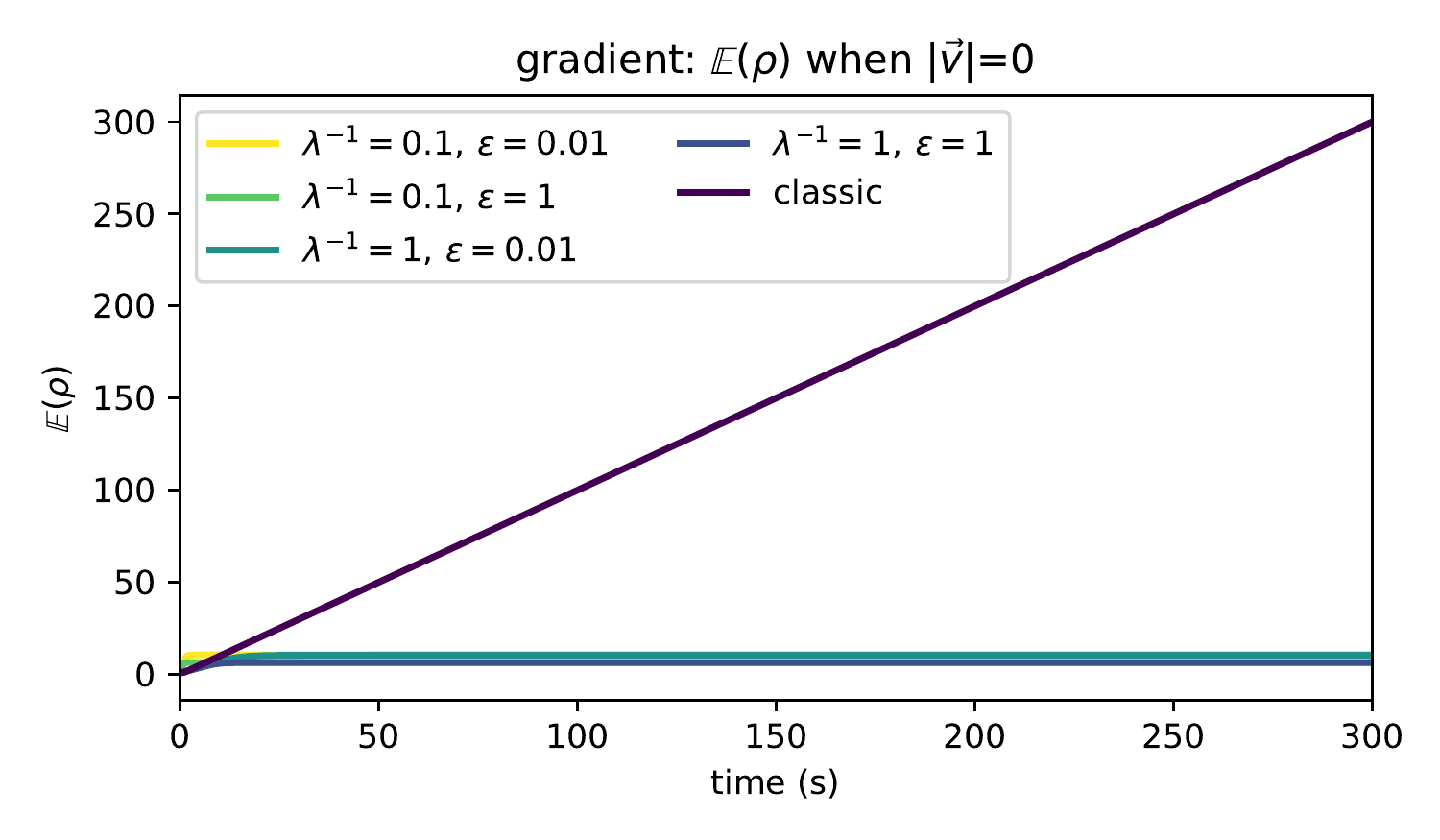}
  \includegraphics[width=.486\textwidth]{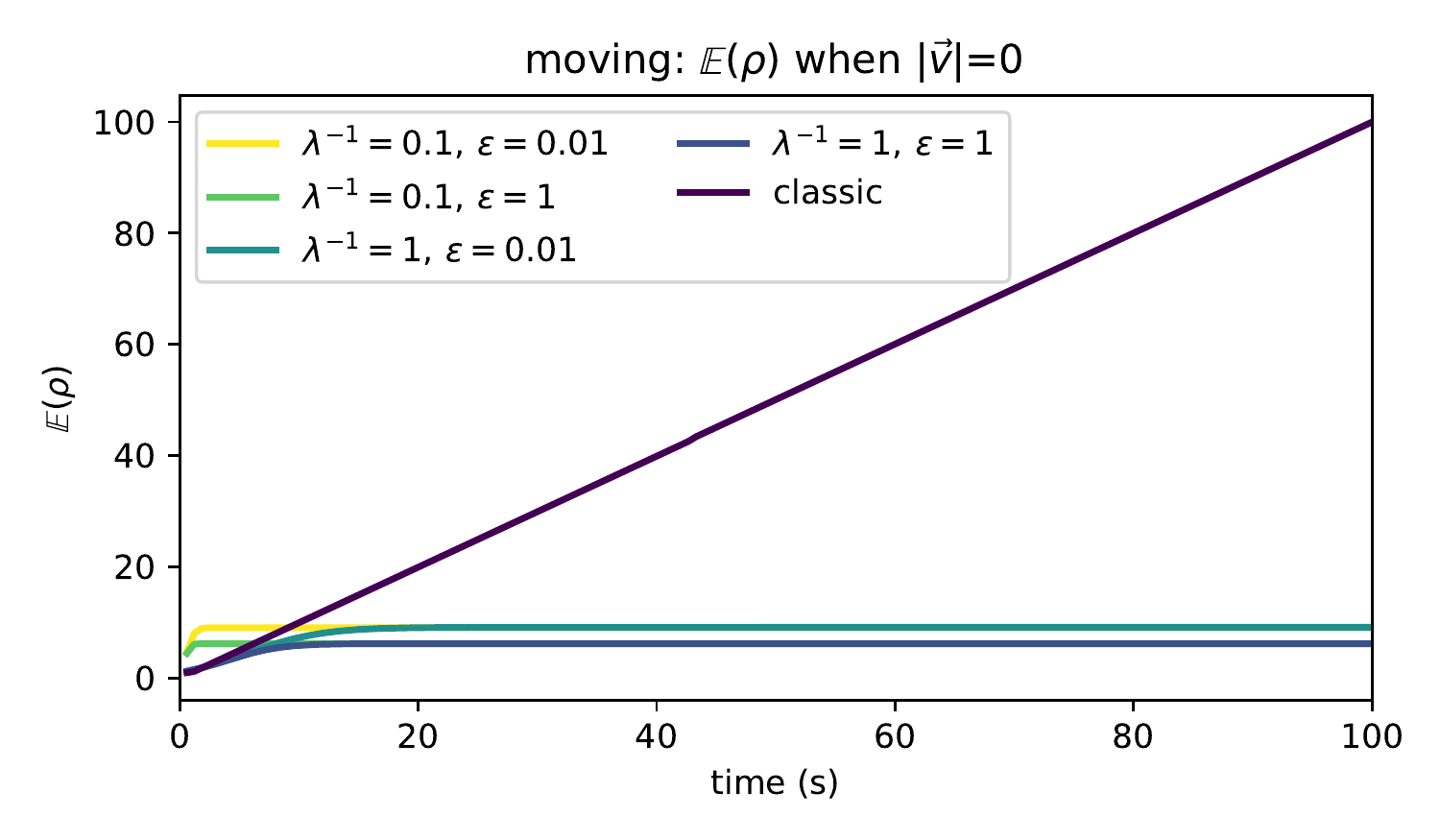}
 \end{center}
 \caption{
Compared cost (estimated using the count of executed rounds $\rho$ as proxy metric) in case of still devices,
for the gradient scenario (left) and the moving scenario (right).
Data shows the ability of the time-fluid version of the system to pause its computation when no relevant changes happen.
Faster devices on faster network with lower tolerance to error initially run more rounds than the classic version.
However, once stabilisation is reached, they stop, while the classic version keeps computing in order to sustain the existence of the computational field.
}
 \label{zerospeed}
\end{figure}

\begin{figure}[t]
 \begin{center}
  \includegraphics[width=.486\textwidth]{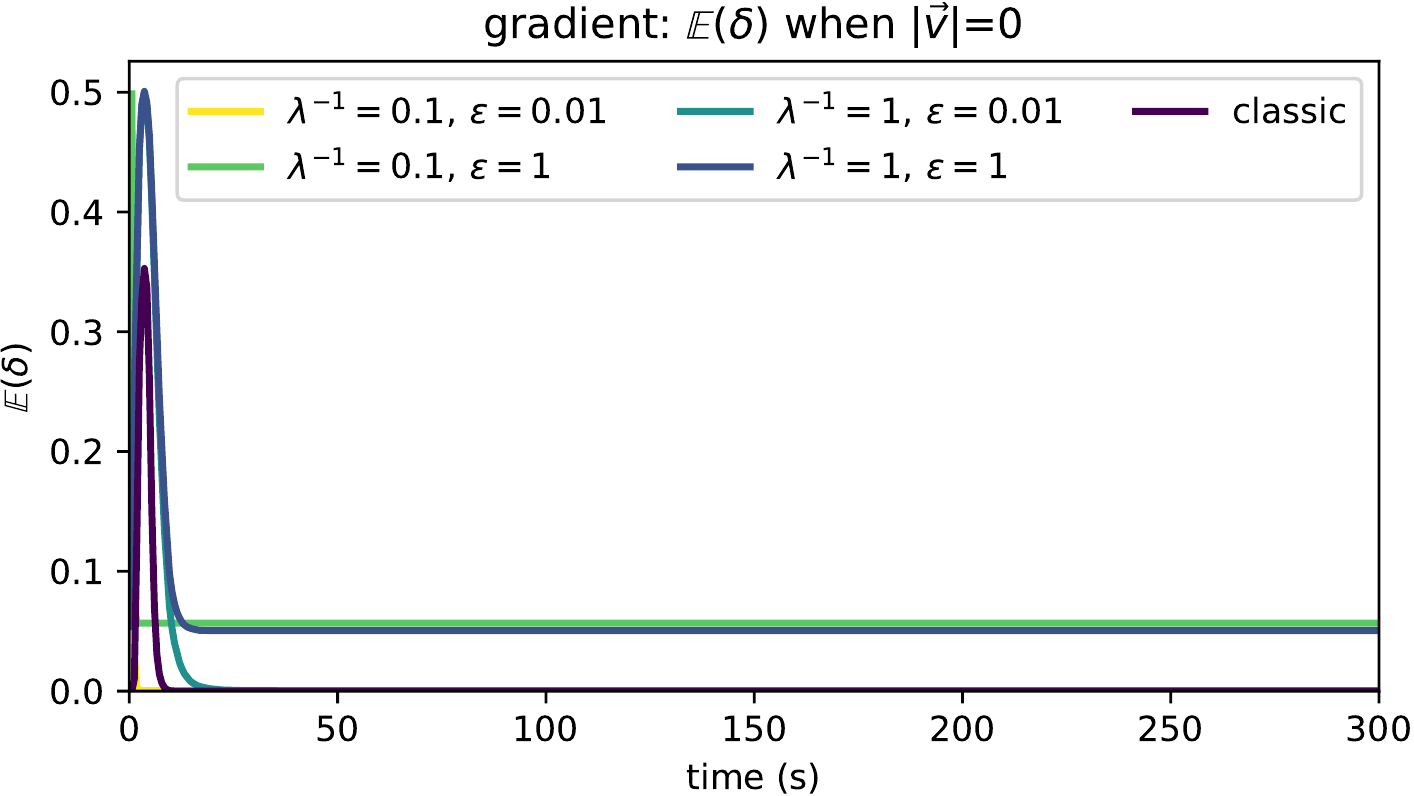}
  \includegraphics[width=.486\textwidth]{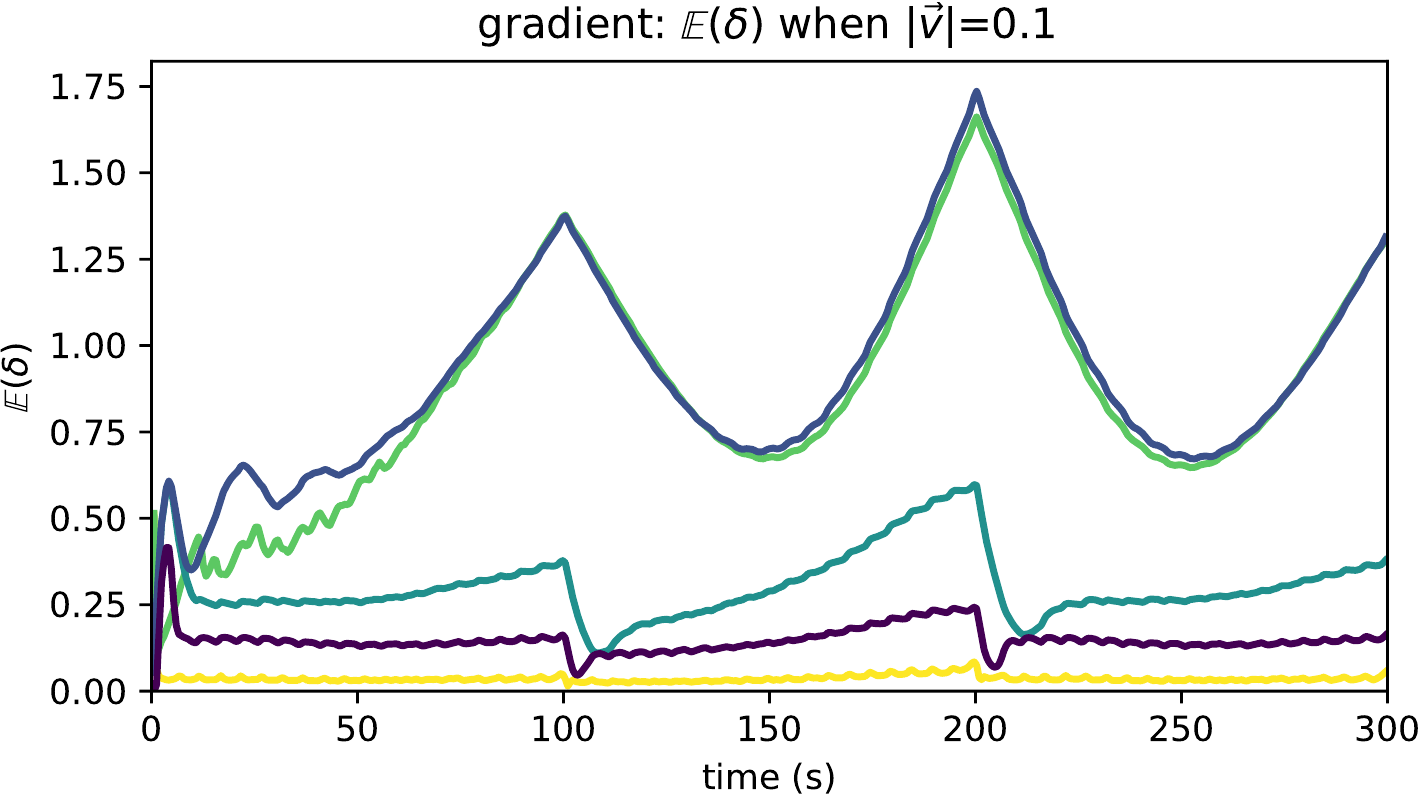}
  \\ \vspace{5pt}
  \includegraphics[width=.486\textwidth]{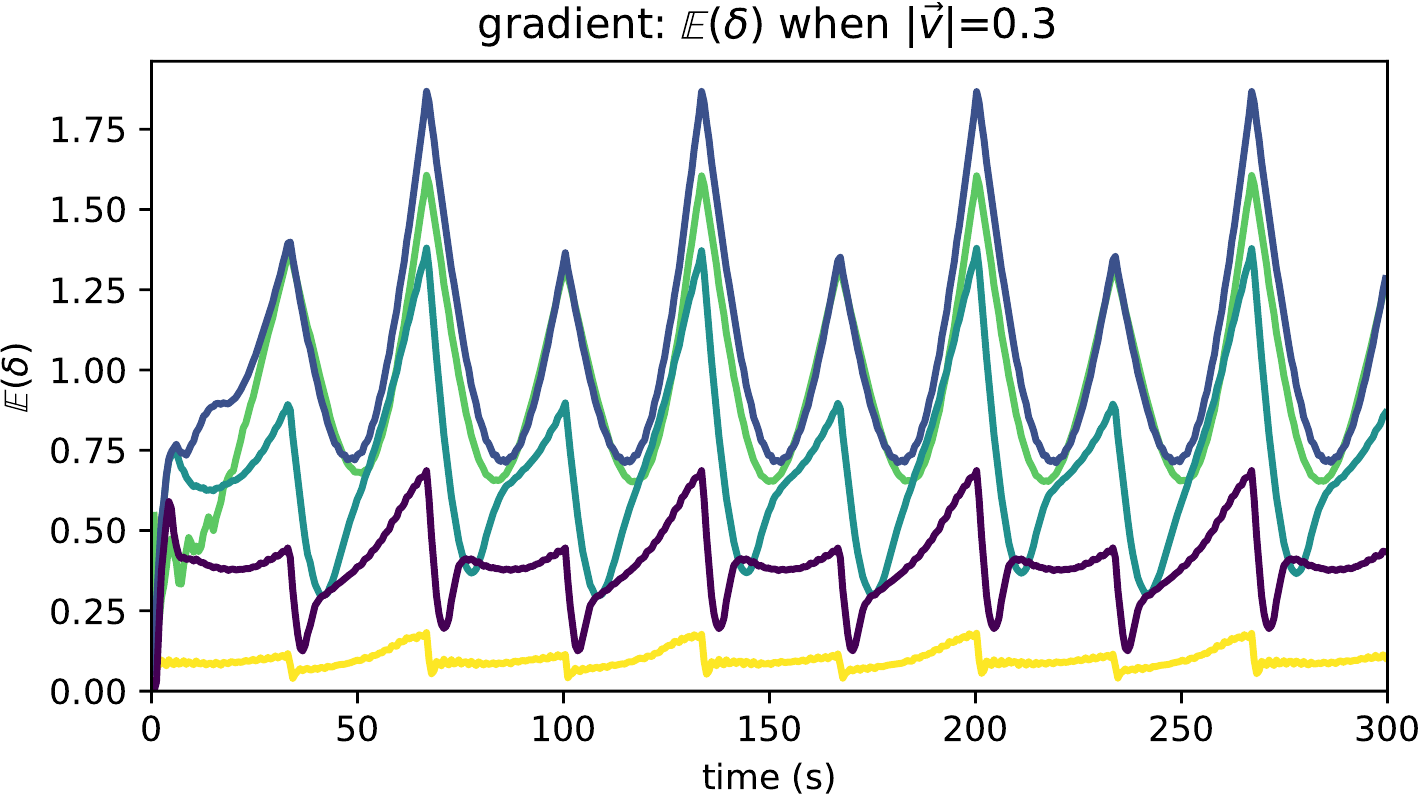}
  \includegraphics[width=.486\textwidth]{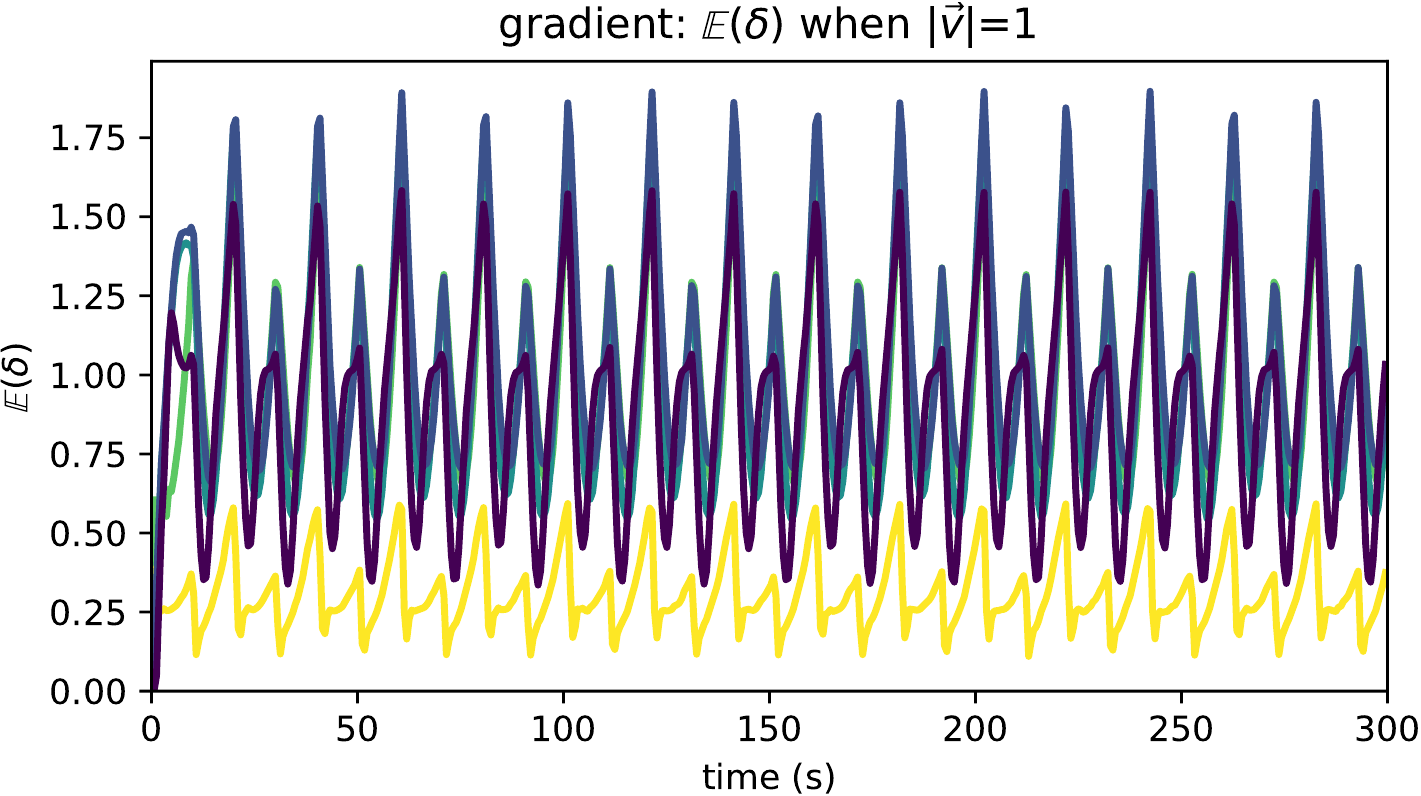}
  \\ \vspace{5pt}
  \includegraphics[width=.486\textwidth]{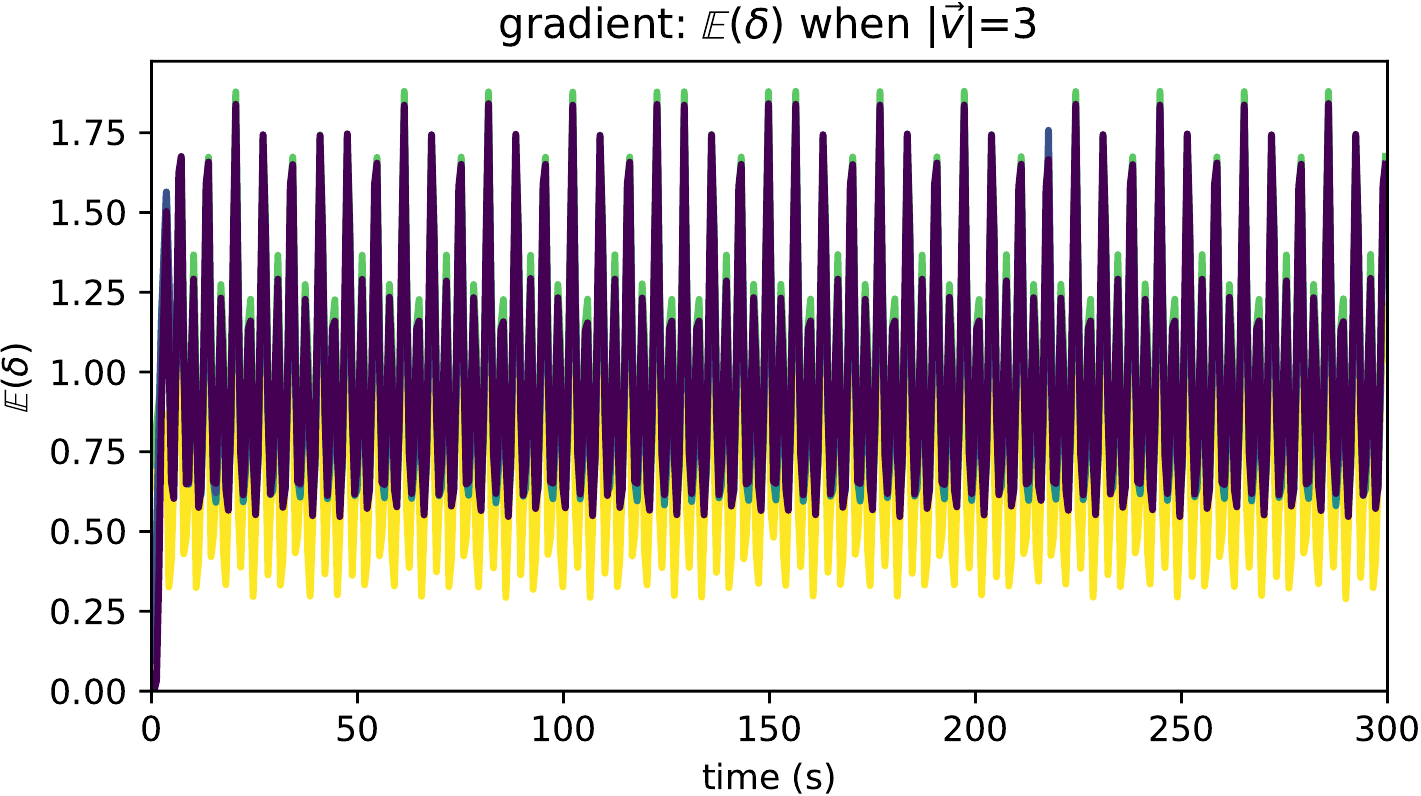}
 \end{center}
 \caption{
Compared error in the \emph{gradient} scenario.
The classic version is the darkest (purple in color) line,
and is compared with four time-fluid variants, with tolerance of 0.01m and 1m,
executed with delays of 0.01s and 1s.
The time fluid version with low tolerance executed on a reasonably fast network consistently outperforms
the classic version of the algorithm.
The time-fluid version is sensible to slow networking:
with a mean network delivery time of 1 ($\lambda^{-1} = 1$),
the time-fluid version accumulates more error on average than the classic version,
at least for low values of $\|\vec{v}\|$.
Interestingly, when the tolerance to error is high,
the system acquires a certain robustness to the network performance as well:
when $\epsilon=1$, the performance between a fast network and a network one order of magnitude
slower is barely noticeable.
Higher values of $\|\vec{v}\|$ progressively flatten the differences between the algorithms,
although the time-fluid version with low tolerance consistently outperforms all other algorithms.
 } 
 \label{chart:error:gradient}
\end{figure}

\begin{figure}[t]
 \begin{center}
  \includegraphics[width=.486\textwidth]{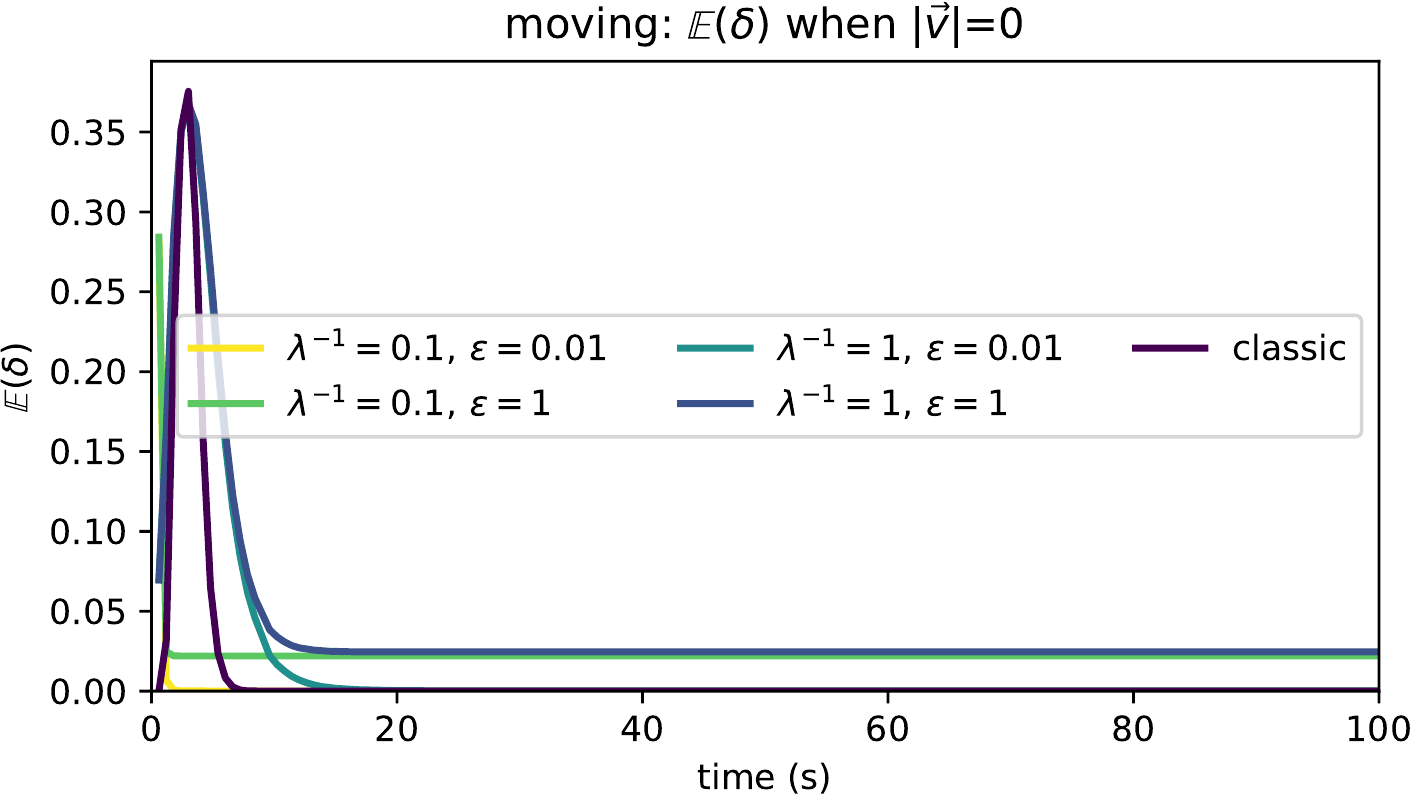}
  \includegraphics[width=.486\textwidth]{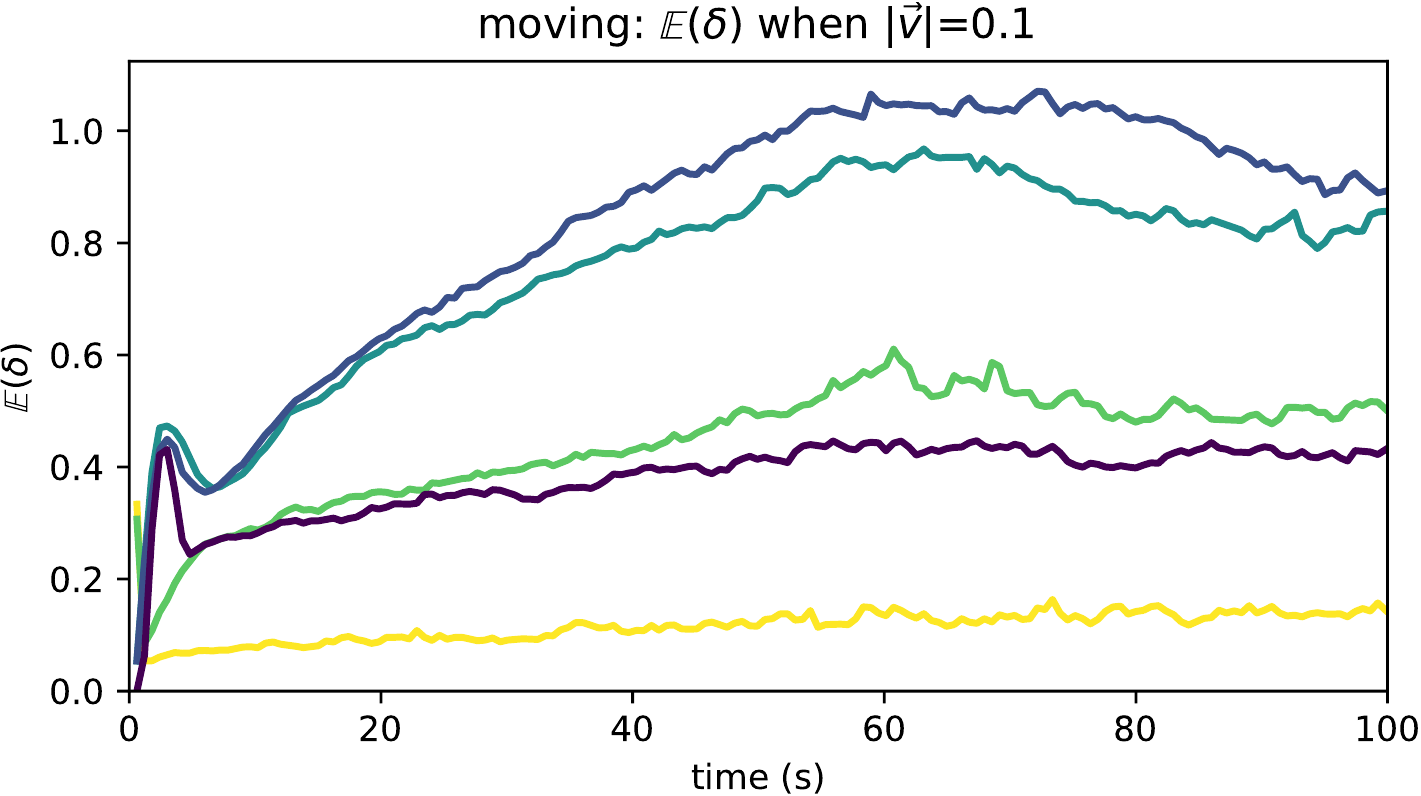}
  \\ \vspace{5pt}
  \includegraphics[width=.486\textwidth]{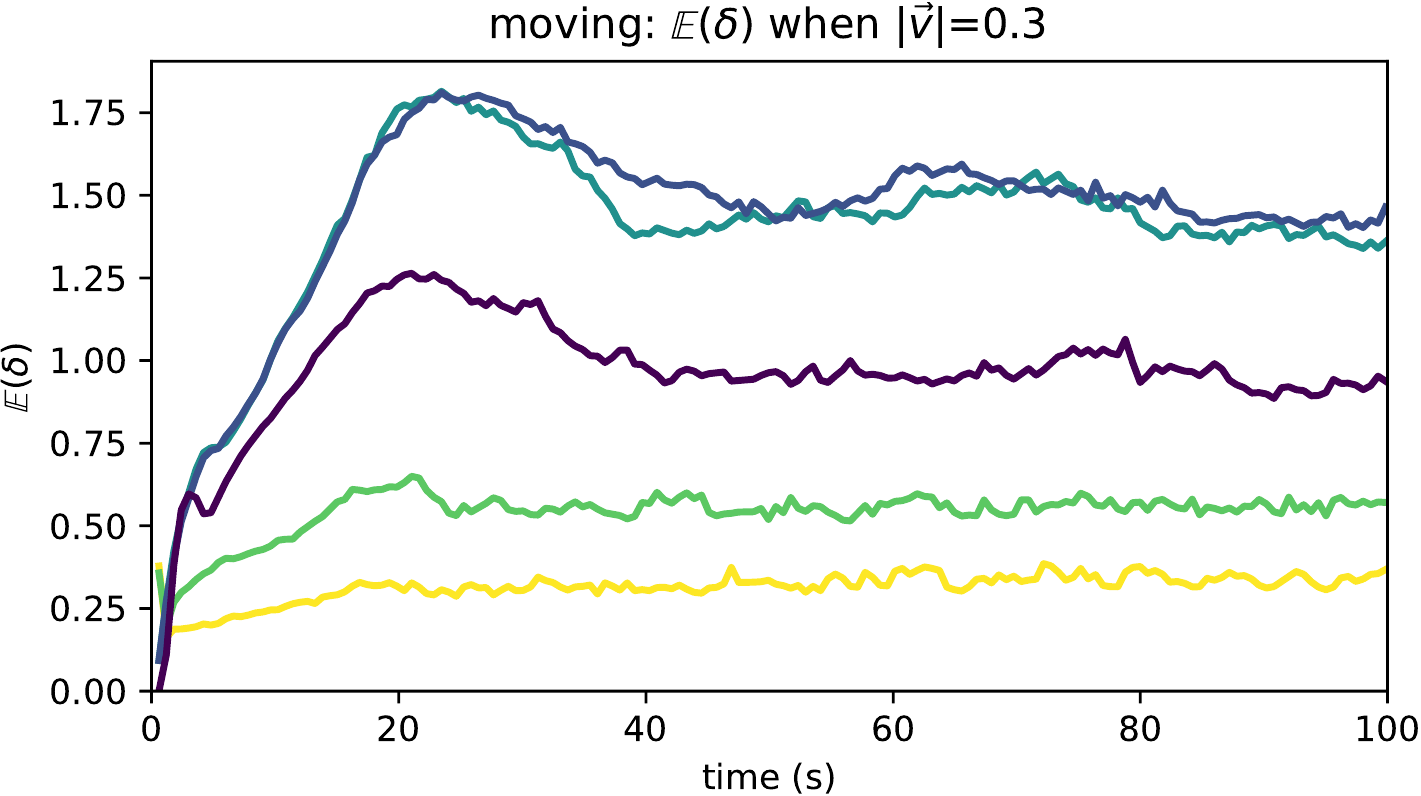}
  \includegraphics[width=.486\textwidth]{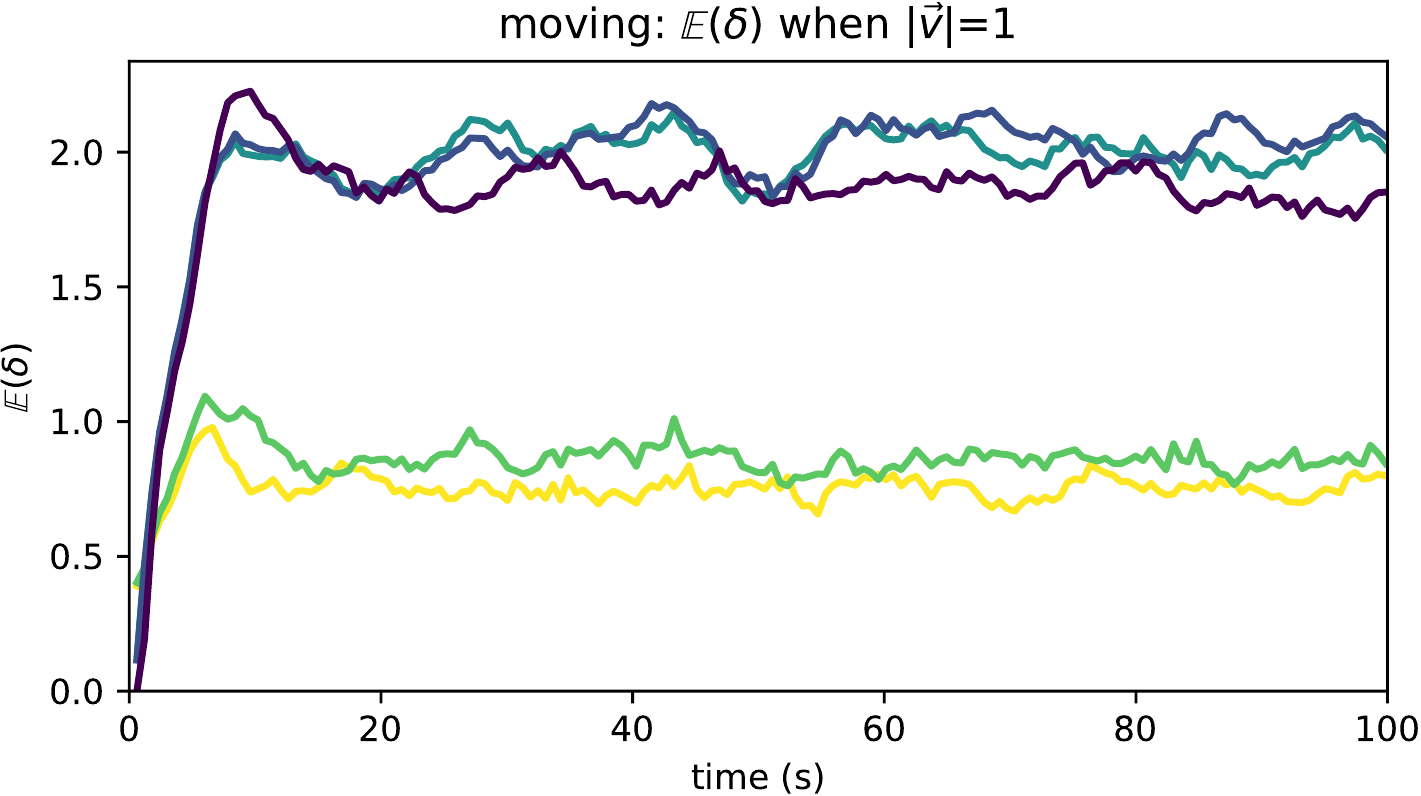}
  \\ \vspace{5pt}
  \includegraphics[width=.486\textwidth]{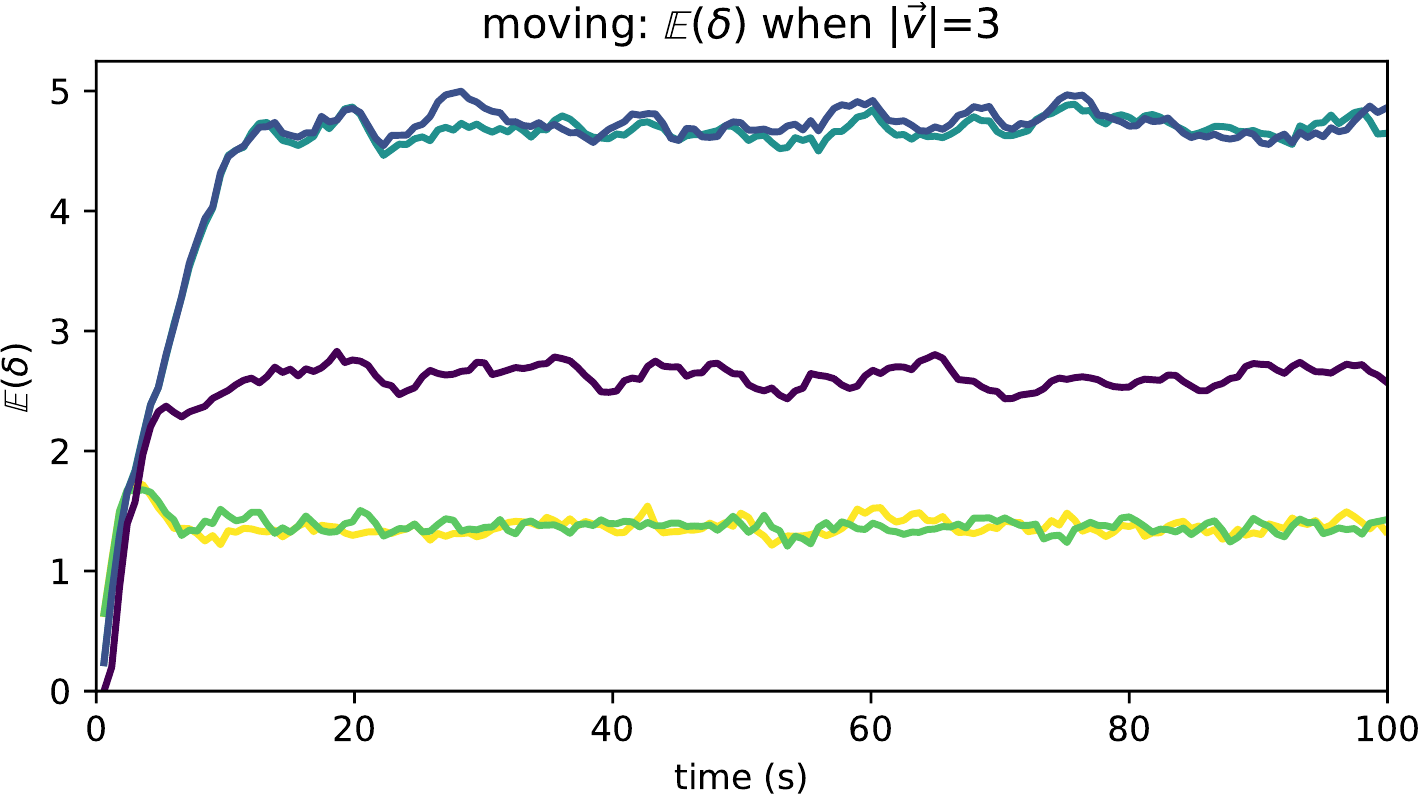}
 \end{center}
 \caption{
Compared error in the \emph{moving} scenario.
The classic version is the darkest (purple in color) line,
and is compared with four time-fluid variants, with tolerance of 0.01m and 1m,
executed with delays of 0.01s and 1s.
Constant movement by all nodes makes the tolerance very often violated
(and hence, time-fluid computation scheduled)
for most of the time, thus producing similar results for both versions running
on a fast network ($\lambda^{-1}=0.1$).
 }
 \label{chart:error:moving}
\end{figure}

\begin{figure}[t]
 \begin{center}
  \includegraphics[width=.486\textwidth]{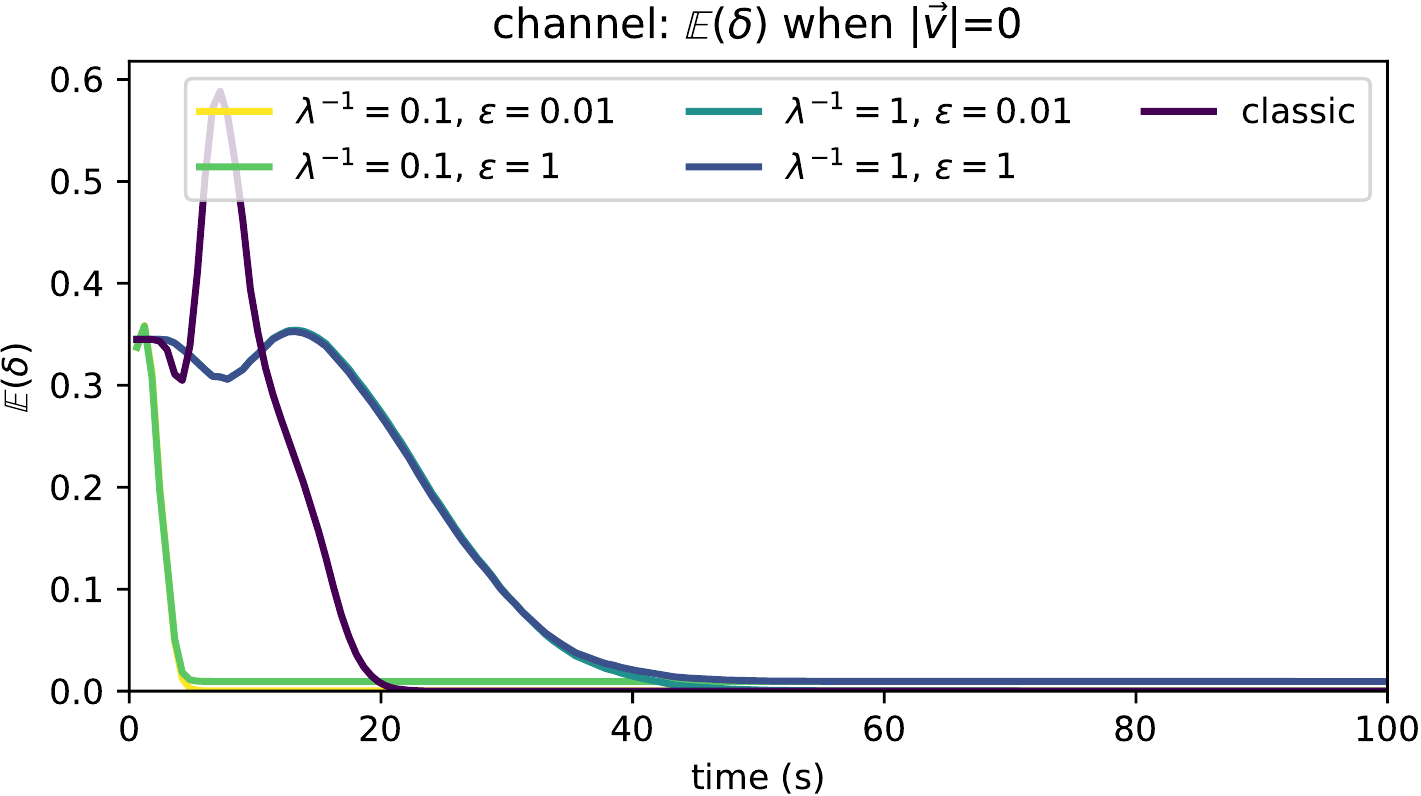}
  \includegraphics[width=.486\textwidth]{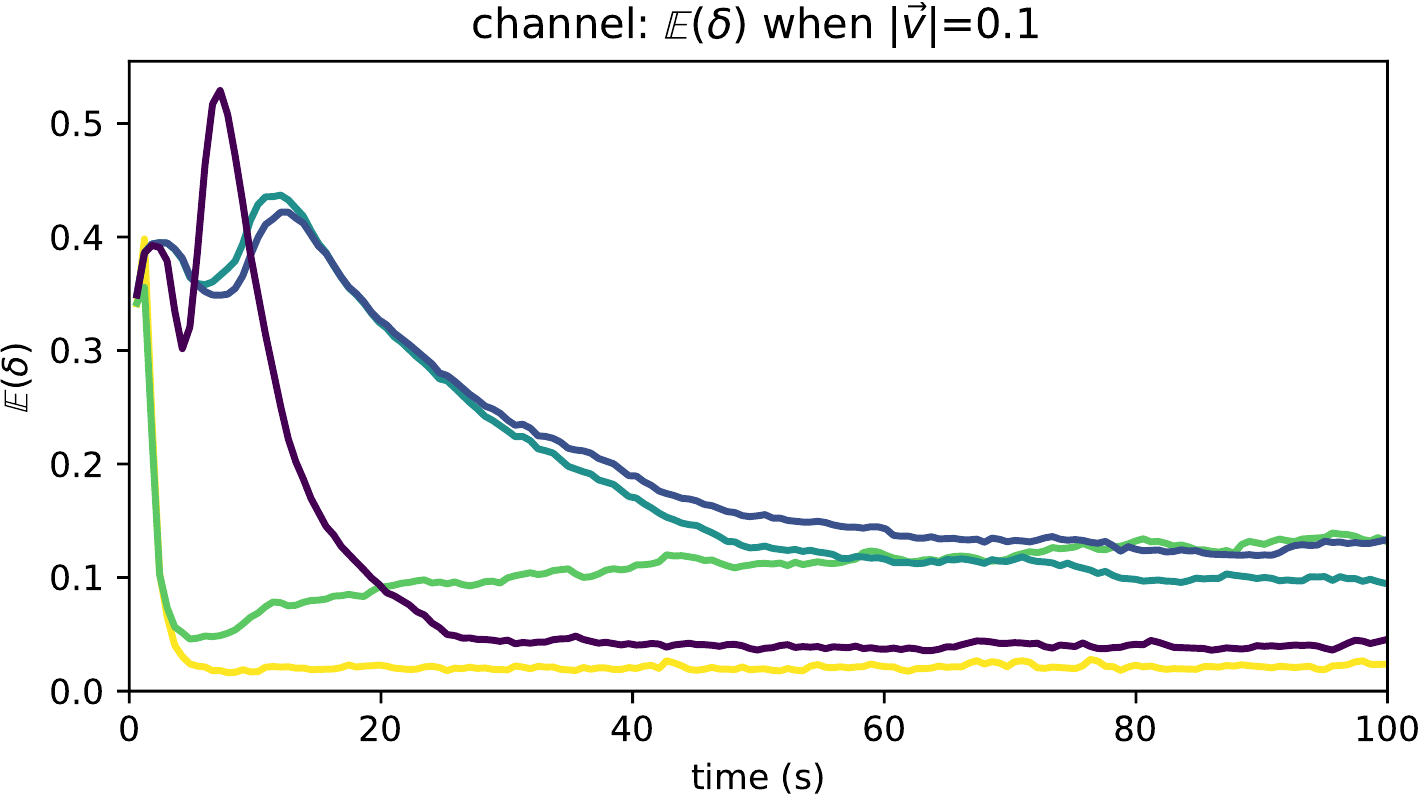}
  \\ \vspace{5pt}
  \includegraphics[width=.486\textwidth]{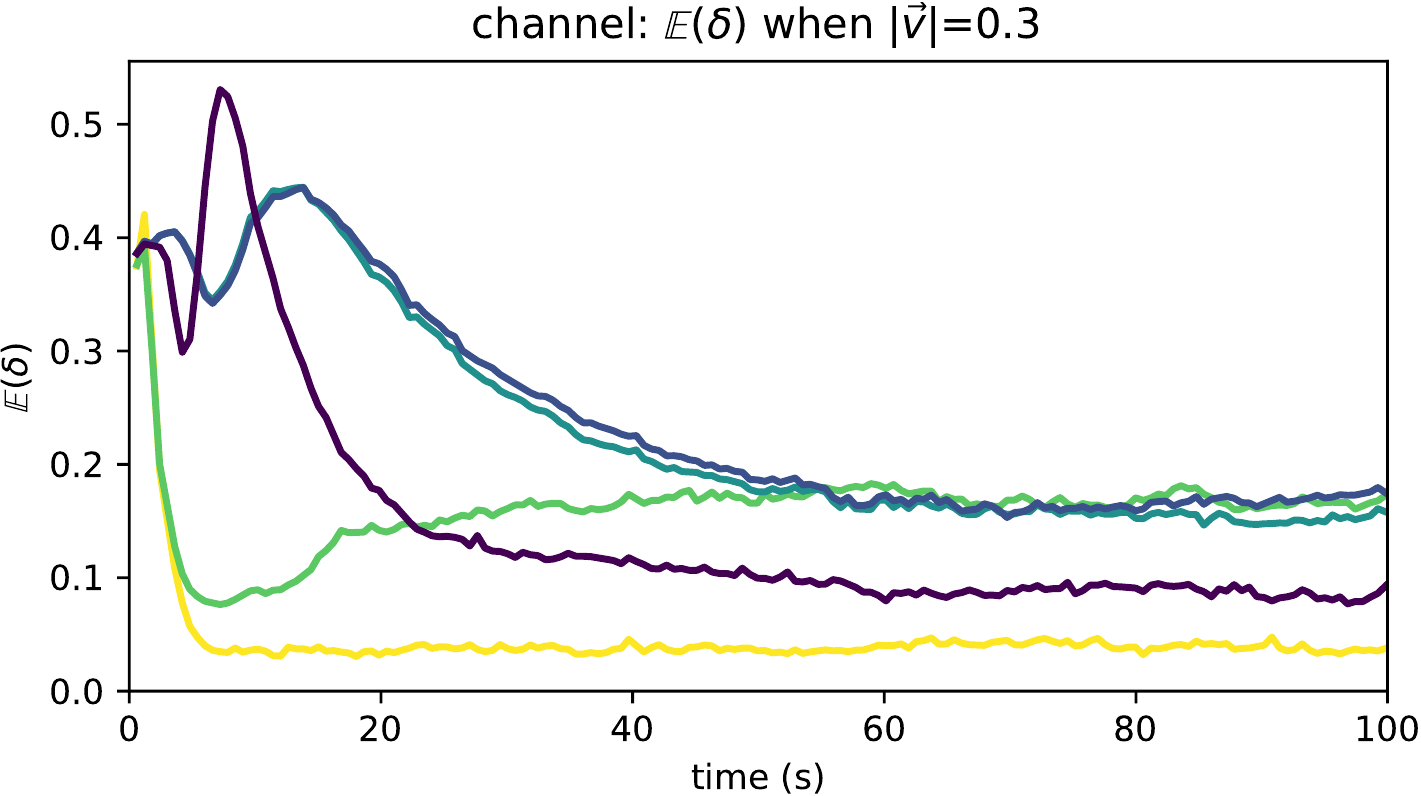}
  \includegraphics[width=.486\textwidth]{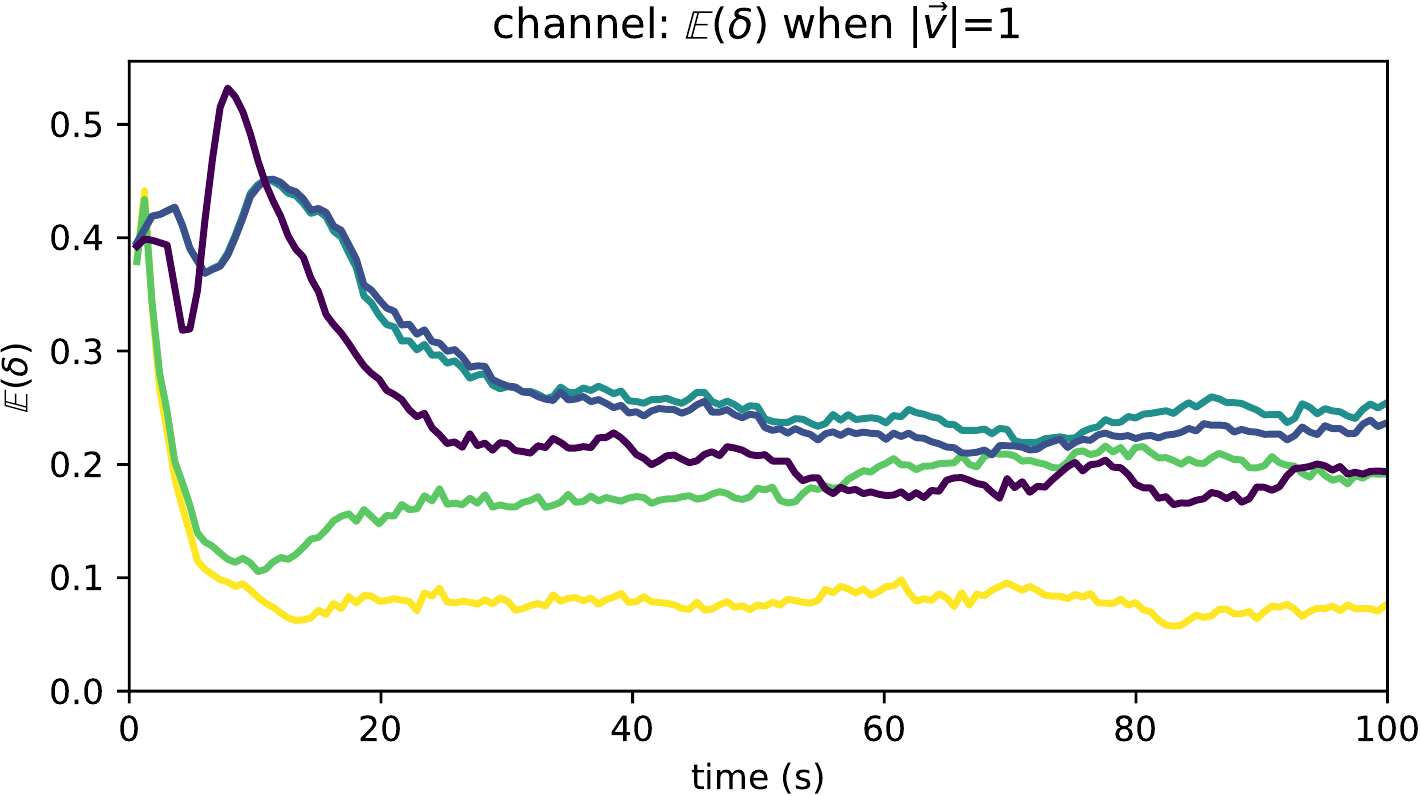}
  \\ \vspace{5pt}
  \includegraphics[width=.486\textwidth]{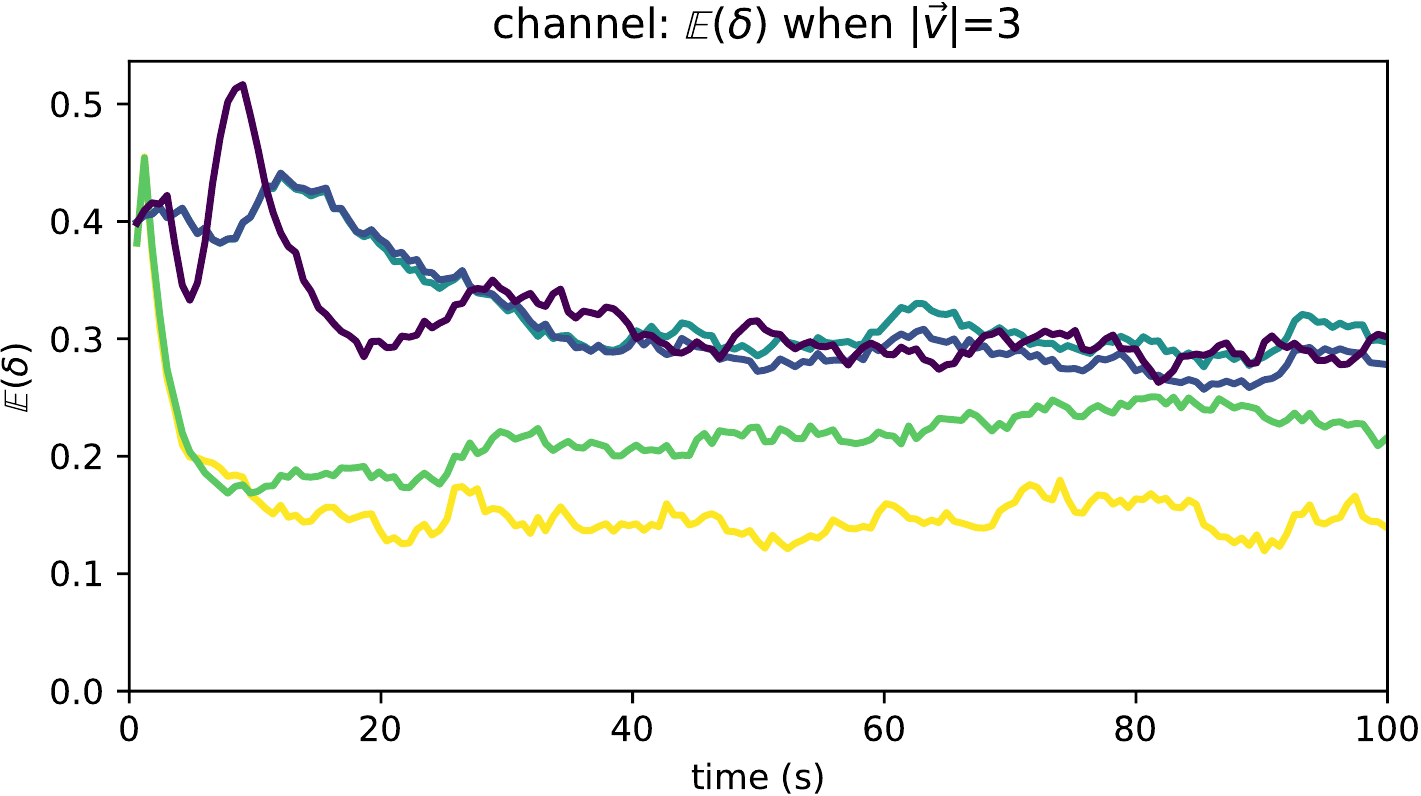}
 \end{center}
 \caption{
Compared error in the \emph{channel} scenario.
The classic version is the darkest (purple in color) line,
and is compared with four time-fluid variants, with tolerance of 0.01m and 1m,
executed with delays of 0.01s and 1s.
The time-fluid versions running on a fast network outperform all the other versions,
while those running on a slower network take more time than the classic one to stabilise,
but once a reasonable stability is reached, performance is comparable.
 }
 \label{chart:error:channel}
\end{figure}

The first expected benefit of a time-fluid version of a field-based computation is its ability to slow down
and even stop altogether if there are no changes in the environment.
This base functionality emerges clearly from \Cref{zerospeed},
where we depict the mean round count $\mathbb{E}(\rho)$
(our proxy metric for resource usage),
and we set $\|\vec{v}\|$=0, making the nodes stand still.
While in the classic implementation $\mathbb{E}(\rho)$ grows linearly with time,
hence consuming resources just to ``maintain'' the coordination fields,
the time-fluid versions stop computing altogether after a stabilisation time.
Interestingly, the time-fluid versions that execute on a faster network
($\lambda^{-1}=0.1s$)
have a quicker transient, and initially show a higher resource usage compared to the classic version,
which is bounded by its fixed working frequency.
In these cases, the time fluid version both converges more quickly and saves resources in the long run.

The error analysis results are depicted
in \Cref{chart:error:gradient} for the gradient scenario,
in \Cref{chart:error:moving} for the moving scenario,
and in \Cref{chart:error:channel} for the channel scenario.
The periodicity of the gradient experiment clearly appears in the experimental data,
in particular when compared to the more chaotic behaviour of the other two versions,
moving a much larger part of the devices with Lévy walks.
The time-fluid implementation with low tolerance consistently outperforms all the other proposals under test.
The other combinations (fast network, low tolerance; and low tolerance with fast and slow network)
are usually competitive with the classic implementation.
Data shows several interesting behaviours:
\begin{itemize}
 \item In the gradient case, low speeds are tolerated much less than in the classic version.
 Since both versions with high tolerance perform similarly despite a much different network performance,
 we can say that this behaviour is due to the tolerance being high enough to account for the amount of error,
 which in fact remains pretty much stable across different values of $\|\vec{v}\|$,
 while all other versions accumulate progressively more error at higher speeds.
 \item In the moving case, the continuous disruption induced by the movement of the whole network
 induces a lower impact of the tolerance: the network is continuously prompted to recalculate,
 and hence the network performance assumes greater relevance
 (as they ultimately dictate how quickly new information can propagate).
 As a consequence, both the time-fluid versions (low and high tolerance) running on a fast network
 outperform the classic version, while the ones limited by the network performance perform on average worse.
\end{itemize}


\begin{figure}[t]
 \begin{center}
  \includegraphics[width=.486\textwidth]{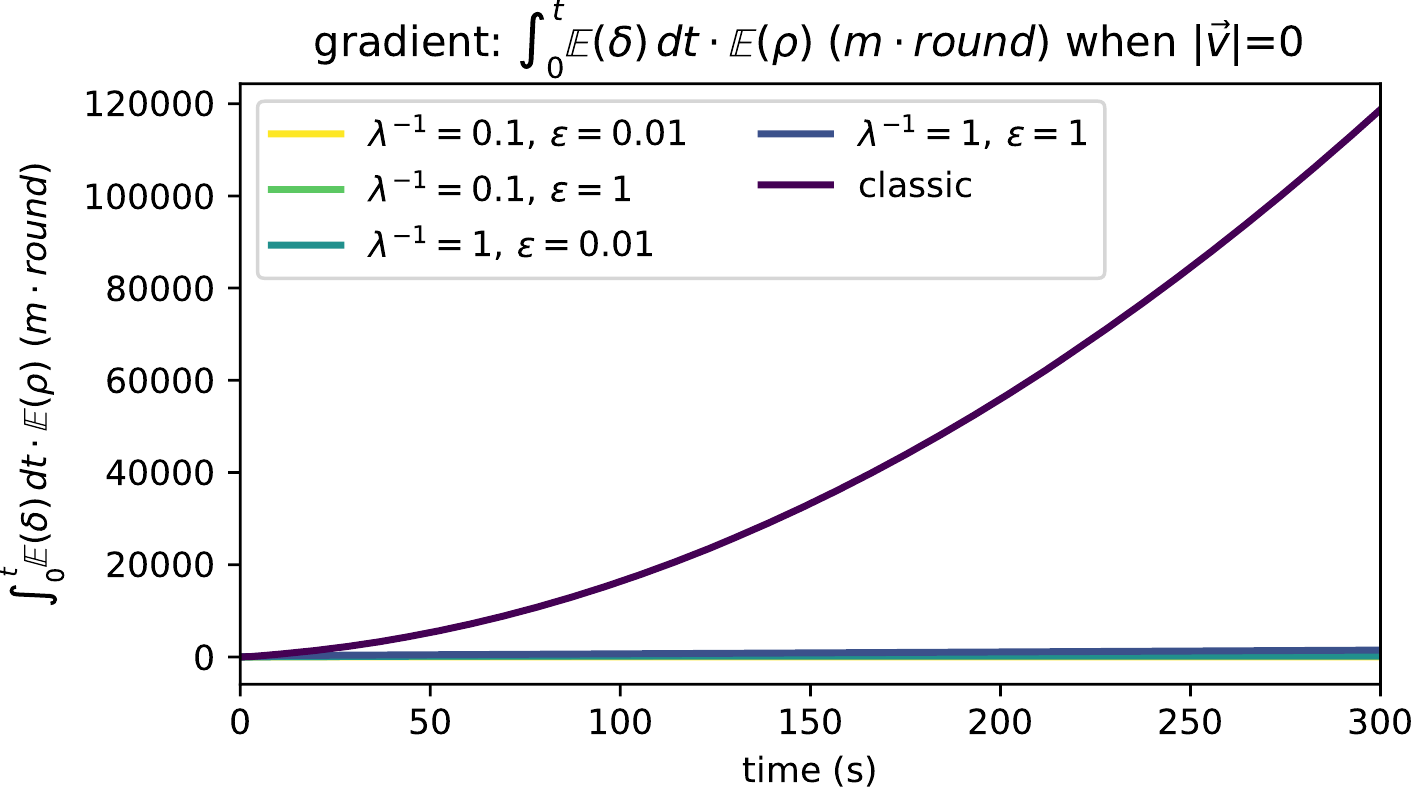}
  \includegraphics[width=.486\textwidth]{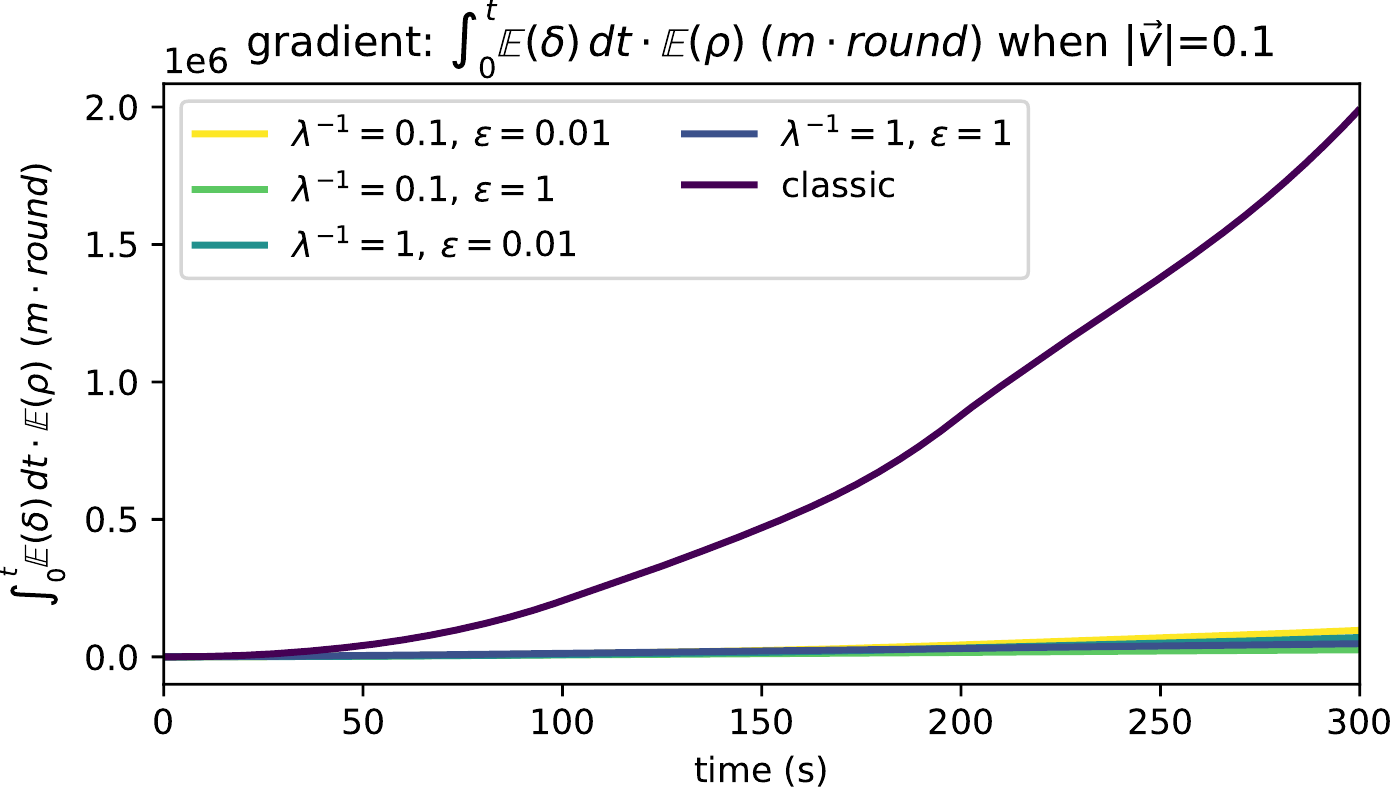}
  \\ \vspace{5pt}
  \includegraphics[width=.486\textwidth]{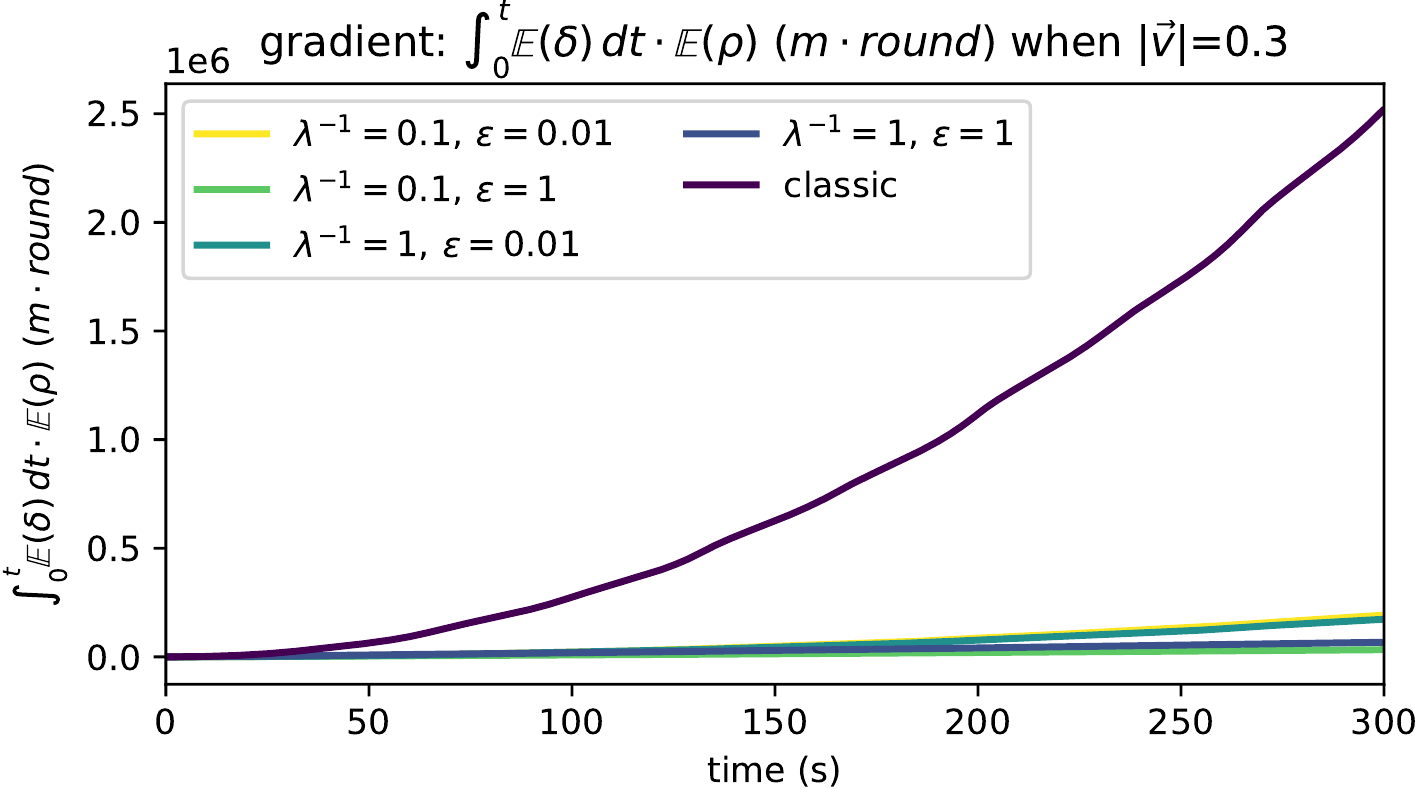}
  \includegraphics[width=.486\textwidth]{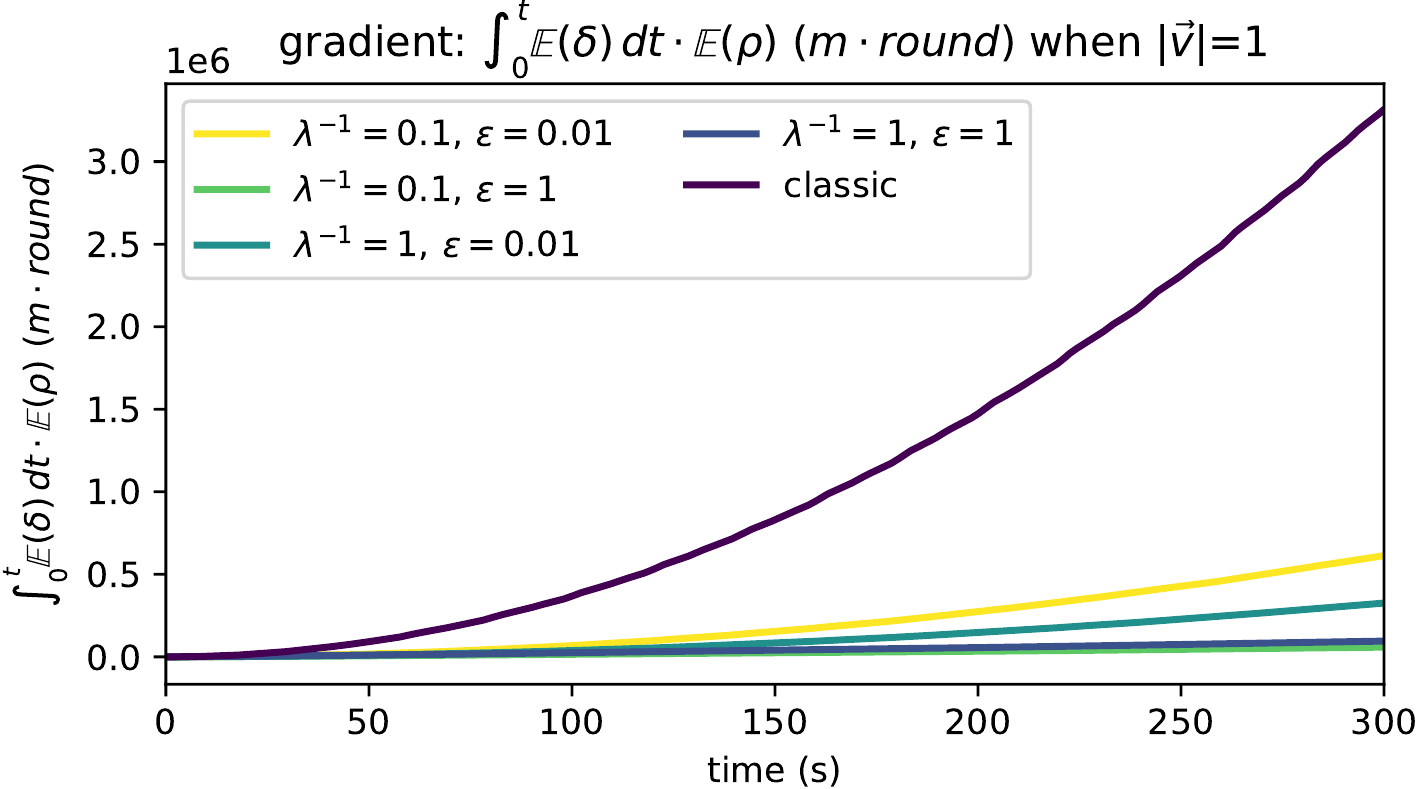}
  \\ \vspace{5pt}
  \includegraphics[width=.486\textwidth]{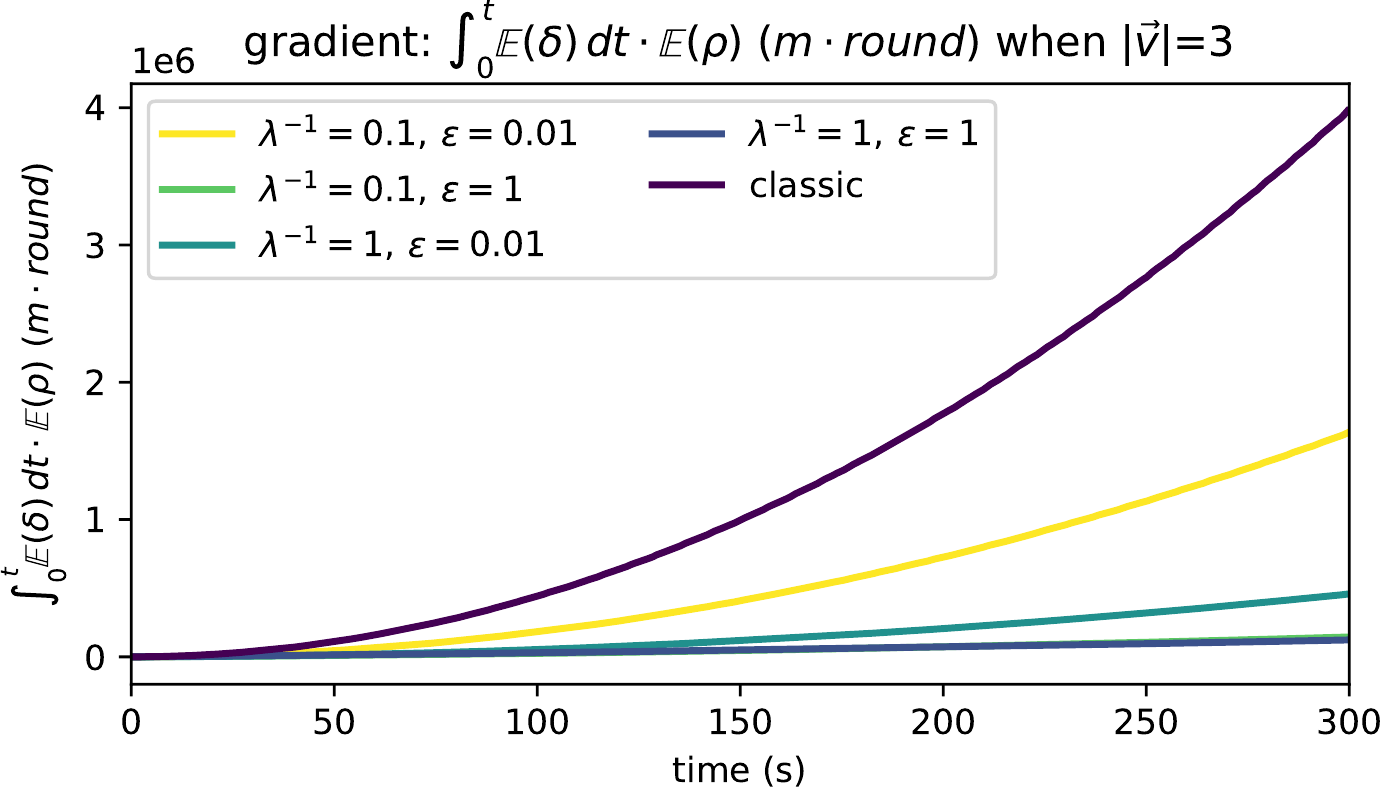}
 \end{center}
 \caption{
Compared product between cumulative error and cumulative round count
in the gradient scenario,
as a proxy metric for efficiency.
Lower values indicate better efficiency.
The classic version is the darkest (purple in color) line,
and is compared with four time-fluid variants, with tolerance of 0.01m and 1m,
executed with delays of 0.01s and 1s.
Time-fluid versions strictly perform better than the classic version,
due to their ability to cope with changes happening in a sub-portion of the whole network.
 }
 \label{chart:efficiency:gradient}
\end{figure}

\begin{figure}[t]
 \begin{center}
  \includegraphics[width=.486\textwidth]{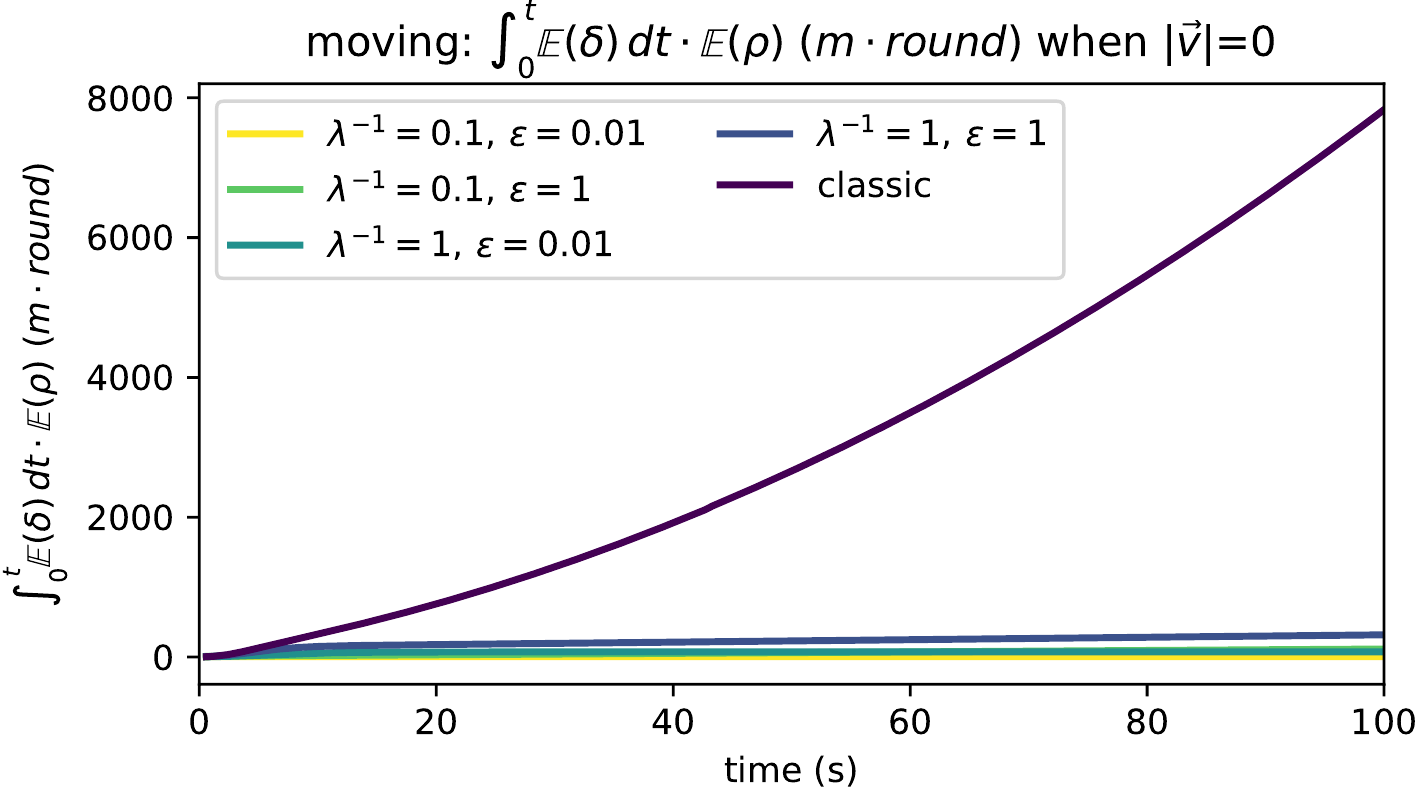}
  \includegraphics[width=.486\textwidth]{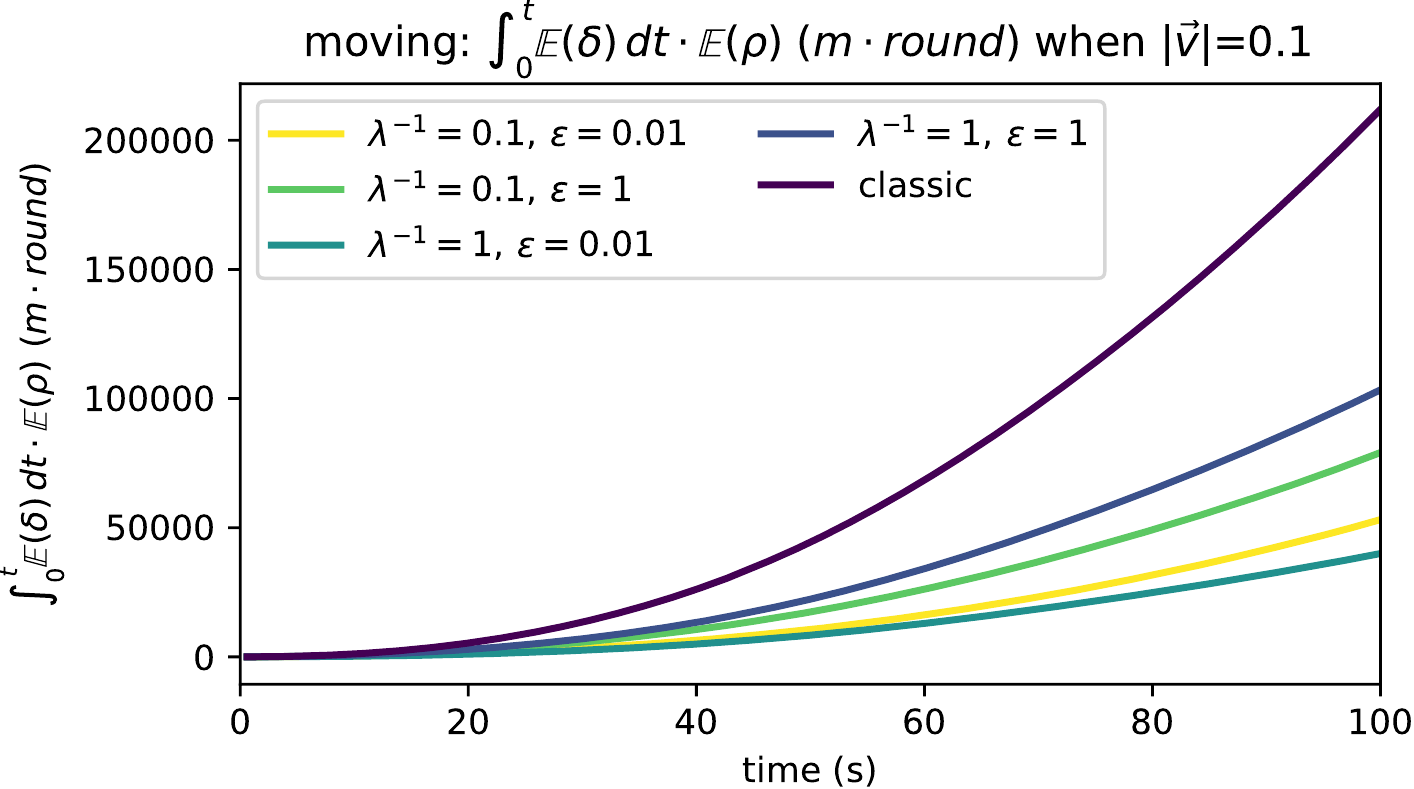}
  \\ \vspace{5pt}
  \includegraphics[width=.486\textwidth]{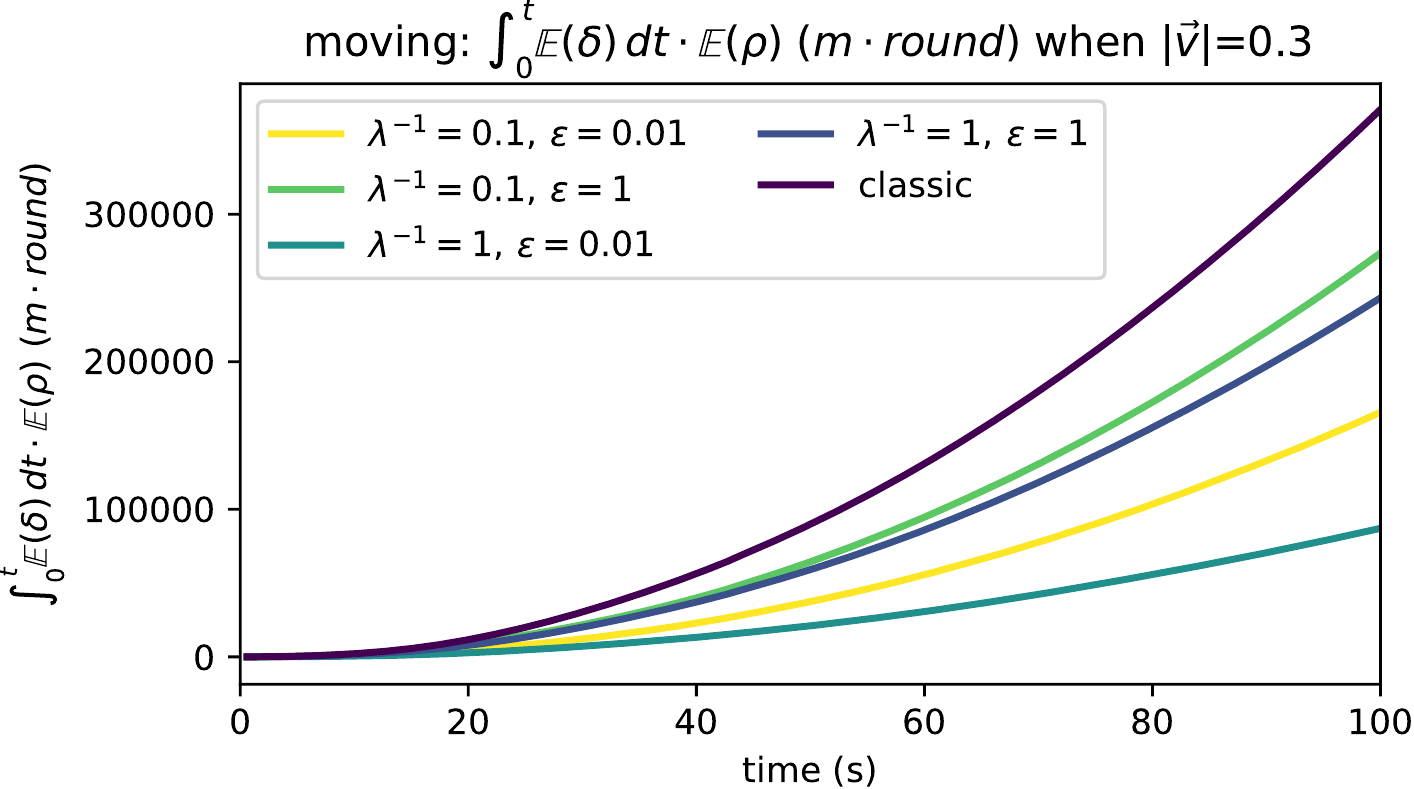}
  \includegraphics[width=.486\textwidth]{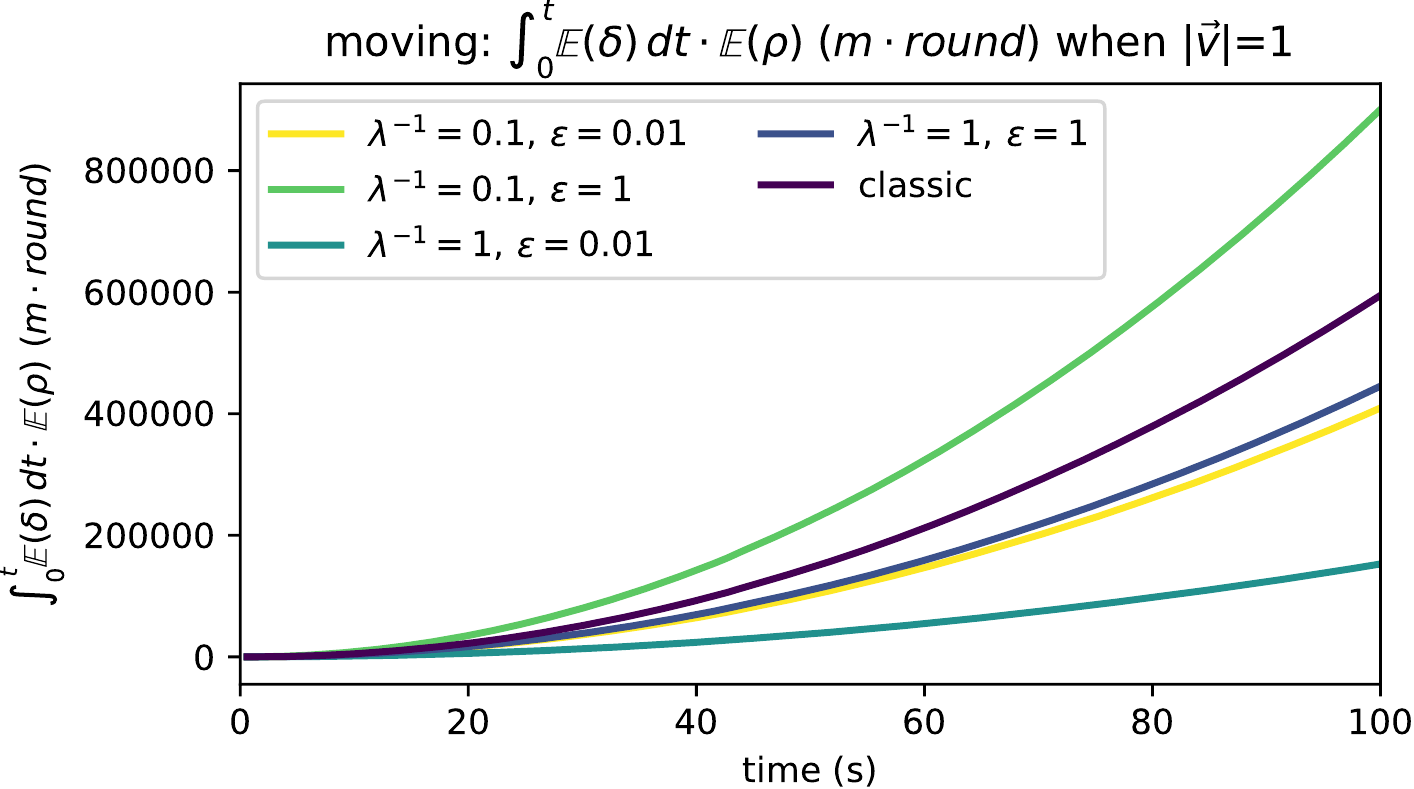}
  \\ \vspace{5pt}
  \includegraphics[width=.486\textwidth]{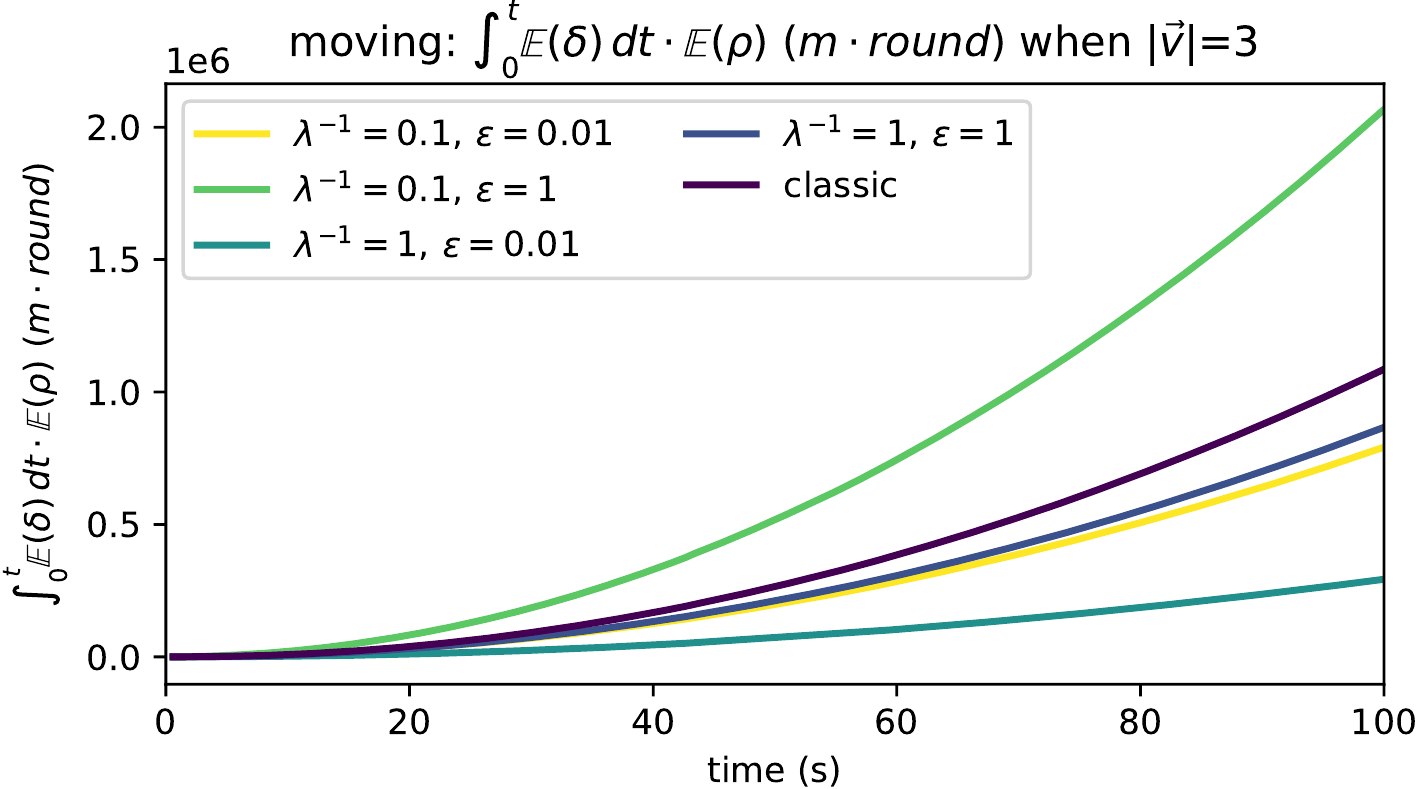}
 \end{center}
 \caption{
Compared error in the channel scenario.
The classic version is the darkest (purple in color) line,
and is compared with four time-fluid variants, with tolerance of 0.01m and 1m,
executed with delays of 0.01s and 1s.
Time-fluid versions of all fashions outperform the baseline for $\|\vec{v}\| \leq 0.3\nicefrac{m}{s}$.
At higher speeds, unfortunate combination of tolerance and network latency may produce values
whose error is high if compared to the amount of computational resources required to perform the computation.
In our experiments, this happens for instance when $\lambda^{-1}=0.1s$ and $\epsilon=1m$:
frequent network messages mean frequent re-adjustments, but the high tolerance does not bring the process
to the point that error gets low enough to justify the expense.
Even though the version with very low tolerance $\epsilon=0.1m$ runs more rounds (hence, costs more)
it reduces the error so much that the final cost metric is lower overall.
 }
 \label{chart:efficiency:moving}
\end{figure} 

\begin{figure}[t]
 \begin{center}
  \includegraphics[width=.486\textwidth]{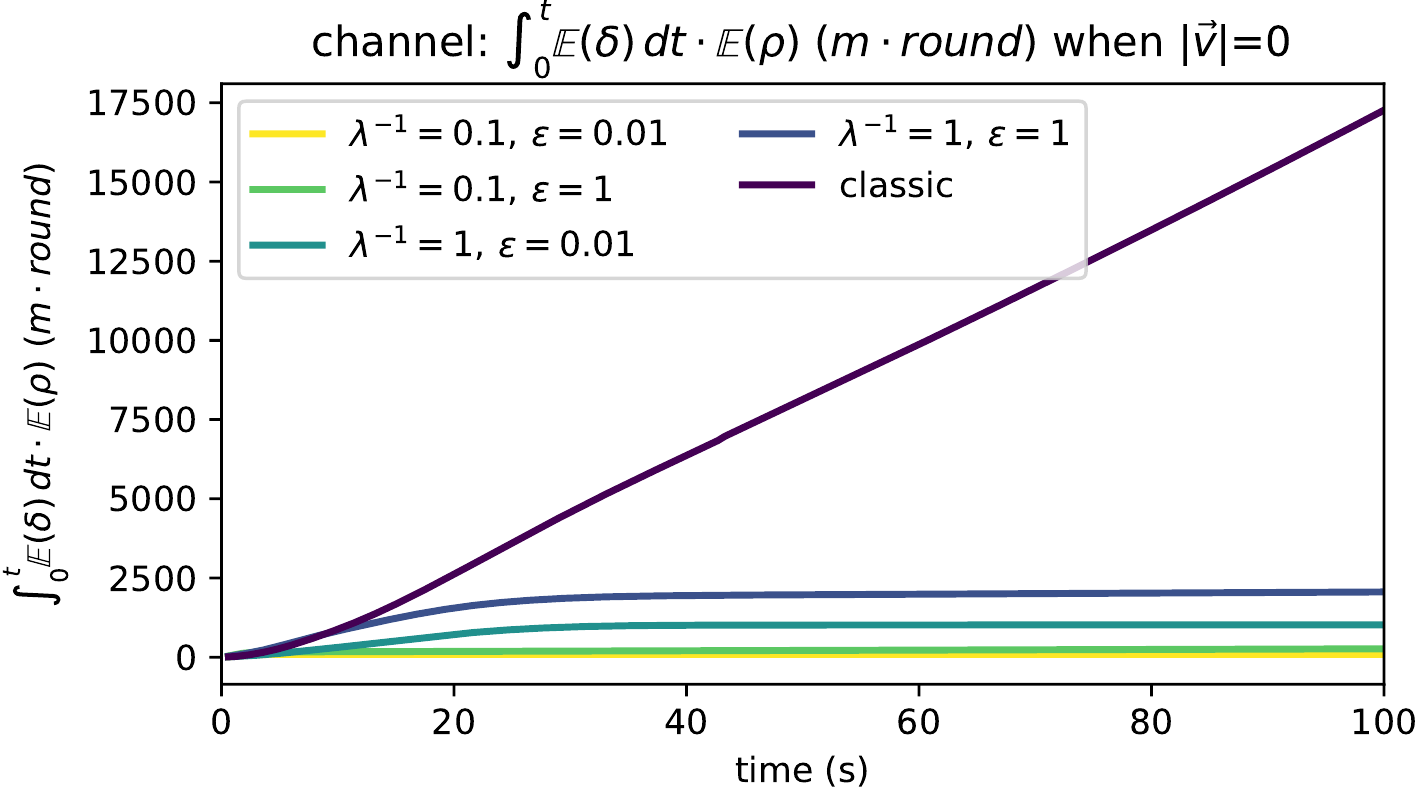}
  \includegraphics[width=.486\textwidth]{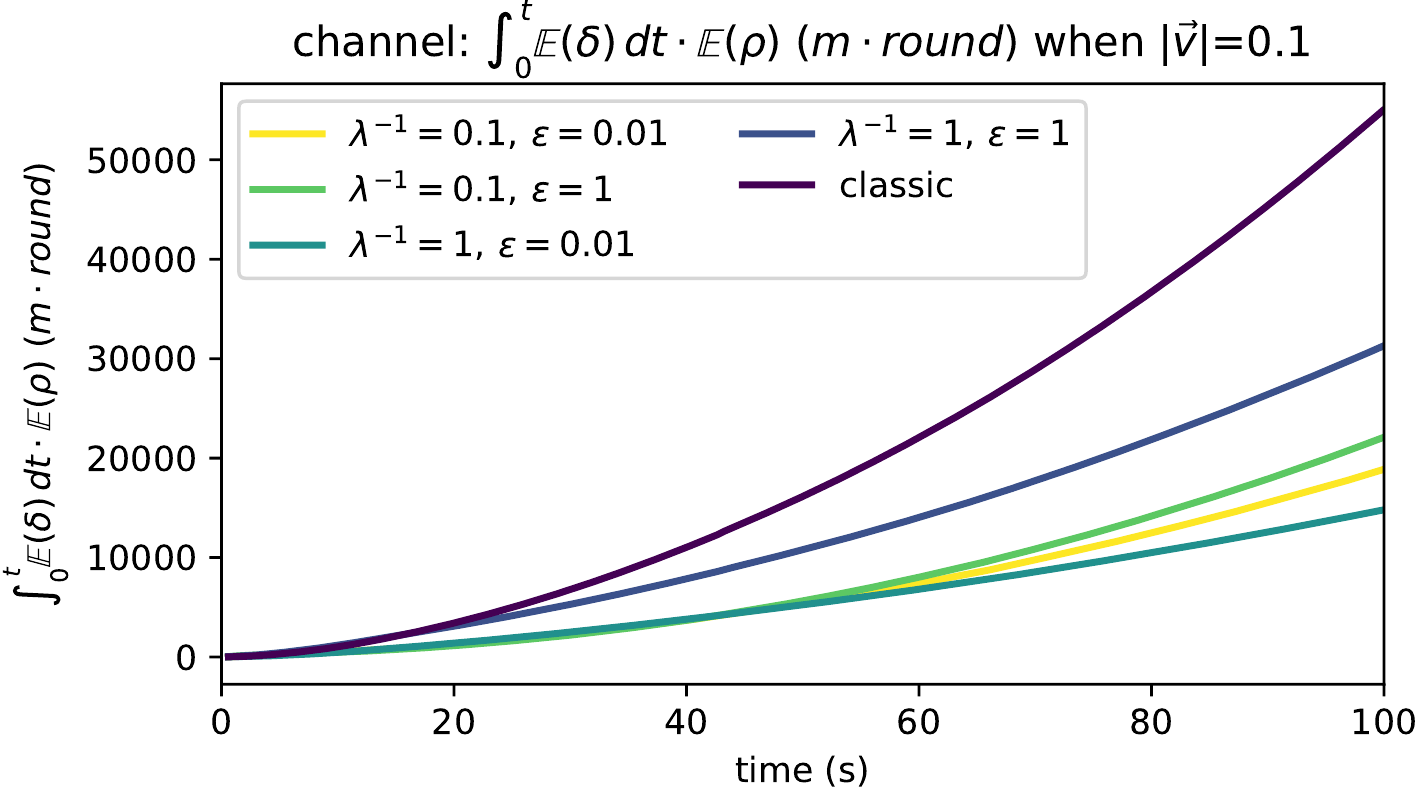}\\
  \vspace{5pt}
  \includegraphics[width=.486\textwidth]{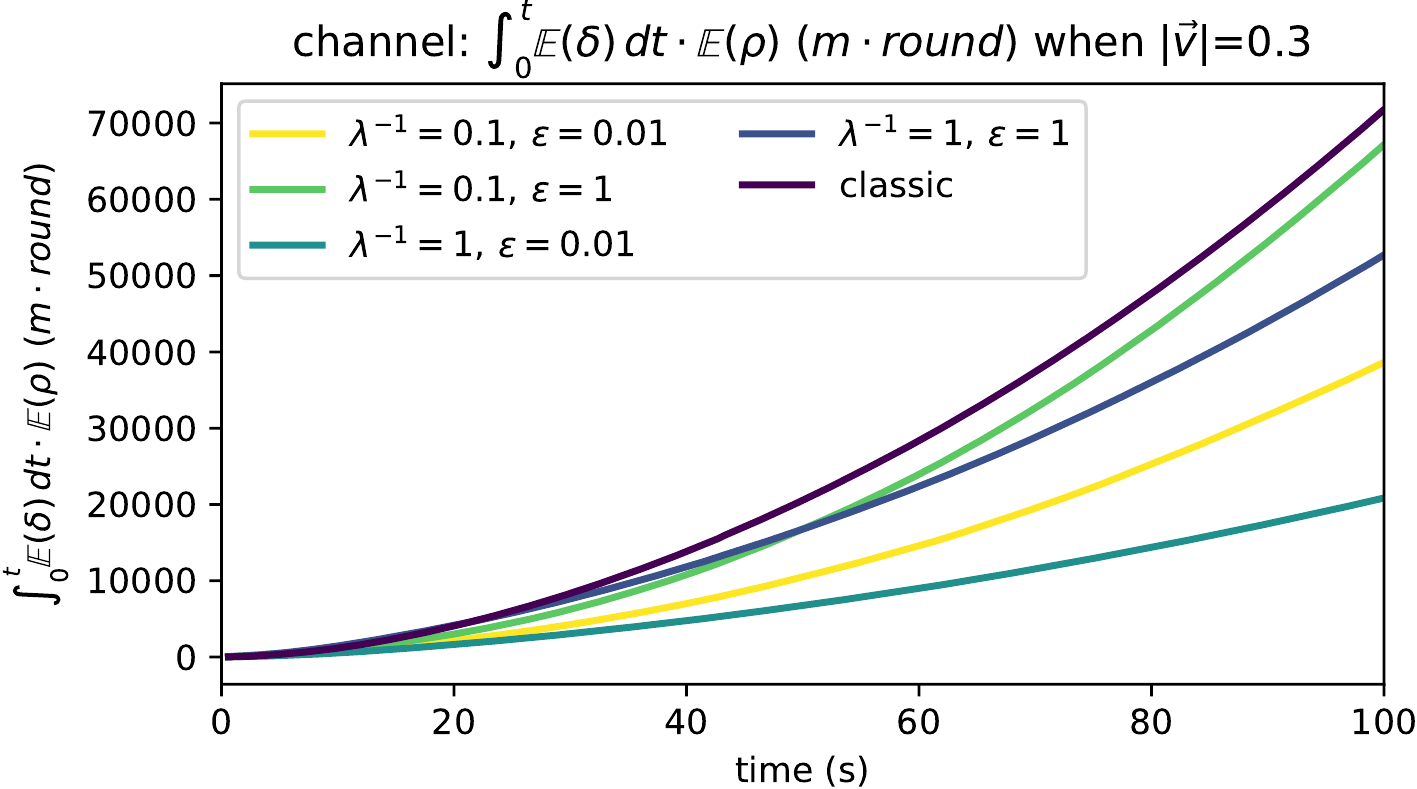}
  \includegraphics[width=.486\textwidth]{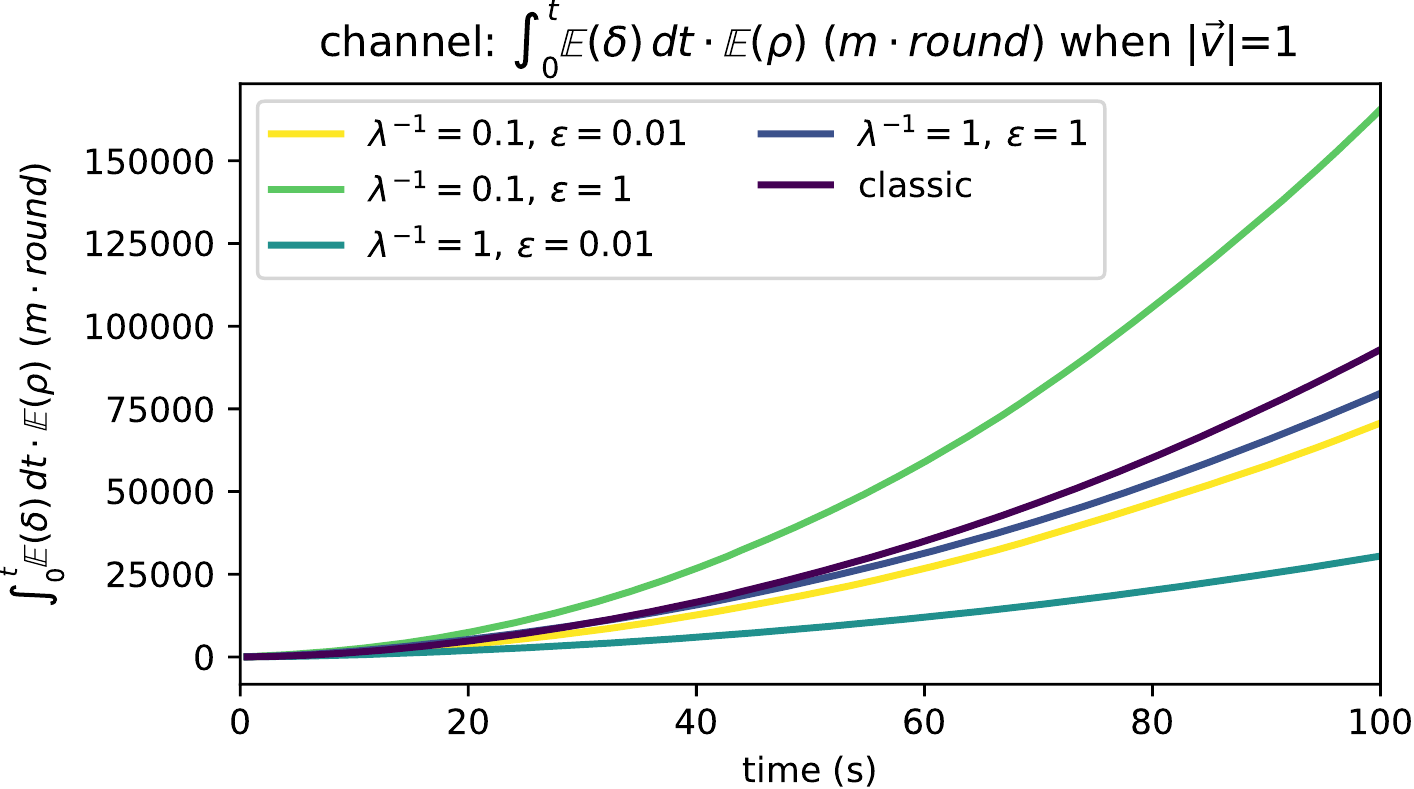}
  \\ \vspace{5pt}
  \includegraphics[width=.486\textwidth]{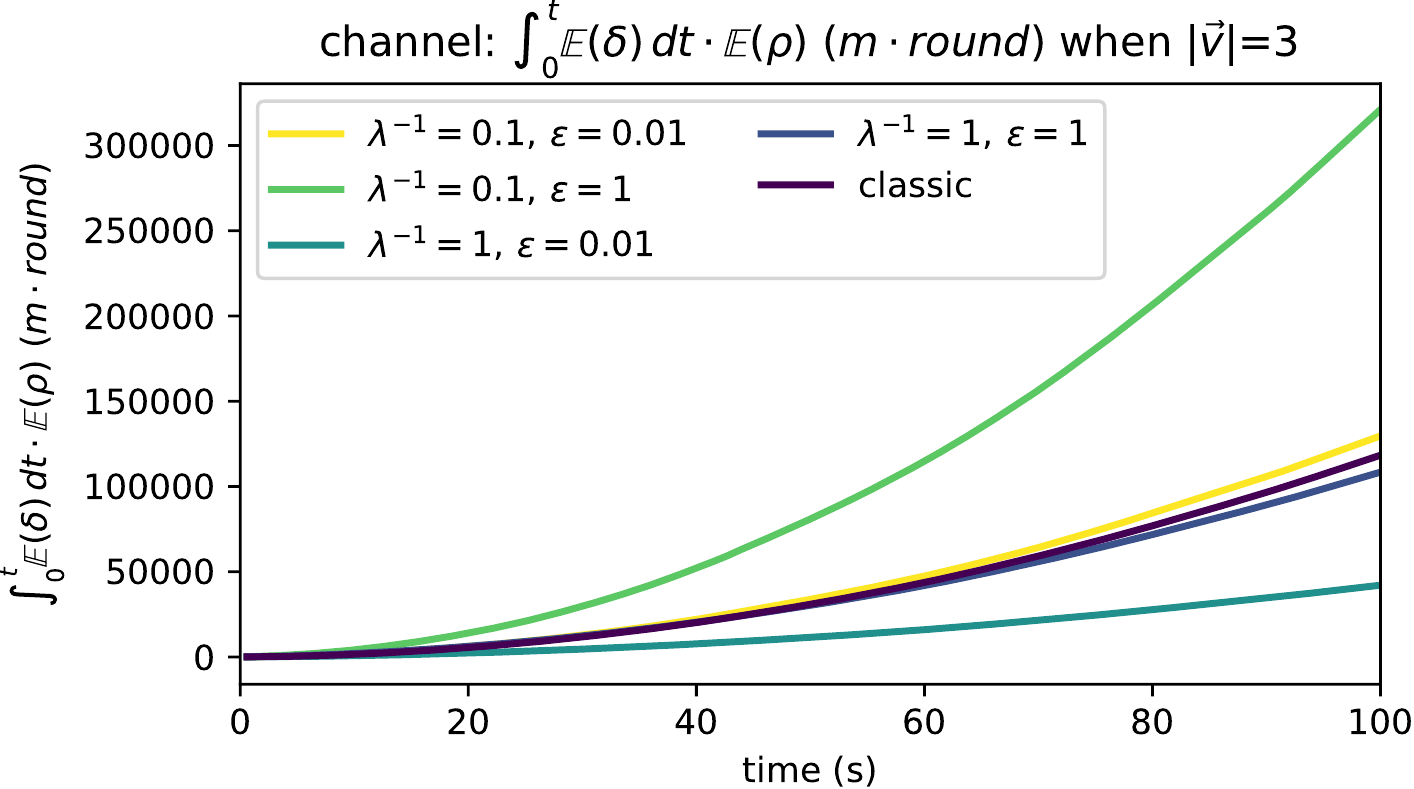}
 \end{center} 
 \caption{
Compared error in the channel scenario.
The classic version is the darkest (purple in color) line,
and is compared with four time-fluid variants, with tolerance of 0.01m and 1m,
executed with delays of 0.01s and 1s.
Time-fluid versions of all fashions outperform the baseline for $\|\vec{v}\| \leq 0.3\nicefrac{m}{s}$.
At higher speeds, unfortunate combination of tolerance and network latency may produce values
whose error is high if compared to the amount of computational resources required to perform the computation.
In our experiments, this happens in particular when $\lambda^{-1}=1s$ and $\epsilon=1m$.
Other versions achieve a ration between performance and cost which is similar to the one of the classic version.
However, we note that there are some circumstances under which the time-fluid version of the channel
achieves a much better performance / cost ratio even at high speeds ($\lambda^{-1}=1s$ and $\epsilon=0.01m$).
 }
 \label{chart:efficiency:channel}
\end{figure}

\begin{figure}[t]
 \begin{center}
  \includegraphics[width=.486\textwidth]{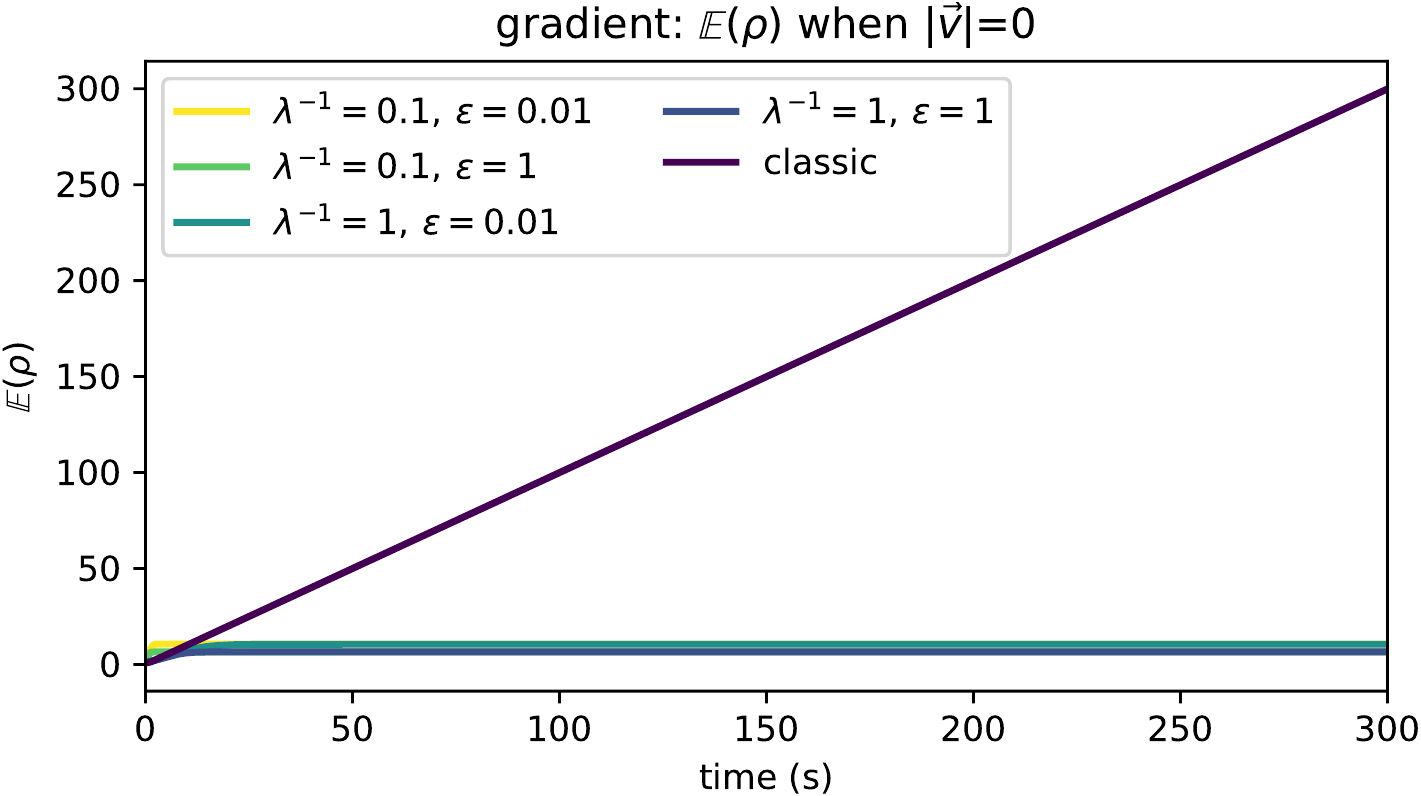}
  \includegraphics[width=.486\textwidth]{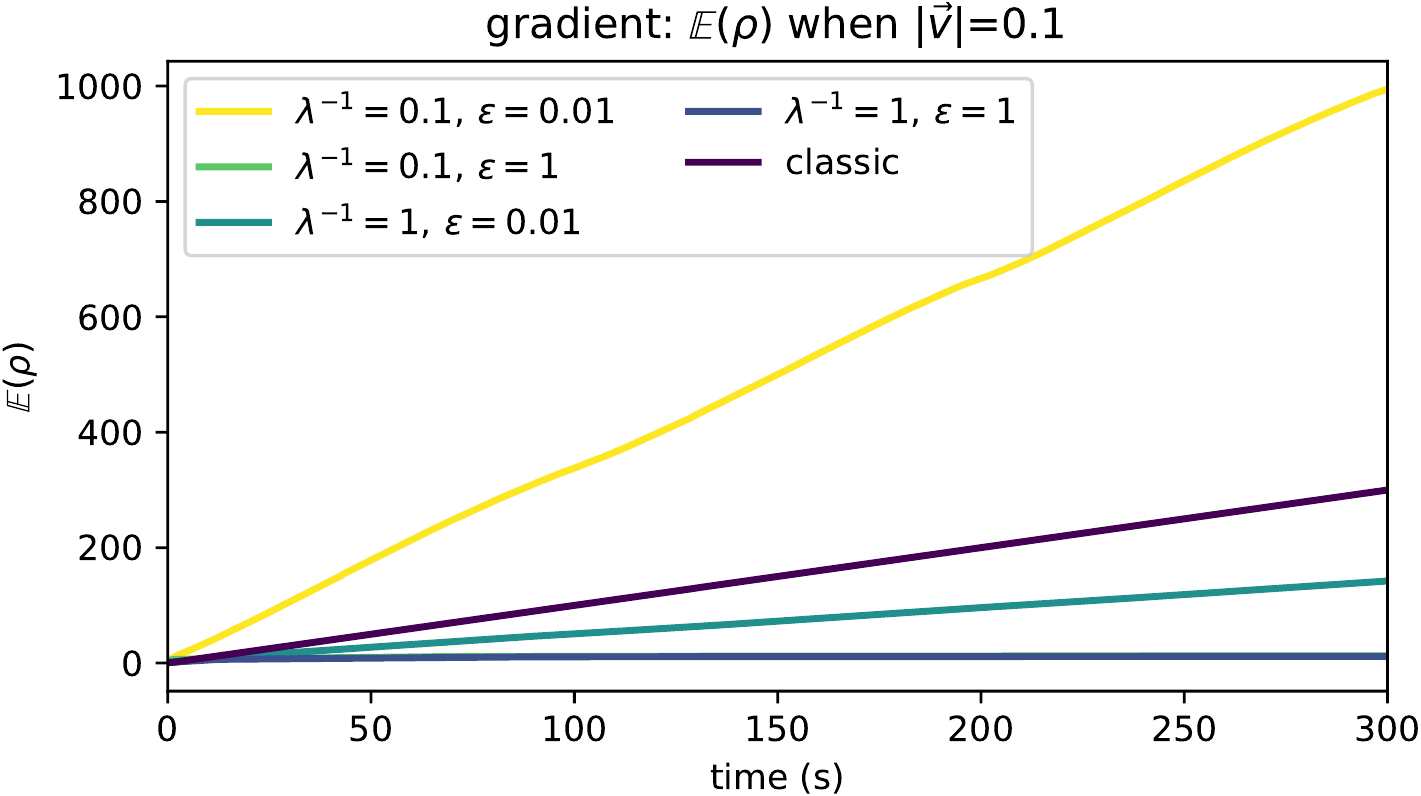}\\
  \vspace{5pt}
  \includegraphics[width=.486\textwidth]{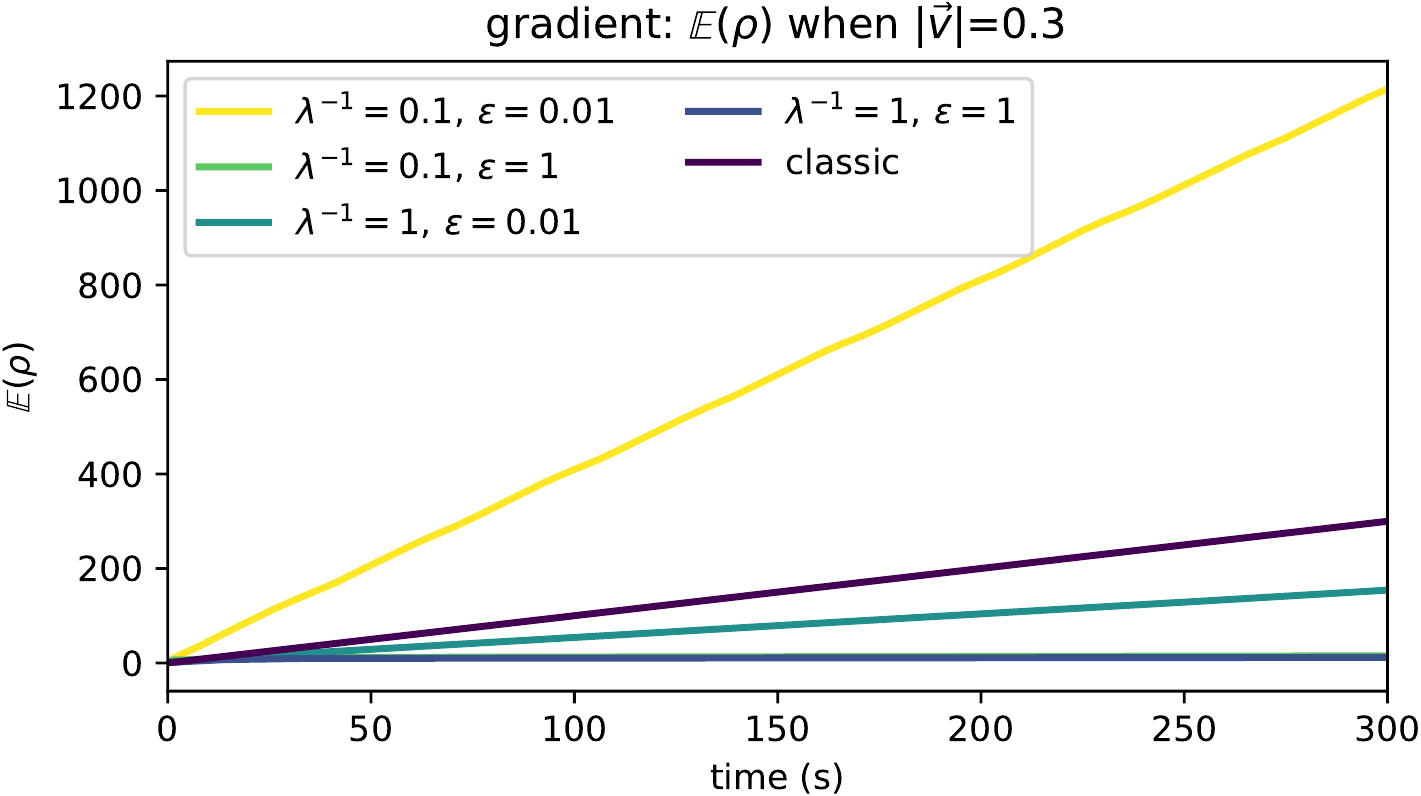}
  \includegraphics[width=.486\textwidth]{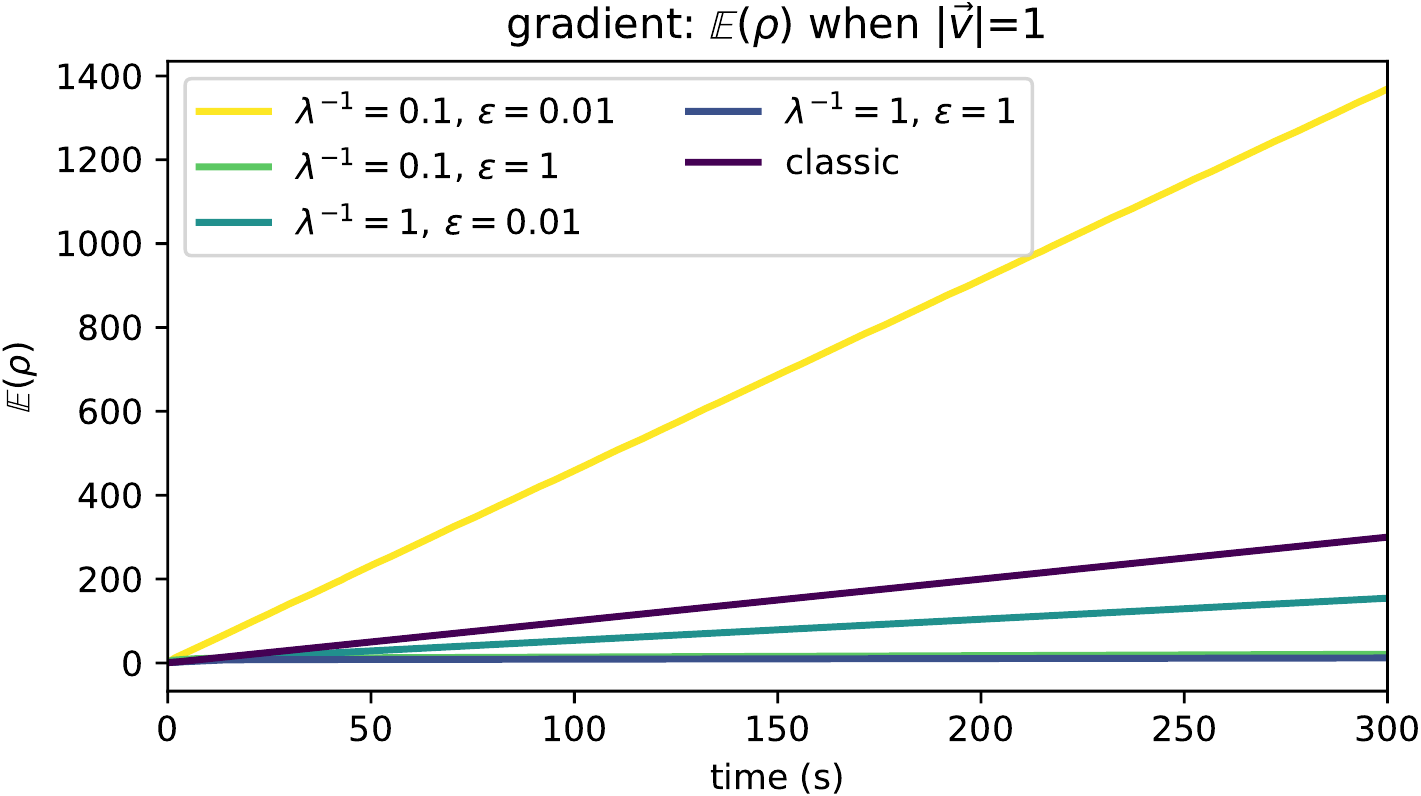}
  \\ \vspace{5pt}
  \includegraphics[width=.486\textwidth]{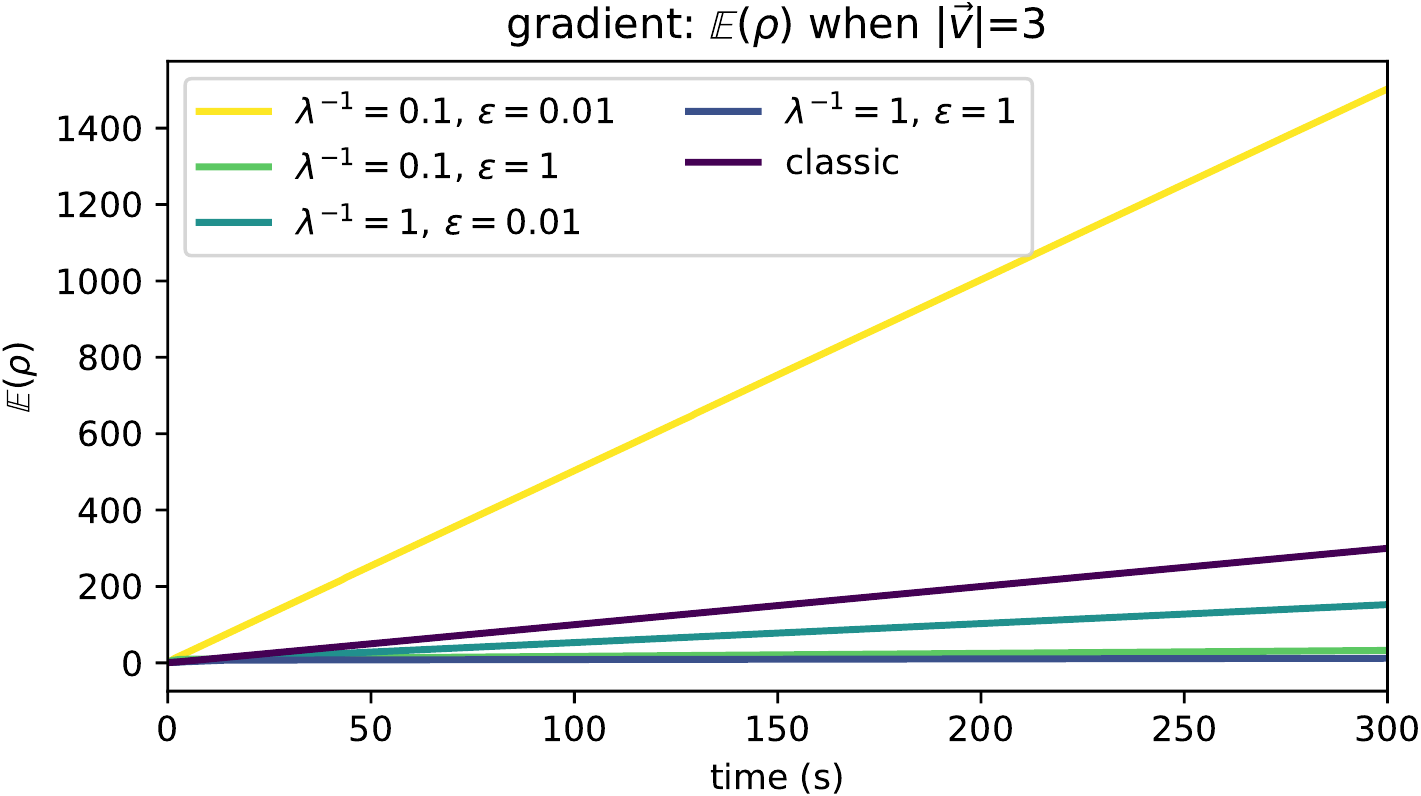}
 \end{center} 
 \caption{
Overall round count in the gradient scenario.
Both network performance and tolerance influence consistently the overall count of rounds performed by the time-fluid versions.
When $\|\vec{v}\| = 0$,
all time-fluid versions stop very quickly compared to the classic version.
Even low movement speeds impact noticeably on the count of rounds executed by the time-fluid version with low tolerance and a low-latency network ($\lambda^{-1}=0.01s$, $\epsilon = 0.01m$),
while both a larger tolerance and higher network latencies are very effective at trimming down the overall count of executed rounds.
Note that the scales on the y-axes differ.
}
 \label{chart:rounds:gradient}
\end{figure}

\begin{figure}[t]
 \begin{center}
  \includegraphics[width=.486\textwidth]{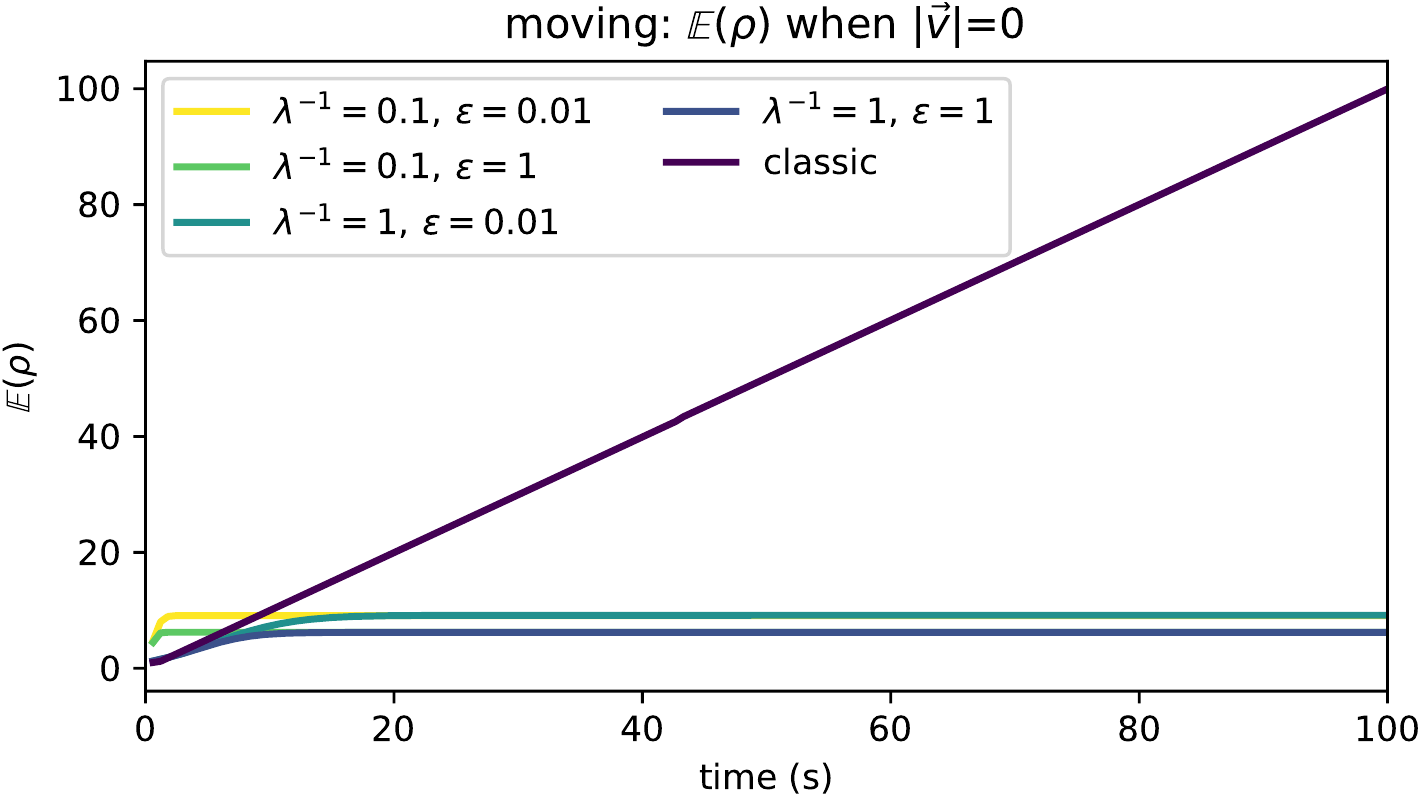}
  \includegraphics[width=.486\textwidth]{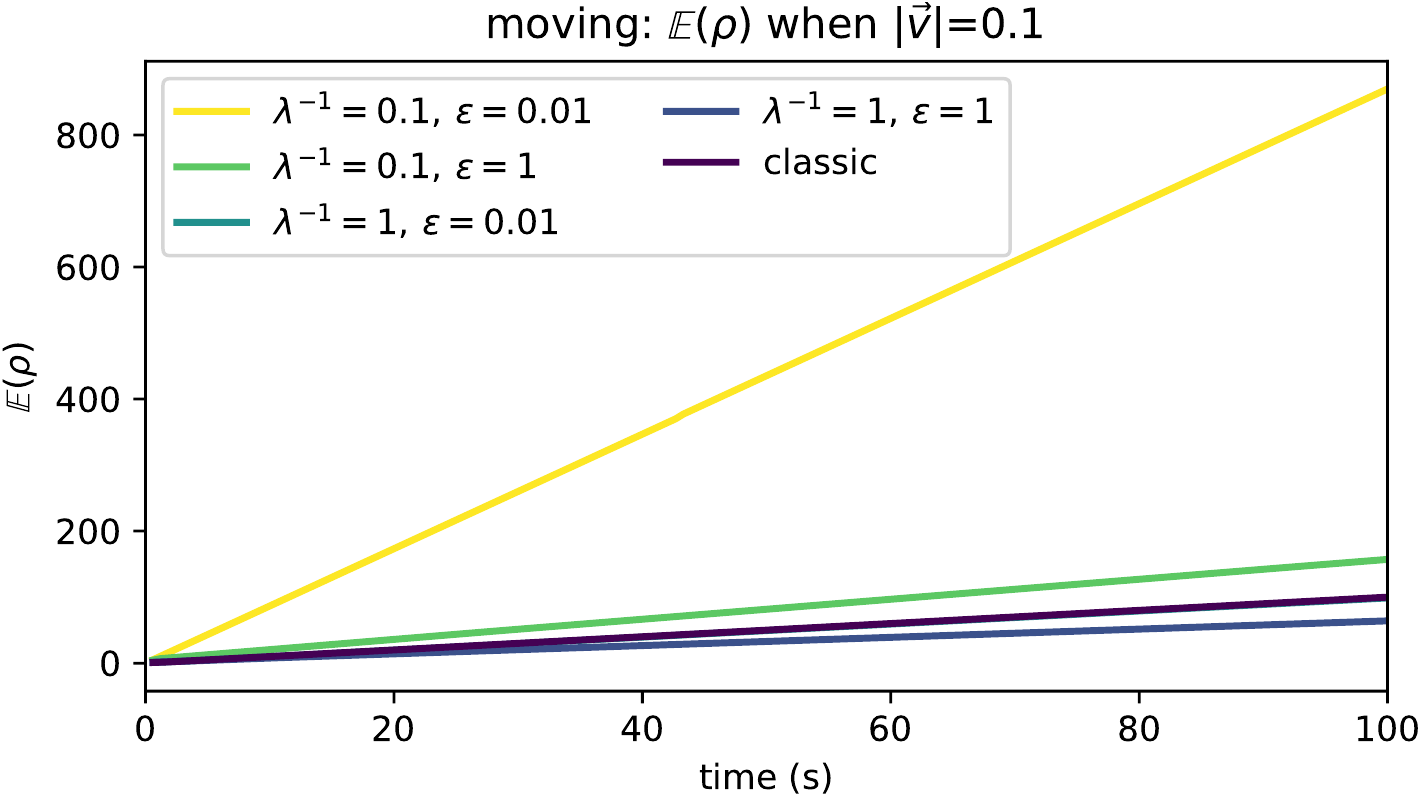}\\
  \vspace{5pt}
  \includegraphics[width=.486\textwidth]{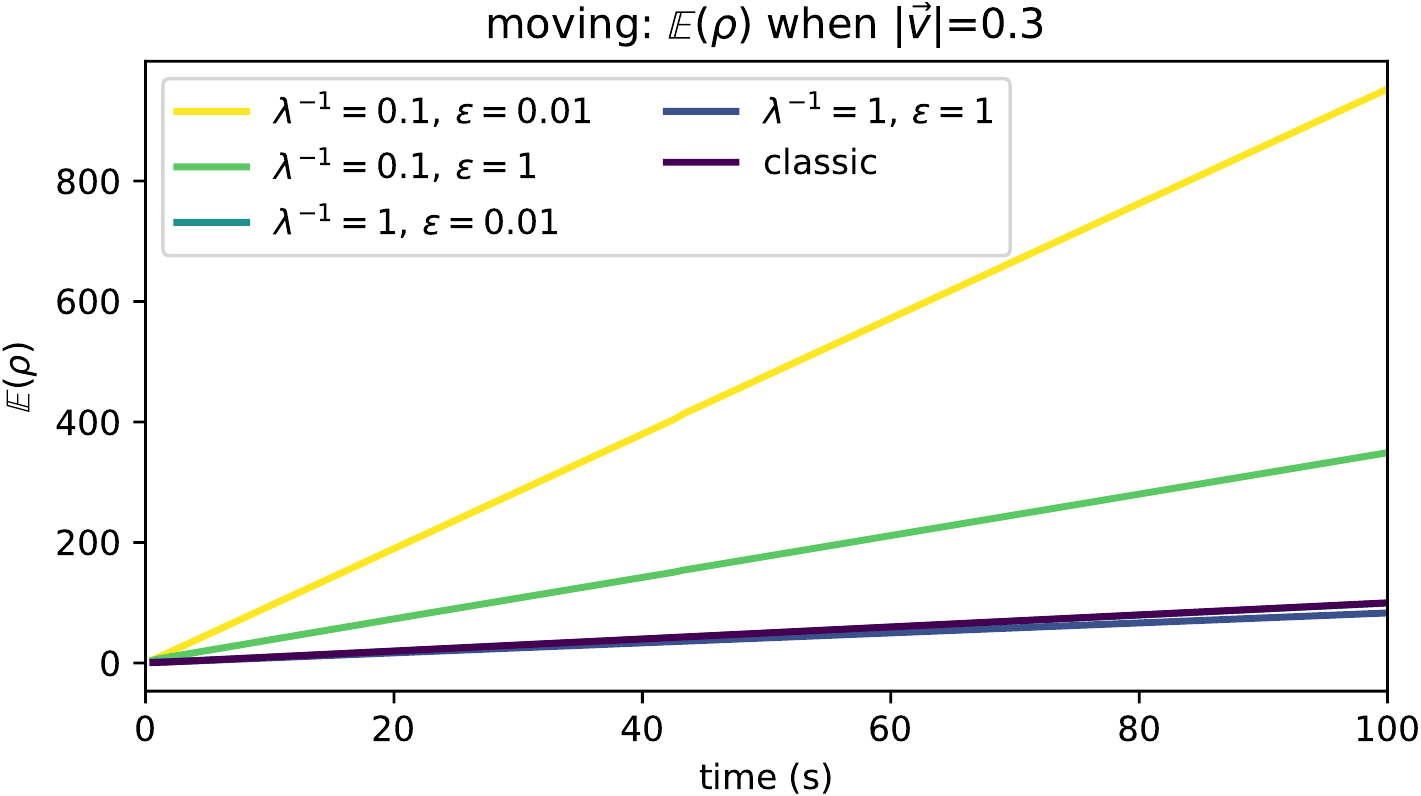}
  \includegraphics[width=.486\textwidth]{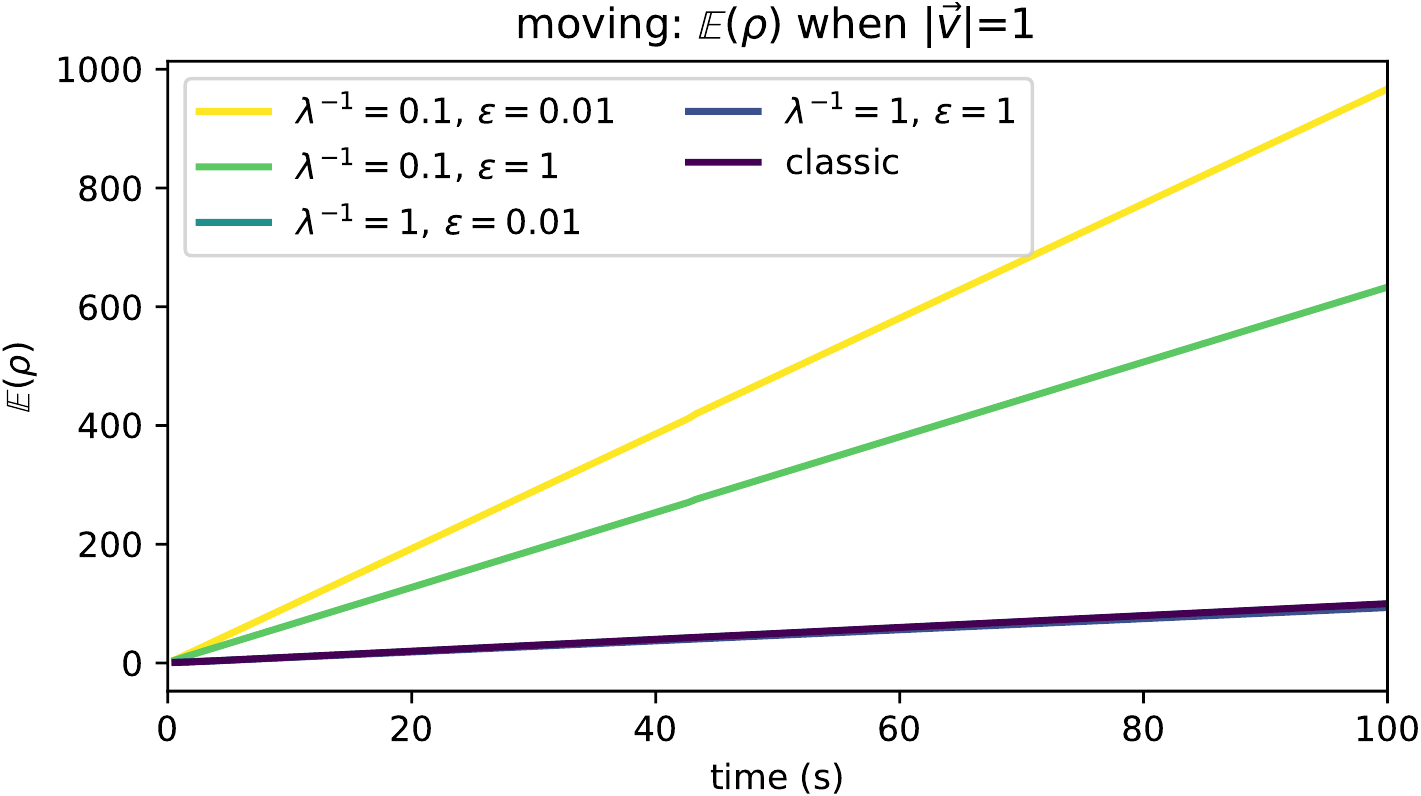}
  \\ \vspace{5pt}
  \includegraphics[width=.486\textwidth]{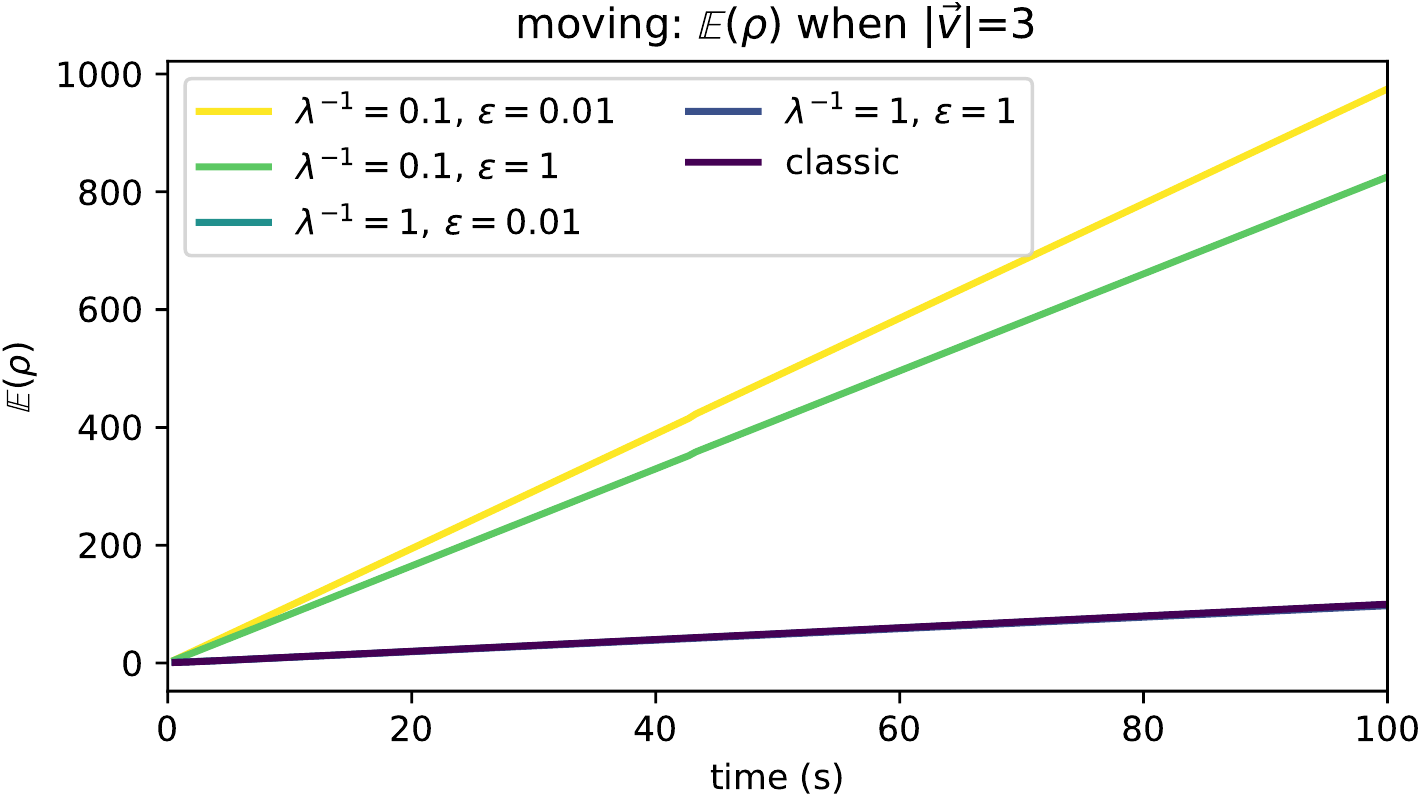}
 \end{center} 
 \caption{
Overall round count in the moving scenario.
Both network performance and tolerance influence consistently the overall count of rounds performed by the time-fluid versions.
When $\|\vec{v}\| = 0$,
all time-fluid versions stop after a transient whose length is shorter for low-latency networks (lower values of $\lambda^{-1}$)
and whose overall round count depends on the accepted error (tolerance $\epsilon$).
In this scenario, network latency is the factor that mostly influences the overall round count for time-fluid versions,
as global widespread movement with higher speed quickly makes devices get past the fixed tolerance threshold.
Note that the scales on the y-axes differ.
}
 \label{chart:rounds:moving}
\end{figure}

\begin{figure}[t]
 \begin{center}
  \includegraphics[width=.486\textwidth]{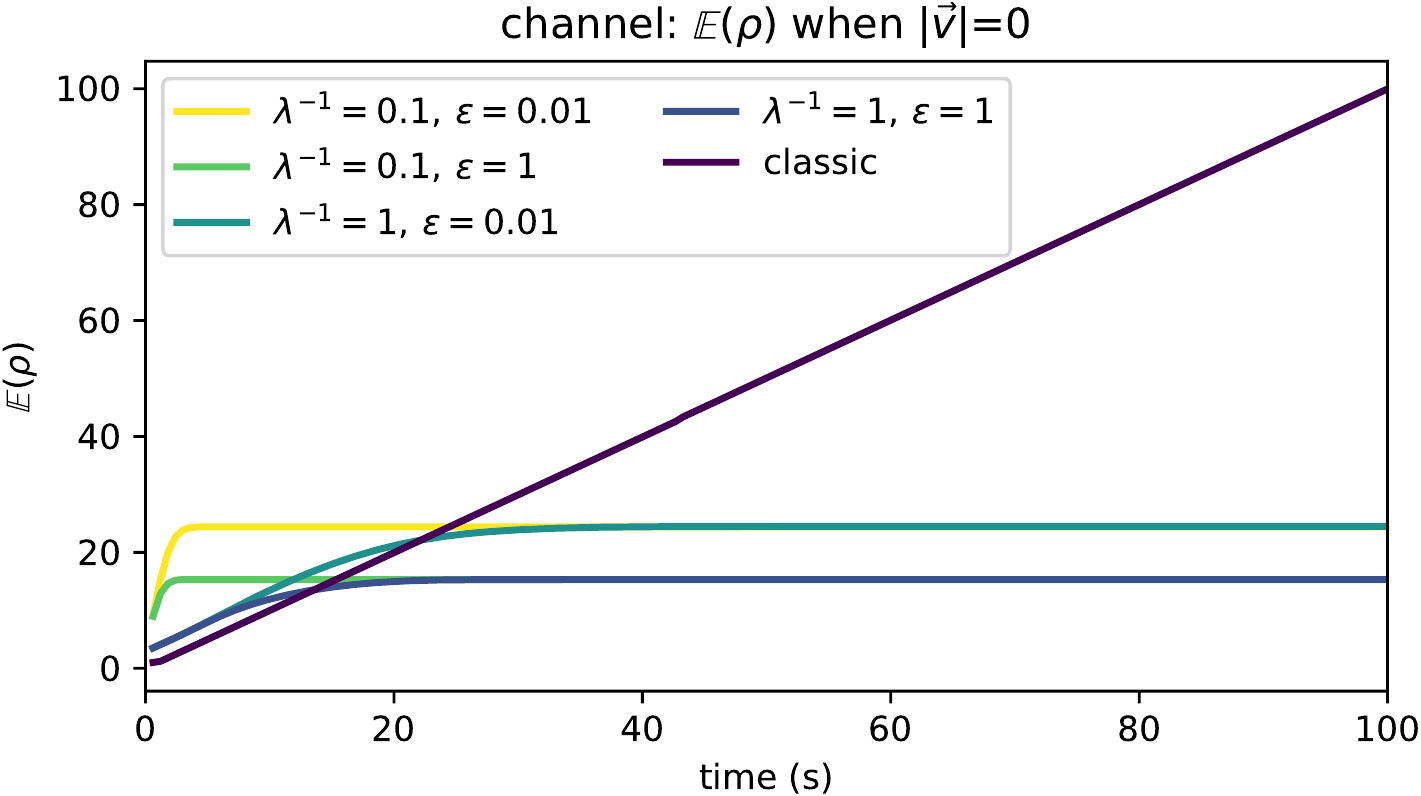}
  \includegraphics[width=.486\textwidth]{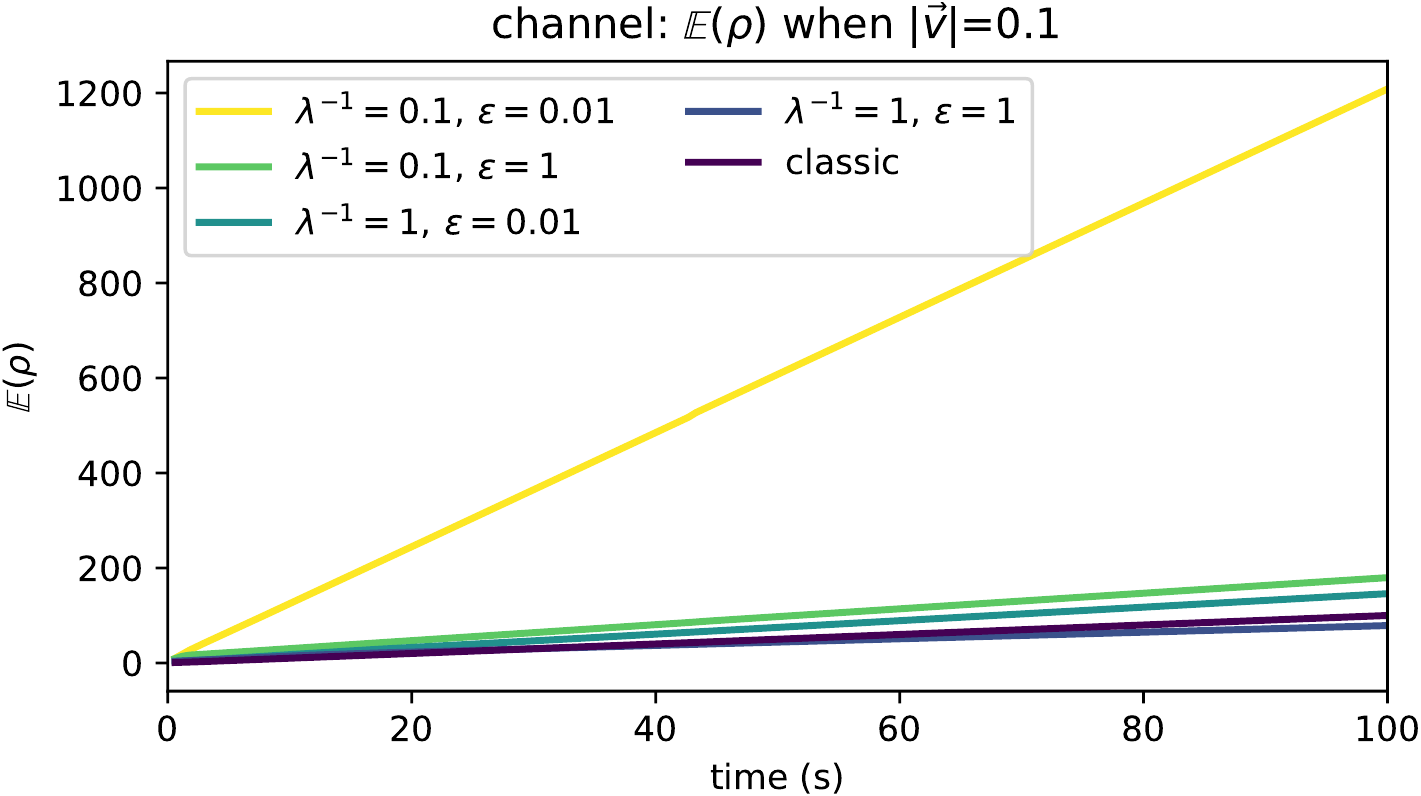}\\
  \vspace{5pt}
  \includegraphics[width=.486\textwidth]{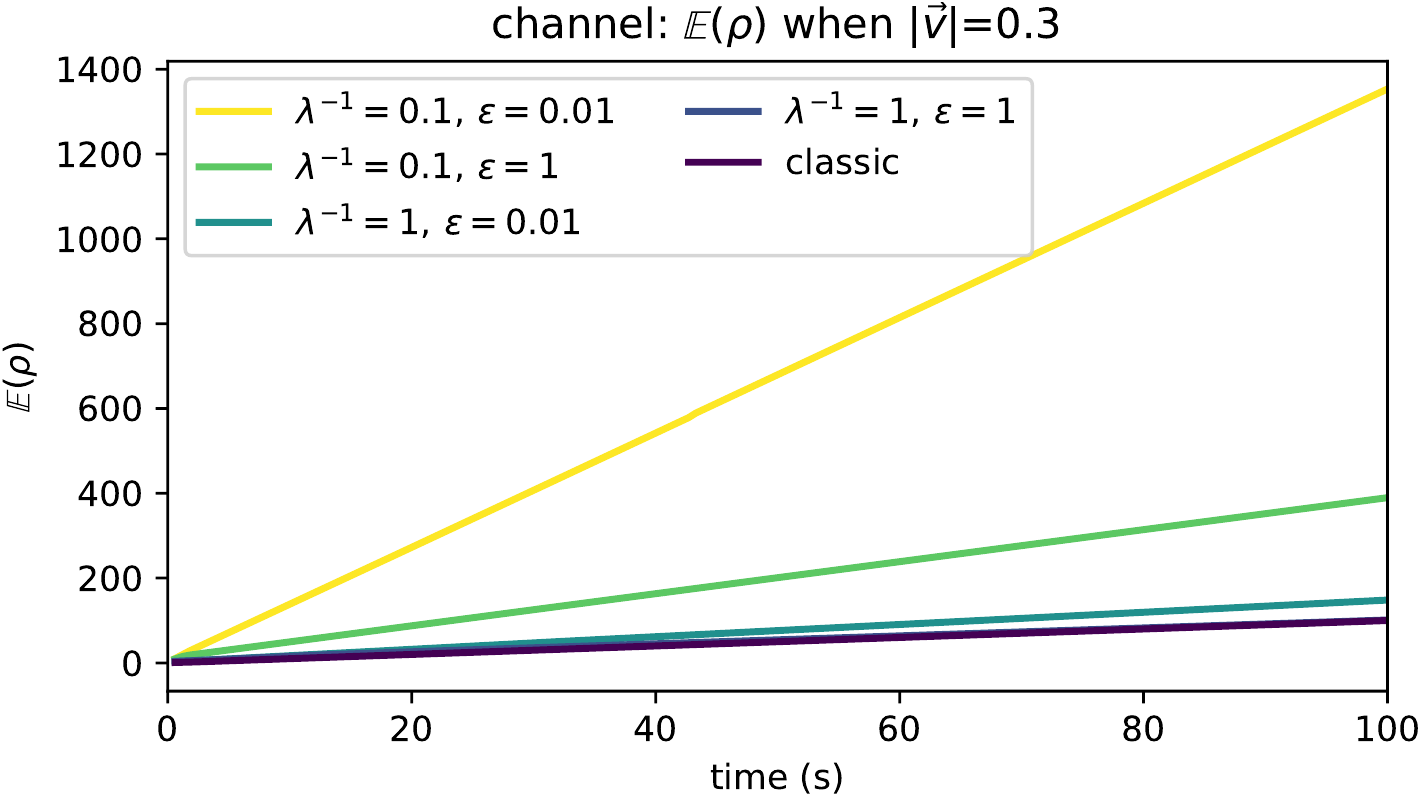}
  \includegraphics[width=.486\textwidth]{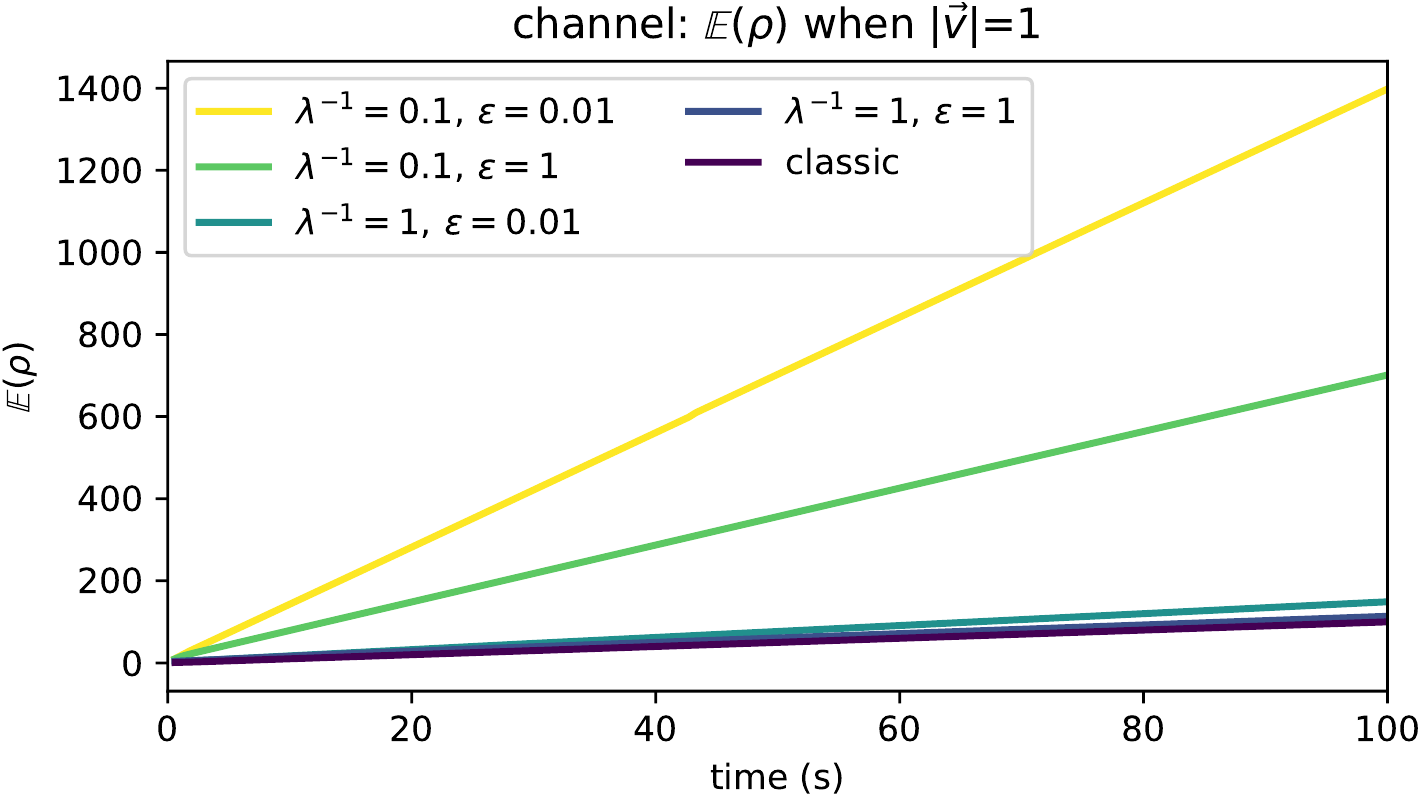}
  \\ \vspace{5pt}
  \includegraphics[width=.486\textwidth]{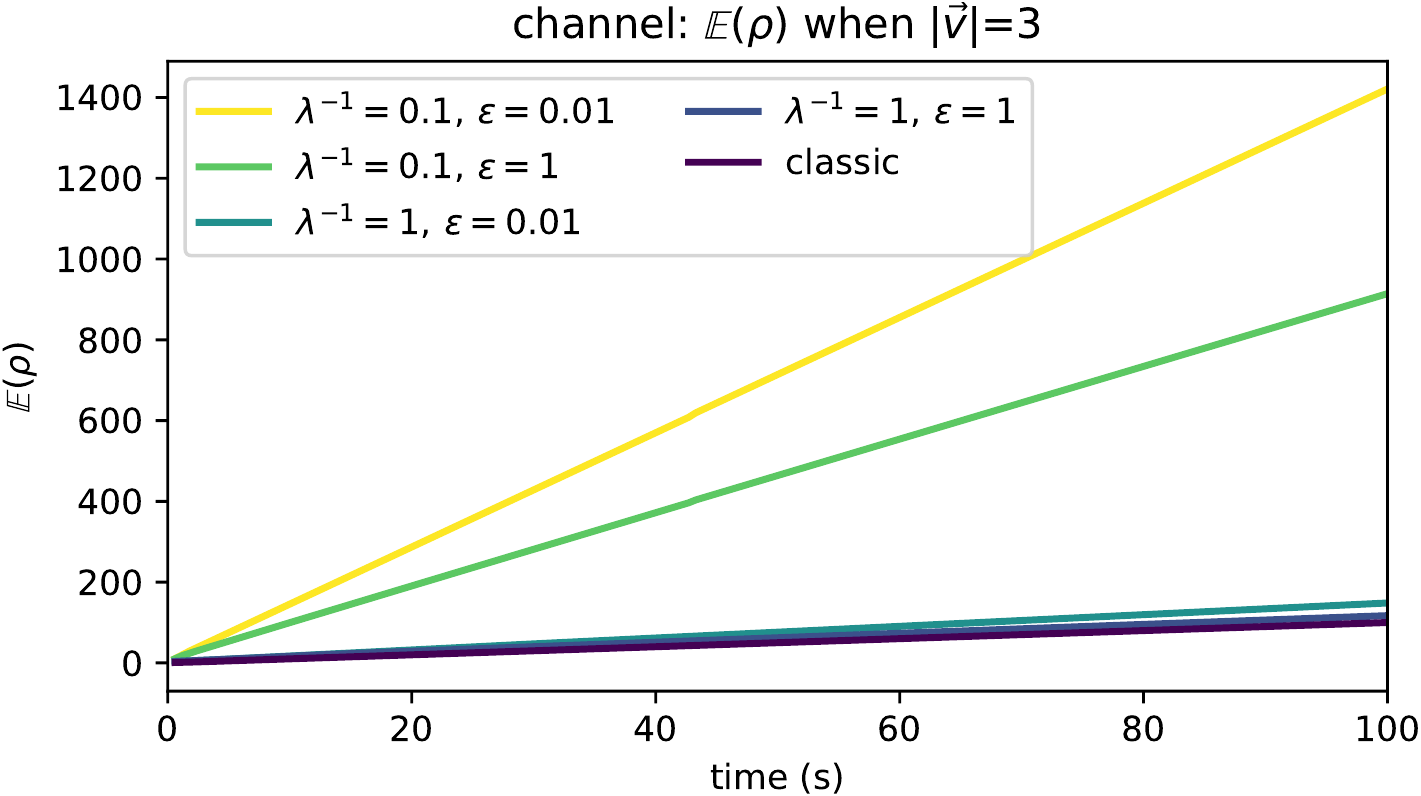}
 \end{center} 
 \caption{
Overall round count in the channel scenario.
Both network performance and tolerance influence consistently the overall count of rounds performed by the time-fluid versions.
When $\|\vec{v}\| = 0$,
all time-fluid versions stop after a transient whose length is shorter for low-latency networks (lower values of $\lambda^{-1}$)
and whose overall round count depends on the accepted error (tolerance $\epsilon$).
Note that the scales on the y-axes differ.
 }
 \label{chart:rounds:channel}
\end{figure}

Pure performance, however, is only a part of the story.
\Cref{chart:efficiency:gradient}, \Cref{chart:efficiency:moving}, and \Cref{chart:efficiency:channel}
show our efficiency proxy metric results for, respectively, the gradient, moving, and channel scenarios.
This simple metric is obtained by multiplicating the cumulative error the system produced for the duration
of its operation ($\int_0^t\mathbb{E}(\delta)\,dt\cdot\mathbb{E}(\rho)$) by the total number of rounds performed up to that time
($\mathbb{E}(\rho)$).
This metric provides an estimate of the operating cost relative to the precision required by the system.
To better understand how much of the overall cost metric is due to the executed round,
we also present separated results for the total number of rounds perfomed ($\mathbb{E}(\rho)$)
in \Cref{chart:rounds:gradient}, \Cref{chart:rounds:moving}, and \Cref{chart:rounds:channel} for the gradient, moving, and channel scenarios respectively.
Data shows that the time-fluid version of the field coordination system is generally more efficient
than the time-driven counterpart.
It can be more expensive to maintain,
but this price delivers lower error;
and on the other hand, it can perform comparably, but with a much lower cost.
This is especially true in cases in which there are localised changes whose propagation is limited.
In fact, while there are circumstances in the moving and channel experiments when
the classic version is more efficient than some configurations of the time-fluid ones,
in the case of the gradient it is consistently more efficient no matter how the conditions are changed.
The reason for this behaviour are to be found in the ability of the time-fluid version
to shut down the computation where there is no need of updates.
\FloatBarrier
\begin{figure}[ht]
 \begin{center}
  \includegraphics[width=.243\textwidth]{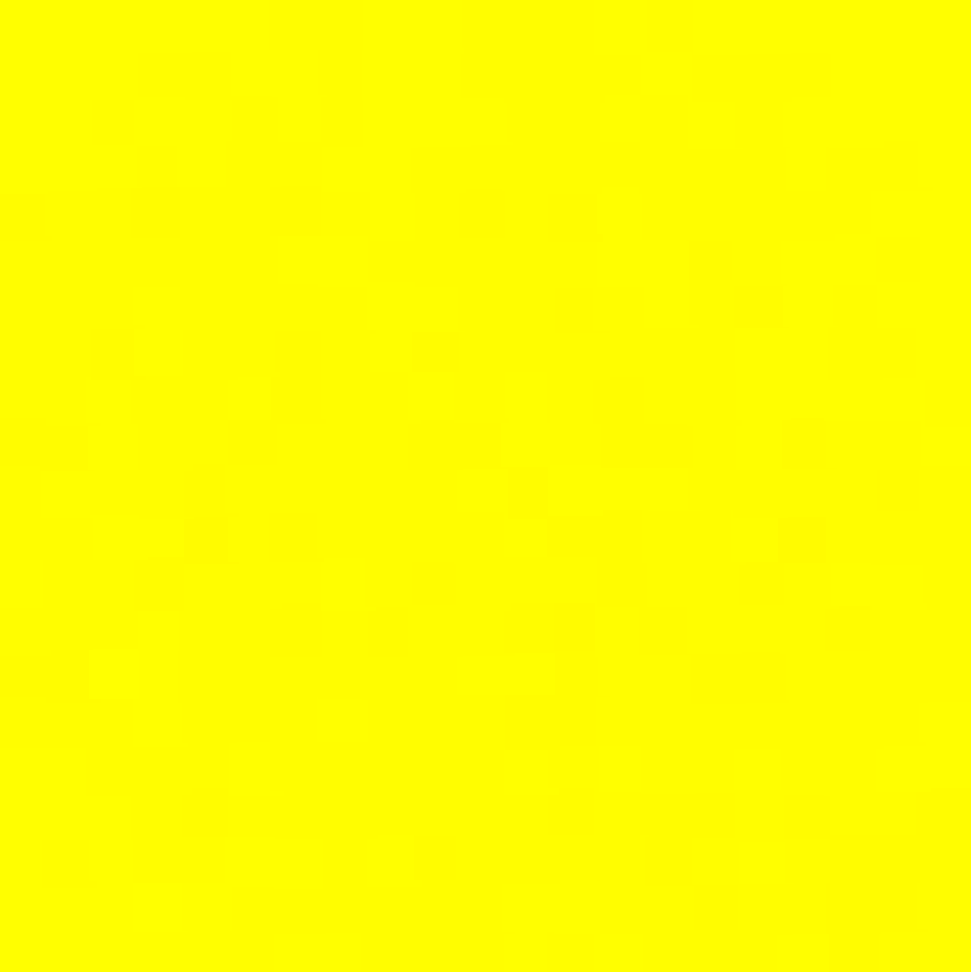}
  \includegraphics[width=.243\textwidth]{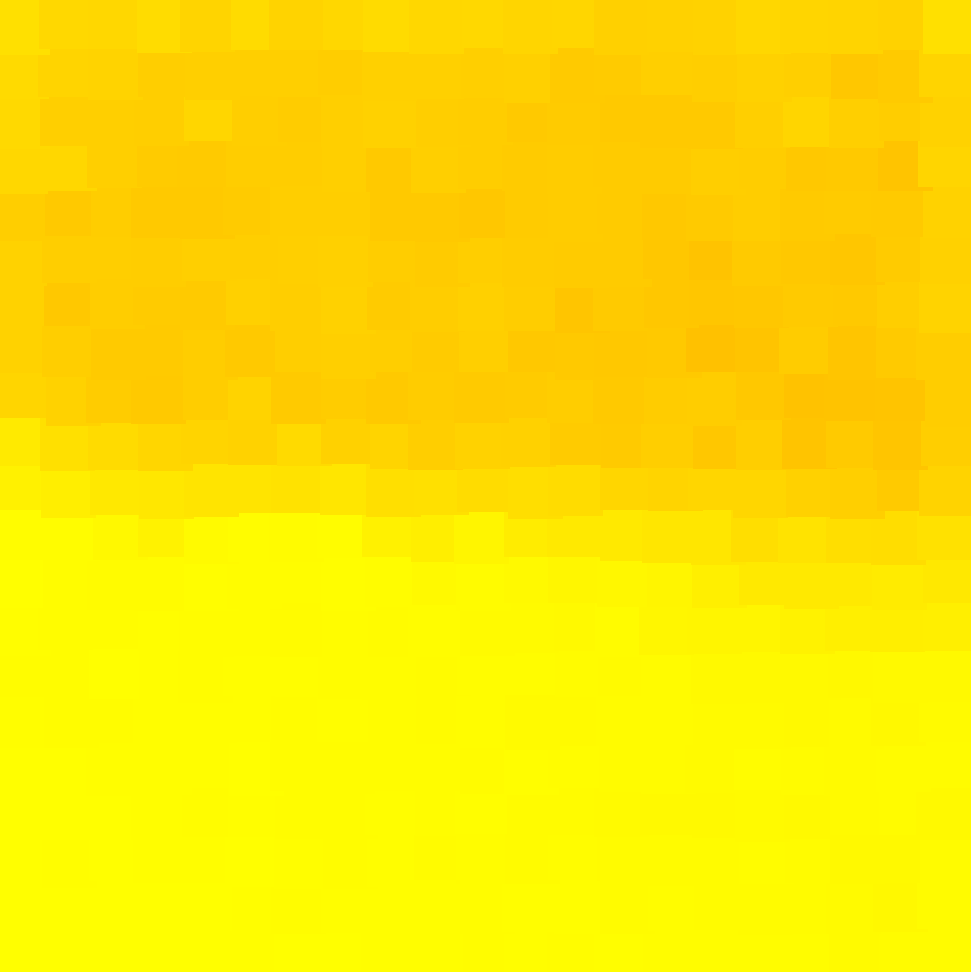}
  \includegraphics[width=.243\textwidth]{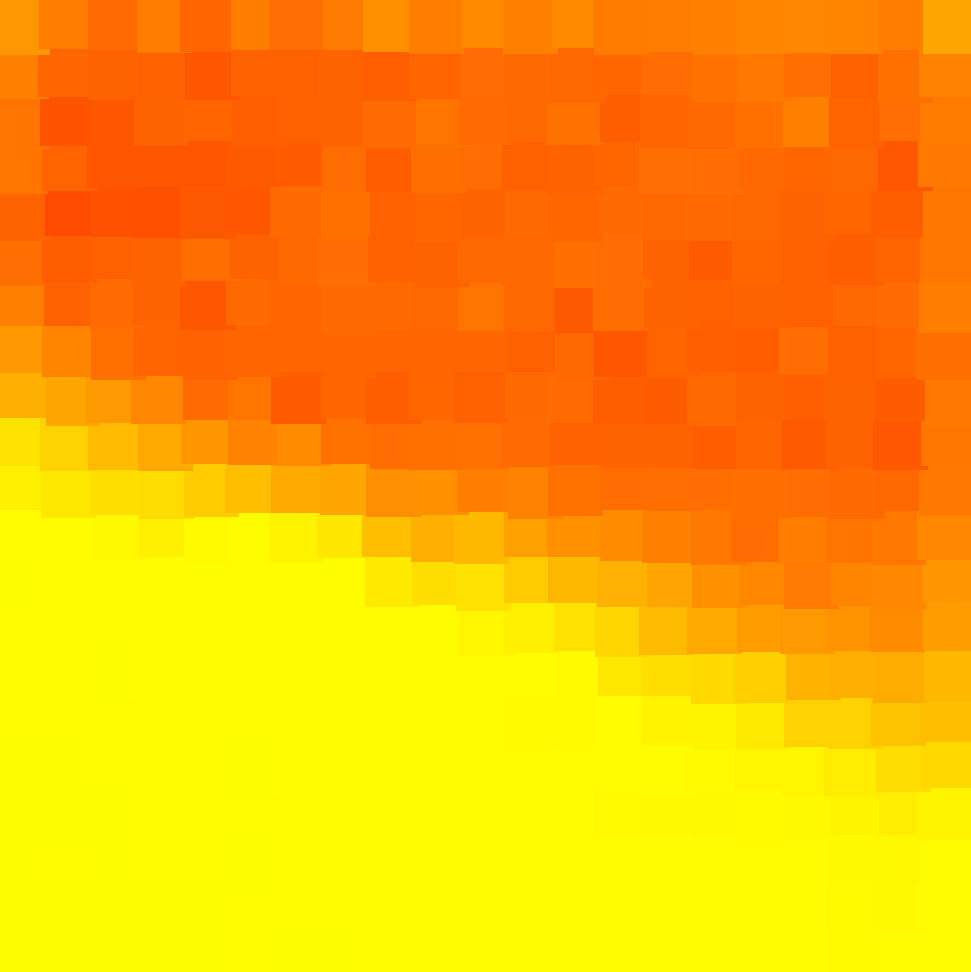}
  \includegraphics[width=.243\textwidth]{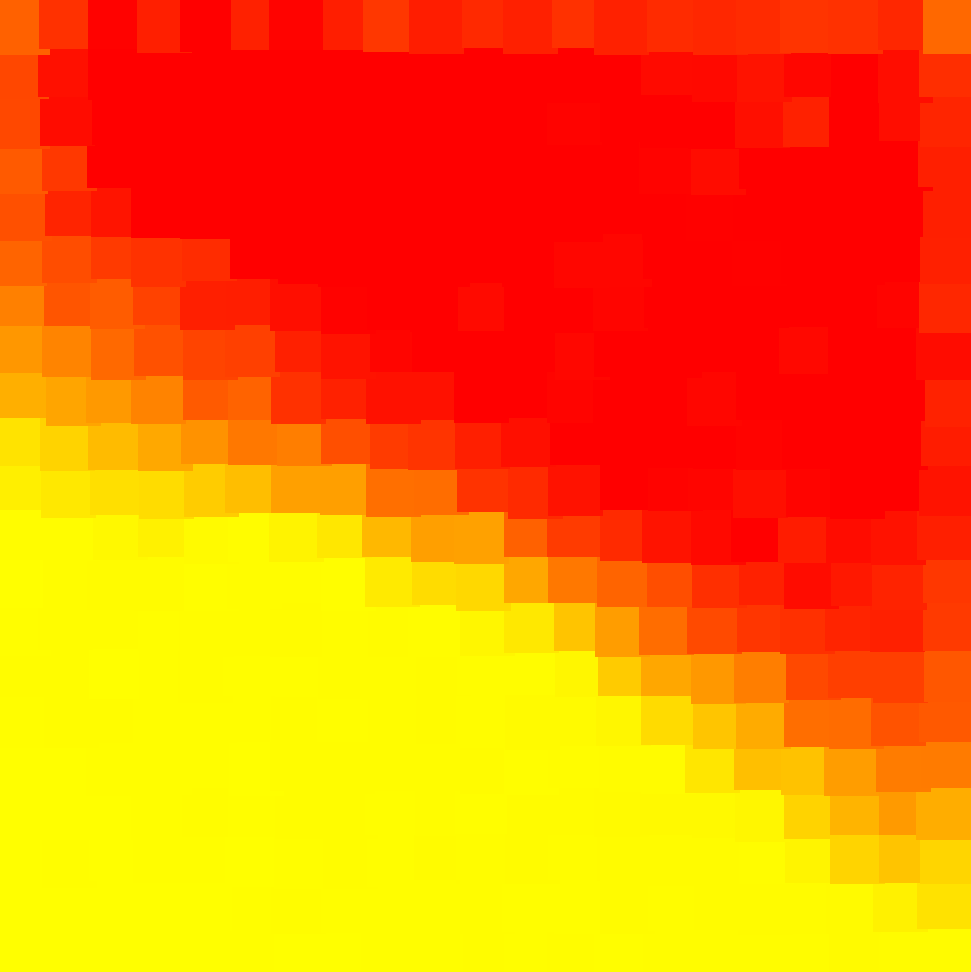}
 \end{center}
 \caption{
 Heat-map representation of the number of executed rounds along time.
 Each device is depicted as a point located on its actual coordinates.
 Time progresses from left to right.
 Devices start (left) with no round executed (yellow) and,
 with the simulation progression (left to right),
 execute rounds, changing their color to red.
 Devices closer to the static source (on the bottom left of the scenario)
 execute fewer rounds than those closer to the moving source,
 hence saving resources.
 }
 \label{heatmap}
\end{figure}
This is especially clear by looking at \Cref{heatmap},
which depicts with increasingly red colour the devices in the grid
based on the count of rounds they performed:
only those devices that come to get closer to the moving source than
to the static sources become red, while others, once stabilised,
stop their computation.

\subsection{Impact of tolerance to change and network latency}
\begin{figure}[thb]
 \begin{center}
  \includegraphics[width=.325\textwidth]{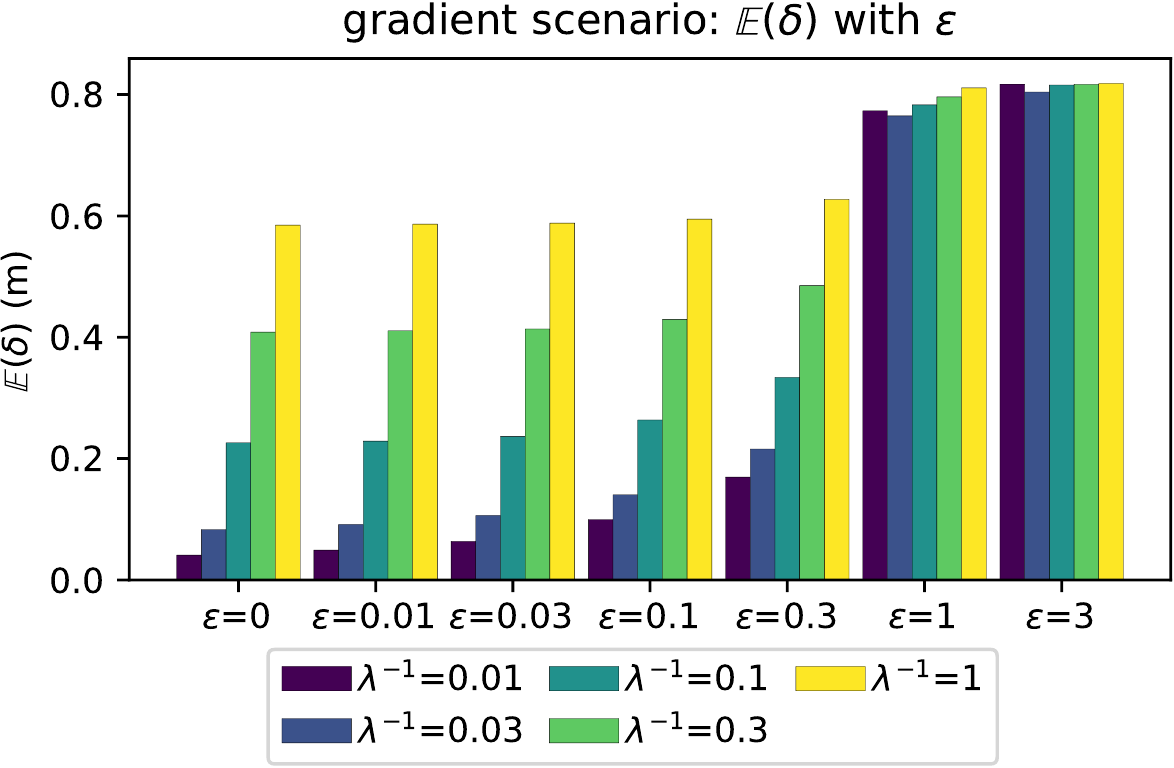}
  \includegraphics[width=.325\textwidth]{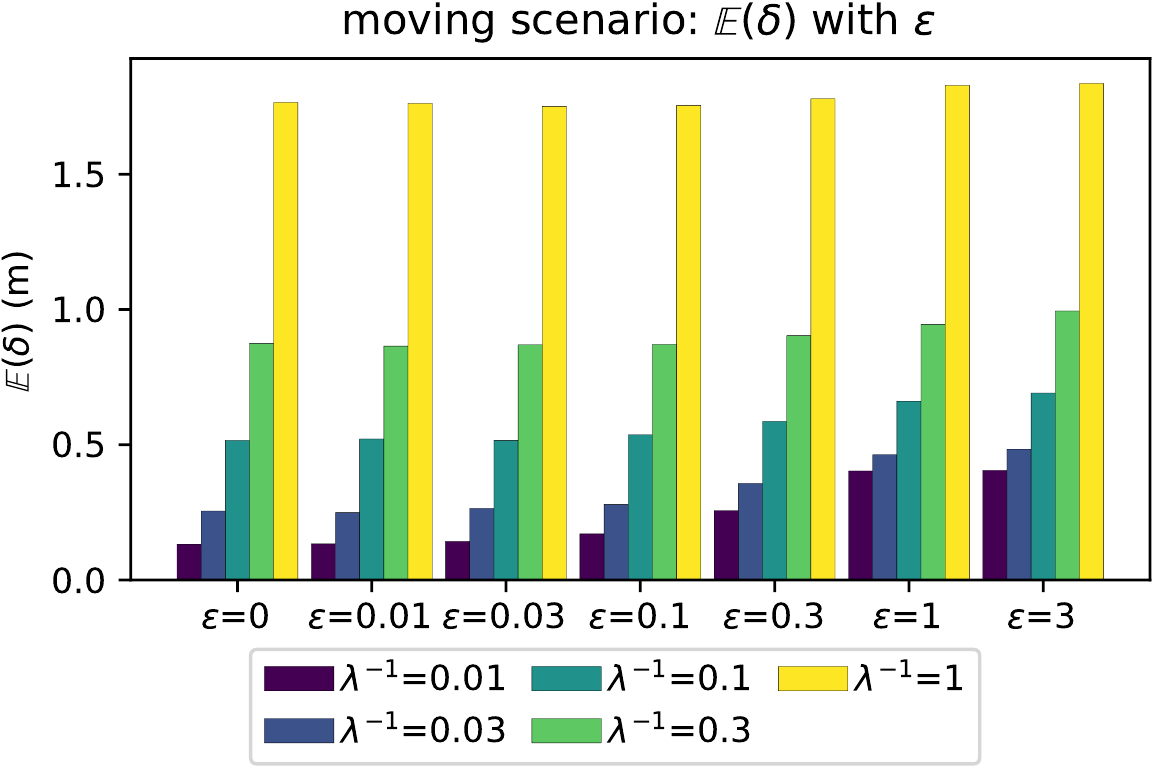}
  \includegraphics[width=.325\textwidth]{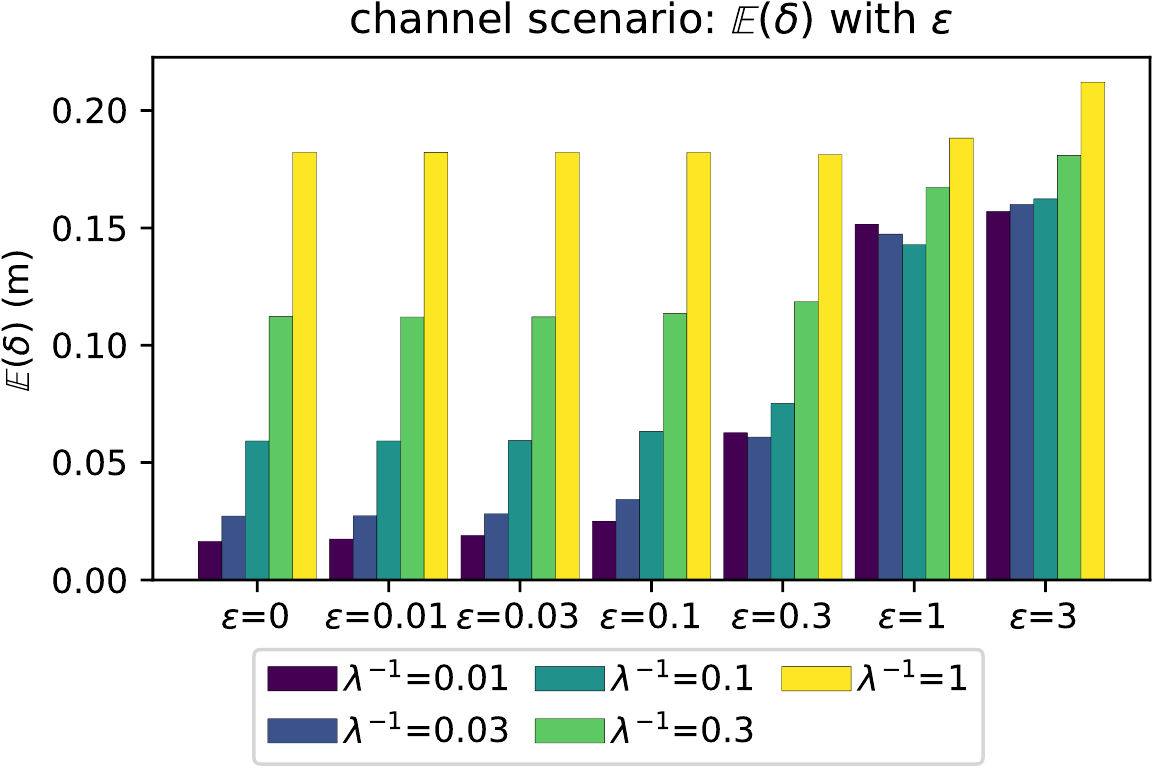}\\
  \vspace{10pt}
  \includegraphics[width=.325\textwidth]{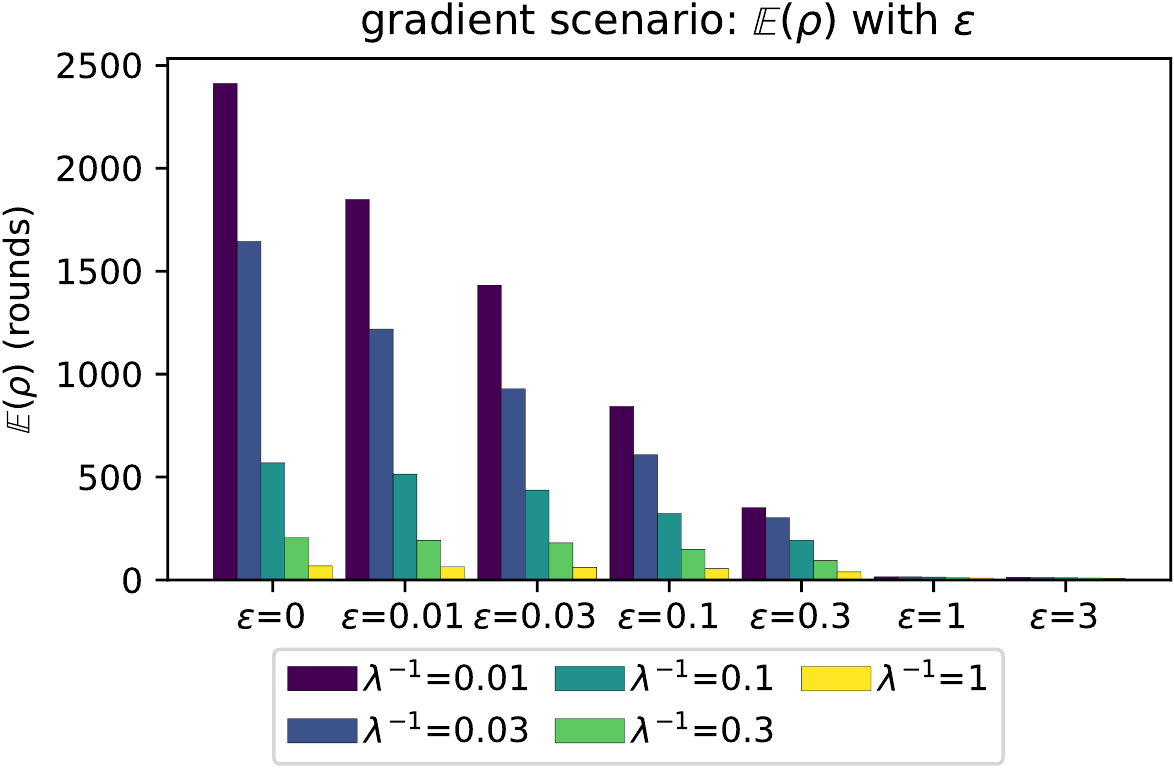}
  \includegraphics[width=.325\textwidth]{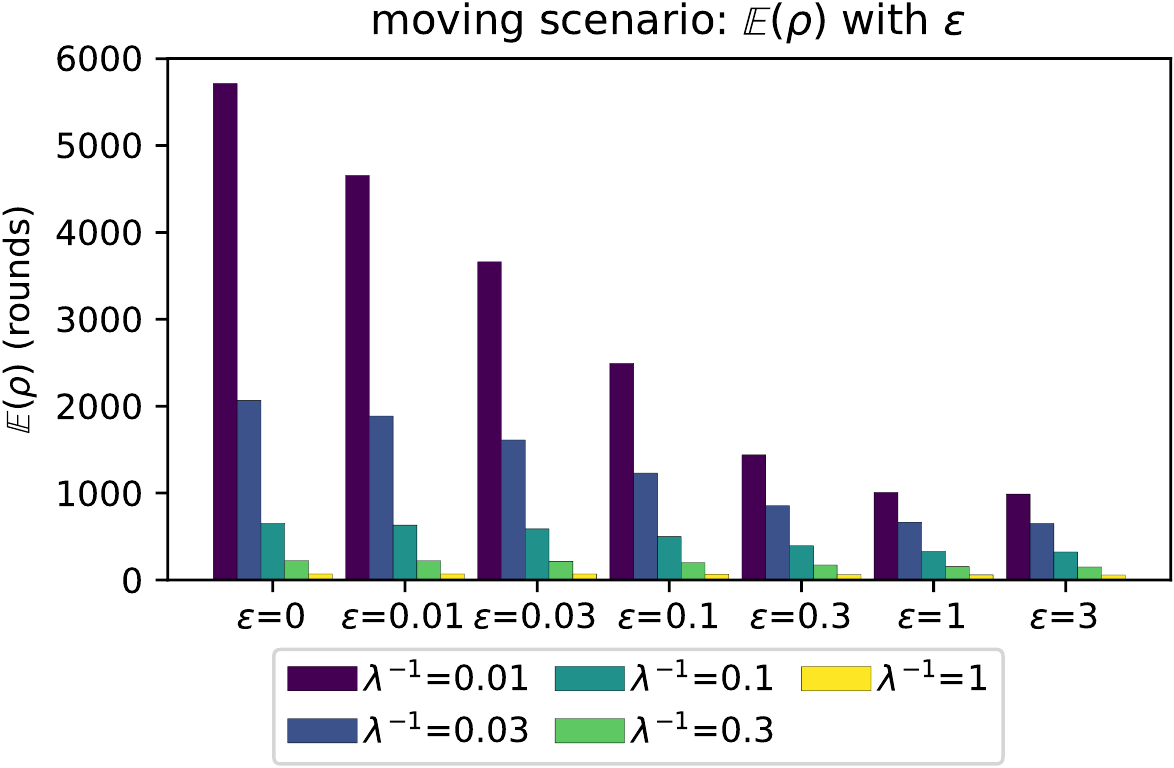}
  \includegraphics[width=.325\textwidth]{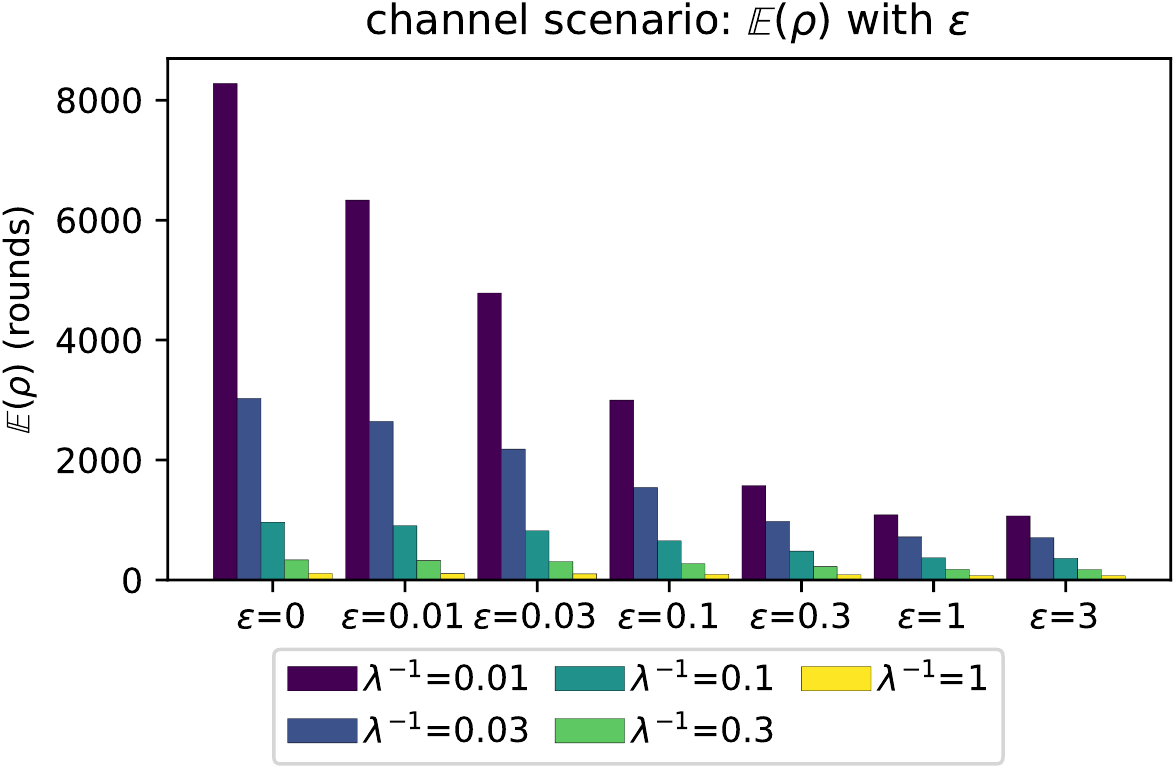}\\

  \vspace{10pt}
  \includegraphics[width=.325\textwidth]{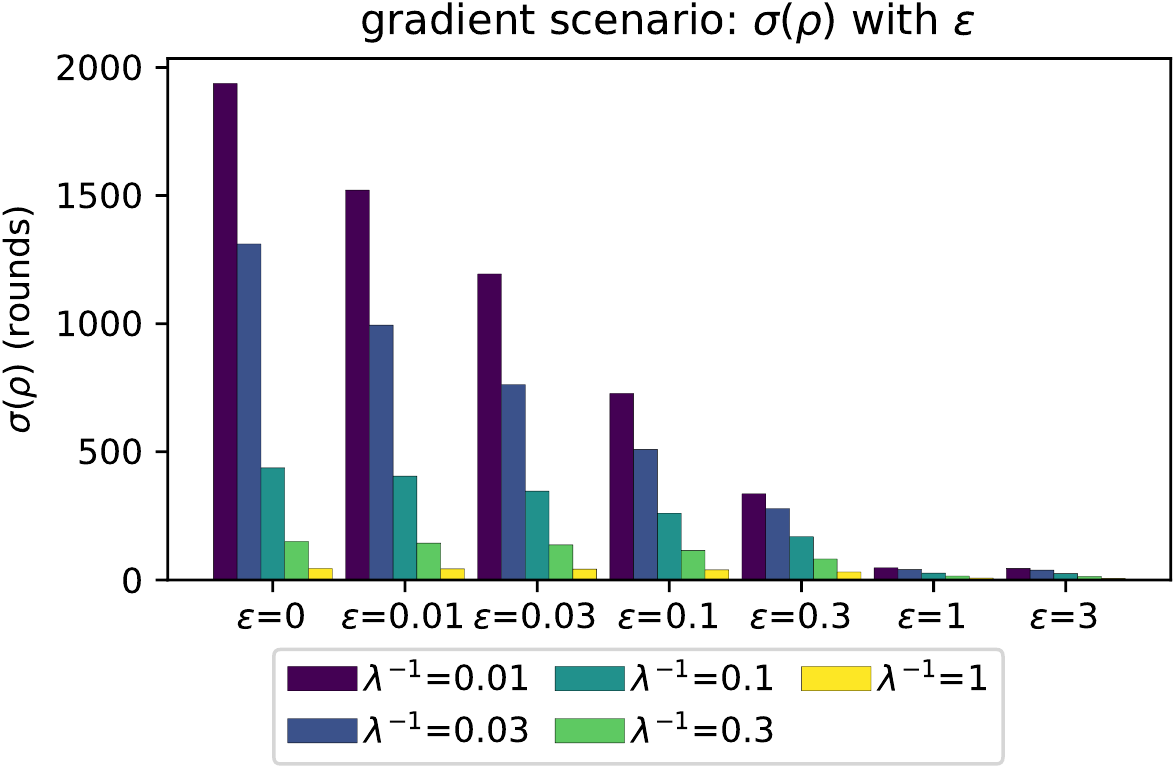}
  \includegraphics[width=.325\textwidth]{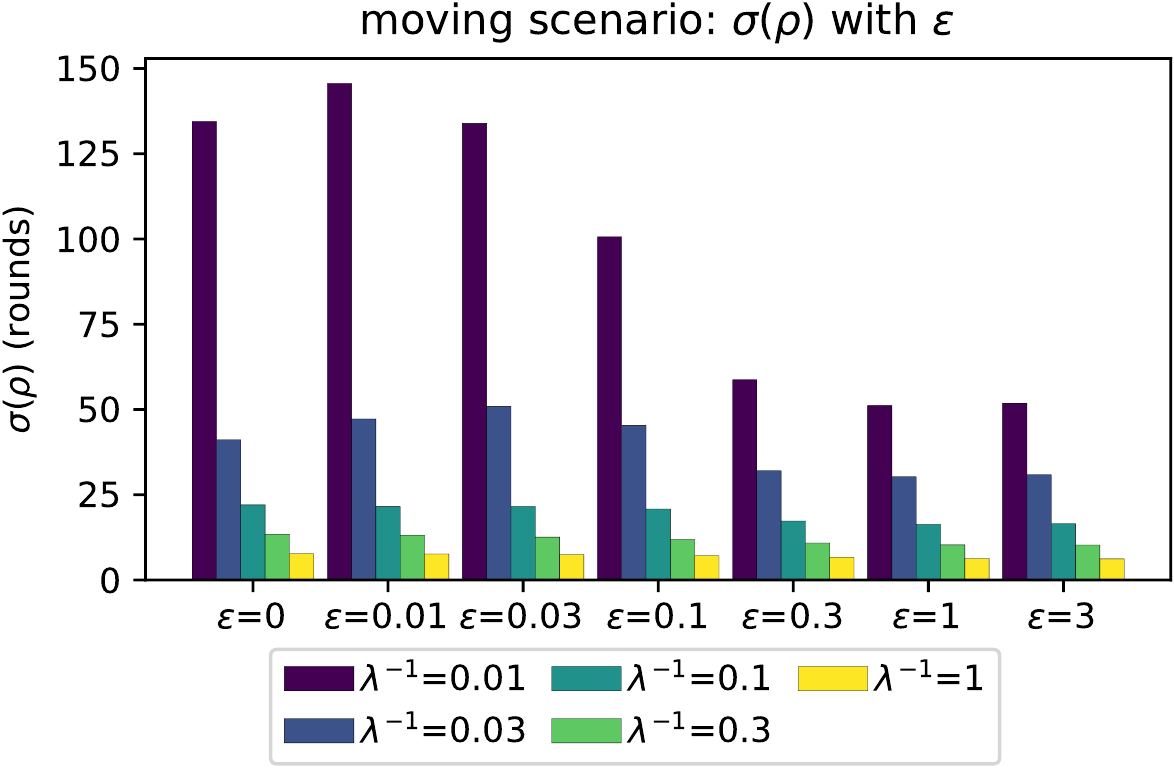}
  \includegraphics[width=.325\textwidth]{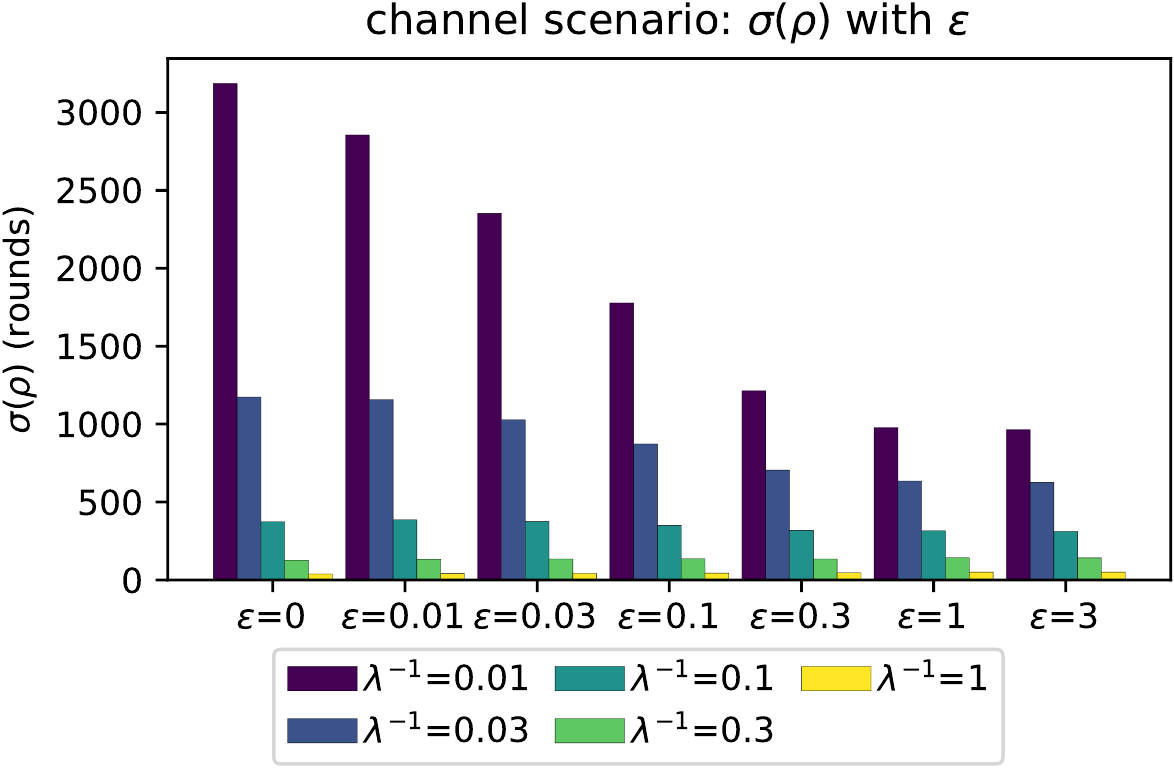}
 \end{center}
 \caption{
Time-fluid algorithms comparison for different values of
tolerance $\epsilon$ (abscissa)
and mean network delivery time $\lambda^{-1}$ (colour, the darker the lower).
The bar charts show error (top line, lower values are better),
cost (middle line, lower values are better),
and standard deviation of cost (bottom line, higher standard deviations indicate larger differences between the round frequencies of different devices);
gradient, moving, and channel scenarios are depicted respectively in the first, second, and third column.
Values are averaged along all the values of $\|\vec{v}\|$ under test.
 }
 \label{barcharts}
\end{figure}

Data showing the overall impact on the time-fluid versions of the algorithms for the scenarios under test
are depicted in \Cref{barcharts}.
Network latency and tolerance interplay in an interesting way:
the latter can be used to limit the ``reactivity'' of a system
in which a change may trigger a long chain of reactions.
This is especially clear by looking at the error for the gradient scenario:
at some point ($\epsilon \geq 1$) the error stabilises for every value of $\lambda$:
the tolerance is dominating the error.
The designer can thus design the scheduling of the field-based coordination
processes by considering what is the maximum level of reactivity that should be supported.
This level can be tuned using a form of tolerance similar to the one used in this work,
in order to obtain a system that follows the correct value as most as fast as
necessary in order to keep the error under a threshold defined by the system architect.
A similar consideration can be made by looking at the other side of the coin:
the ability of a system to promptly react to changes is only limited by the
physical slowdowns related to the network communication times
(or, from a more philosophically sound perspective, by the time required to events to be perceived).
The practical consequences of this fact are that a time-fluid system is able to
autonomously adapt to working into a different infrastructure: if there is no form
of tolerance to error included by the designer, then the time-fluid system will
potentially try to converge as quickly as possible.
Moreover, there is a large potential for time-fluid coordination systems to be successful
in saving global resources: since the system is able to stop computing where and when unnecessary,
it can lead to resource savings.
On the other hand, however, time-fluidity introduces \emph{asymmetry} in computation frequency:
some devices, located in ``hot spots'' of the computation, will need more resources than others,
and this might potentially complicate the prediction of the resource usage by each device.

\section{Conclusion and future work}\label{sec:conclusion}

In this work we developed a different way of conceptualising time in field-based coordination systems.
Inspired by causal models of space-time in physics,
we leverage a novel concept of \emph{causality field}
 and accordingly introduce a model where 
 field computations may be caused by
 other field computations
 or platform-level triggers capturing changes in the computation context.
We formalise the approach by extending the field calculus operational framework 
 with a notion of \emph{time-fluid scheduling}
 over a tree of causally-related field-calculus programs.
The model is implemented in the Alchemist-Protelis simulation framework, and an API is proposed to program
 the scheduling together with the application logic itself.
Finally, by means of simulation-based experiments,
 we show that the time-fluid approach
 provides significant performance and efficiency benefits
 with respect to the classical time-driven approach.
%
%
%

Future work will be devoted to provide more in-depth insights by evaluating the impact of the approach
in realistic setups, both in terms of scenarios (e.g. using real world data) and evaluation precision
(e.g. by leveraging network simulators such as Omnet++ or NS3).
%
%
The relationship between adaptive scheduling and adaptive deployment is also interesting,  in particular when considering dynamic architectures and ``pulverisation'' approaches for application partitioning~\cite{DBLP:conf/soco/BozgaJMS12,casadei2020pulverization}.

On a more foundational perspective of aggregate computing and field-based coordination, there are several research direction.
First, it would be interesting to investigate how
 time-fluid scheduling approach relates
 to aggregate processes~\cite{EAAI2020-processes},
 a recent construct proposal for modelling dynamic numbers of concurrent
 field computations which dynamically spread in the system.
Second, it is interesting to consider the possibility of supporting time-fluidness purely at the linguistic level, by a scheduling construct properly defining the domain of events of another field computation.
Finally, other than time-fluidness one shall consider space-fluidness as well, that is, the possibility of defining computations that affect the shape of perceived space (which devices are to be considered neighbours), with the goal of addressing performance of communication but also to opportunistically cover space in an effective way.

\subsection*{Acknowledgements}

This work has been supported by the MIUR PRIN 2017 Project N. 2017KRC7KT ``Fluidware''.
This work has been partially supported by the ``Alma Attrezzature 2017'' University of Bologna Project ``Area 4.0''.

\bibliographystyle{alpha}
\bibliography{bibliography}

\end{document}